\documentclass[a4paper,fleqn]{cas-sc}

\usepackage[authoryear]{natbib}

\usepackage{graphicx, caption}

\usepackage{amsmath}
\usepackage{siunitx}
\usepackage{diagbox}
\usepackage{multirow}
\usepackage{rotating}
\usepackage{moresize}

\usepackage{hyperref}
\hypersetup{colorlinks=true, citecolor=blue, linkcolor=blue}
\usepackage{color}

\def \vv#1{\mathbf{#1}}

\def \degree {^{\circ}}
\def \inc {I}
\def \xxx {x}

\def \angrot {\omega}
\def \mmi {\overline{n}_1}     
\def \mmj {\overline{n}_2}     

\def \ii {{\rm i}}
\def \ep {{\rm e}}

\def \lambdai {\lambda_1}
\def \lambdaj {\lambda_2}
\def \Lambdai {\Lambda_1}
\def \Lambdaj {\Lambda_2}
\def \sigmaa {\sigma}
\def \Sigmaa {\Sigma}
\def \sigmai {\sigmaa_1}
\def \sigmaj {\sigmaa_2}
\def \sigmak {\sigmaa_k}
\def \Sigmai {\Sigmaa_1}
\def \Sigmaj {\Sigmaa_2}
\def \psia {\varphi}
\def \Psia {\Phi}
\def \psii {\psia_1}
\def \psij {\psia_2}
\def \psik {\psia_k}
\def \Psii {\Psia_1}
\def \Psij {\Psia_2}

\def \Ka    {\mathcal{K}_1}
\def \Kb    {\mathcal{K}_2}
\def \Oa    {\mathcal{O}_1}
\def \Ob    {\mathcal{O}_2}
\def \Oc    {\mathcal{O}_3}
\def \Od    {\mathcal{O}_4}
\def \Sa    {\mathcal{S}_1}
\def \Sb    {\mathcal{S}_2}
\def \Sc    {\mathcal{S}_3}
\def \Sd    {\mathcal{S}_4}
\def \Se    {\mathcal{S}_5}
\def \Sf    {\mathcal{S}_6}
\def \Sg    {\mathcal{S}_7}
\def \Ra    {\mathcal{R}_1}
\def \Rb    {\mathcal{R}_2}
\def \Rc    {\mathcal{R}_3}
\def \Rd    {\mathcal{R}_4}
\def \Re    {\mathcal{R}_5}
\def \Rf    {\mathcal{R}_6}

\def \cartxi    {x_1}
\def \cartxbi   {\overline{x}_1}
\def \cartxj    {x_2}
\def \cartxbj   {\overline{x}_2}
\def \cartxk    {x_k}
\def \cartxbk   {\overline{x}_k}
\def \cartyi    {y_1}
\def \cartybi   {\overline{y}_1}
\def \cartyj    {y_2}
\def \cartybj   {\overline{y}_2}
\def \cartyk    {y_k}
\def \cartybk   {\overline{y}_k}

\def \Energinion {\mathcal{H}}
\def \MedH {\bar \Energinion}
\def \SecH {\Energinion_0}

\def \vr {\vv{r}}
\def \ur {\hat \vr}

\def \Ru {R_0}
\def \Grav {\mathcal{G}}
\def \dt {\tau}

\def \klU {k_{2,0}}

\def \kli {k_{2,1}}
\def \klj {k_{2,2}}

\def \epsx {\varepsilon_\pm}

\def \thetid {\theta_\elli}
\def \vthetid {\vartheta_\elli}

\def \ij {{ij}}
\def \elli {j}

\def \Ak    {\mathcal{A}_k}
\def \Bk    {\mathcal{B}_k}
\def \Ck    {\mathcal{C}_k}
\def \Jk    {\mathcal{J}_k}

\begin{document}

\shorttitle{Dynamical evolution of the Uranian satellite system II}    

\shortauthors{S.R.A. Gomes \& A.C.M. Correia}  

\title [mode = title]{Dynamical evolution of the Uranian satellite system II.}  
\title [mode=sub]{Crossing of the 5/3 Ariel--Umbriel mean motion resonance}

\author[lab1]{S\'ergio R. A. Gomes}

\cormark[1]

\ead{sergio.ra.gomes@outlook.com}

\author[lab1,lab2]{Alexandre C. M. Correia}

\ead{acor@uc.pt}

\address[lab1]{CFisUC, Departamento de F\'isica, Universidade de Coimbra, 3004-516 Coimbra, Portugal}
\address[lab2]{IMCCE, UMR8028 CNRS, Observatoire de Paris, PSL Universit\'e, 77 Av. Denfert-Rochereau, 75014 Paris, France}

\begin{abstract}
At present, the main satellites of Uranus are not involved in any low order mean motion resonance (MMR).
However, owing to tides raised in the planet, Ariel and Umbriel most likely crossed the 5/3~MMR in the past.
Previous studies on this resonance passage relied on limited time-consuming $N-$body simulations or simplified models focusing solely on the effects of the eccentricity or the inclination.
In this paper, we aim to provide a more comprehensive view on how the system evaded capture in the 5/3~MMR.
For that purpose, we developed a secular resonant two-satellite model with low eccentricities and low inclinations, including tides using the weak friction model.
By performing a large number of numerical simulations, we show that capture in the 5/3~MMR is certain if the initial eccentricities of Ariel, $e_1$, and Umbriel, $e_2$, are related through $(e_1^2 + e_2^2)^{1/2} < 0.007$.
Moreover, we observe that the eccentricity of Ariel is the key variable to evade the 5/3~MMR with a high probability.
We determine that for $e_1 > 0.015$ and $e_2 < 0.01$, the system avoids capture in at least $60\%$ of the cases.
We also show that, to replicate the currently observed system, the initial inclinations of Ariel and Umbriel must lay within $\inc_1 \le 0.05\degree$ and $0.06\degree \le \inc_2 \le 0.11\degree$, respectively.
We checked these results using a complete $N-$body model with the five main satellites and did not observe any significant differences.
\end{abstract}


\maketitle


\section{Introduction}

The Uranian satellite system is very intricate, with a rich dynamical and geological past. 
Results from the only spacecraft that visited this remote planet, Voyager 2, have shown that, among the five regular moons, Miranda, Ariel, and Titania display signs of resurfacing, while Umbriel and Oberon present very cratered ancient surfaces \citep{Smith_etal_1986, Plescia_1987, Avramchuk_etal_2007,Kirchoff_etal_2022,Bottke_etal_2024}. 
Neither radiogenic nor primordial heat from the formation can solely explain the geological features observed in Miranda, Ariel, and Titania \citep[eg.][]{Peale_1988, Castillo_Rogez_etal_2022}.
Since the formation of the Solar System, $4.5$ Gyr ago, tidal friction is though to induce a slow outward migration of the satellites \citep[eg.][]{Peale_1988}. 
The different radial distances of the moons with respect to the planet translate into distinct migration paces, changing the relative distance between the satellites. 
This probably led to several mean motion resonances (MMRs) encounters \citep[eg.][]{Greenberg_1975, Smith_etal_1986, Strom_1987, Peale_1988}.
Such events are often invoked as an energy source to explain the features observed on some satellites' surfaces.

The currently observed eccentricities of the Uranian satellites are small, but still abnormally high when tidal dissipation is taken into account.
Indeed, tides are expected to damp the eccentricity on a $10^7-10^8$ timescale \citep{Squyres_etal_1985, Gomes_Correia_2024p1}, and the oscillations owing to mutual perturbations between the satellites cannot explain the present values \citep{Dermott_Nicholson_1986, Laskar_1986, Gomes_Correia_2024p1}.
Furthermore, the current inclination of Miranda, $\sim 4.3\degree$, is also notably high \citep{Tittemore_Wisdom_1989,Tittemore_Wisdom_1990,Malhotra_Dermott_1990,Verheylewegen_etal_2013,Cuk_etal_2020}. 
The origin of the Uranian satellite system is not yet completely understood, and it is still under debate, but the main satellites likely formed in a circumplanetary disk \citep[eg.][]{Pollack_etal_1991, Crida_Charnoz_2012, Szulagyi_etal_2018, Ishizawa_etal_2019, Inderbitzi_etal_2020, Ida_etal_2020, Rufu_Canup_2022}.
Regardless of the formation mechanism, the initial eccentricities and inclinations of the main satellites should have been extremely small.
The relatively high eccentricities of the satellites and the inclination of Miranda are then clear indicators of a dynamically rich past, were MMR may have played a preponderant role. 

At present, there is no resonant commensurabilities within the Uranian system \citep{Gomes_Correia_2024p1}.
As a result, we can infer that the system always managed to escape previous encounters with some MMR.
Adopting similar dissipation rates for all satellites, the last commensurability to have been experienced was presumably the 5/3 MMR between Ariel and Umbriel \citep{Peale_1988, Cuk_etal_2020, Gomes_Correia_2023, Gomes_Correia_2024p1}.
This event is reinforced by the younger predicted ages of Ariel's geological features \cite[eg.][]{Zahnle_etal_2003, Cartwright_etal_2020}.  

Previous studies were conducted to study the passage of Ariel and Umbriel through the 5/3 MMR.
Using a resonant two-satellite planar model with small eccentricities, \citet{Tittemore_Wisdom_1988} concluded that if the resonance is approached with eccentricities of both satellites smaller than $\sim 0.01$, then capture is certain.
These eccentricity values are higher than the presently observed ones, and so it remains difficult to explain how the satellites may have acquired such an eccentricity prior to the resonance encounter, in order to avoid capture.

\citet{Cuk_etal_2020} also studied the passage through the 5/3 MMR, but using a $N-$body numerical integrator, which included the five regular satellites with non-planar, eccentric orbits and spin evolution. 
Starting with the current eccentricities and nearly zero inclinations, they confirm that Ariel and Umbriel are captured in resonance, as predicted by \citet{Tittemore_Wisdom_1988}, but posteriorly evaded it due to the chaotic excitation of the eccentricities and inclinations.
These chaotic variations are not limited to the two satellites involved in the 5/3 MMR, they also propagate to the orbits of the remaining three satellites.
This outcome is particularly significant for the inclination of Miranda: it can attain a value higher than $5^\circ$, which is comparable to the presently observed one.
However, the inclinations of the remaining satellites also grow to values around $1^\circ$, which cannot be conciliated with the presently observed low values.
Indeed, contrarily to the eccentricities, tides are not very efficient to damp the inclinations, and so their values are presumably fossilised after the resonance crossing.
\citet{Cuk_etal_2020} additionally show that a spin-orbit resonance between the precession frequency of Umbriel's node and the precession frequency of Oberon's spin axis can decrease Umbriel's inclination, but such resonance is not observed today and this mechanism also fails to damp the inclinations of the remaining satellites \citep{Gomes_Correia_2024p1}.

To better understand the impact of the inclination on the passage of the 5/3 MMR, \citet{Gomes_Correia_2023} revisited this problem both analytically and numerically, using a secular two-satellite circular model with small inclinations.
Despite the chaotic behaviour, it was shown that a non-zero inclination of Umbriel also facilitates the crossing of this resonance.
Moreover, the currently observed inclinations were used to constrain the initial configuration of the system before the resonant encounter.
However, in a non-circular model, the eccentricity terms introduce more resonant angles into the problem, that result in additional libration and chaotic regions \citep{Tittemore_Wisdom_1988}.
Indeed, for the 3/1~MMR between Miranda and Umbriel, it was shown that the coupling between the eccentricity and inclination resonances may cause significant variations in the eccentricity evolution of Miranda \citep{Tittemore_Wisdom_1990}.
Finally, the presence of the remaining three large satellites or Uranus can also introduce three-body resonances that may further excite the eccentricities and the inclinations \citep{Cuk_etal_2020}.

In this paper we aim to understand how the Uranian system evaded the 5/3~MMR between Ariel and Umbriel. 
To this end, in Sect.~\ref{sec:Conservative_Model}, we extend the model presented in \citet{Gomes_Correia_2023} by developing a secular two-satellite model with small eccentricities and small inclinations.
In Sect.~\ref{sec:tidal_evol}, we add the tidal effects to the equations of motion using a constant time-lag model.
In Sect.~\ref{plandyn}, we analyse in detail the planar dynamics, which is a simpler two degree-of-freedom problem.
This allows us to compare with previous studies \citep{Tittemore_Wisdom_1988} and better understand the outcomes of the full problem.
In Sect.~\ref{sec:Numerical_integration}, we consider the complete secular problem and perform a large number of numerical simulations to estimate the possible outcomes of the crossing of the 5/3~MMR. 
In Sect.~\ref{sec:n_body_simulations}, we compare the secular model with the results of $N-$body simulations, which include the five main satellites.
Finally, in the last section, we summarise and discuss our results.
In a companion paper \citep{Gomes_Correia_2024p1}, hereafter Paper~I, we explore the subsequent evolution of the system after the crossing of the 5/3~MMR until the present day.

\begin{table*}
    \centering
    \caption{Physical and mean orbital parameters of the five largest Uranian satellites. The masses and orbital parameters are from \citet{Jacobson_2014}, the radius from \citet{Thomas_1988}, the second order gravity field coefficient, $J_2$, and the tidal Love numbers are from Table~2 in \citet{Chen_etal_2014}. The fluid Love number, $k_{\rm f}$, and the inner structure coefficient, $\zeta$, are from \cite{Gomes_Correia_2024p1}.}
    \label{table:physical_orbital_parameters}
\renewcommand{\arraystretch}{1.1}
    \begin{tabular}{c|c|c|c|c|c|c}
		\hline       
       & Uranus & Miranda & Ariel & Umbriel & Titania & Oberon\\
       \hline
       Mass ($\rm \times 10^{-10} \, M_{\odot}$)    & $\num{4.365628e5}$ & $0.323997$ & $6.291561$ & $6.412118$ & $17.096471$ & $15.468953$\\
       Radius (km)                                 &  $25\,559$  &$235.8$ & $578.9$ & $584.7$ & $788.4$ & $761.2$ \\
       $J_2$                                       &  $\num{3.5107e-3}$  & $\num{6.10e-3}$ & $\num{1.39e-3}$ &$\num{6.13e-4}$&$\num{1.13e-4}$&$\num{1.48e-5}$\\
       $\zeta$                                    & \num{0.225}  &  $\num{0.327}$  & $\num{0.320}$ & $\num{0.342}$ &$\num{0.326}$&$\num{0.310}$\\
       $k_{\rm f}$                                &  $\num{0.356}$  & $\num{0.907}$ & $\num{0.862}$ &$\num{1.016}$&$\num{0.899}$&$\num{0.790}$\\
       $k_2$                                       &  $\num{0.104}$  & $\num{8.84e-4}$ & $\num{1.02e-2}$ &$\num{7.35e-3}$&$\num{1.99e-2}$&$\num{1.68e-2}$\\
       Period (day)                                &  $0.718328$  & $1.413480$ & $2.520381$ & $4.144176$ & $8.705883$ & $13.463254$ \\
       $a \, (\Ru)$ & &$5.080715$ & $7.470167$ & $10.406589$ & $17.069604$ & $22.827536$ \\
       $e \, (\times 10^{-3})$ & & $1.35$ & $1.22$ & $3.94$ & $1.23$ & $1.40$ \\
       $\inc \, (\degree)$ & & $4.4072$ & $0.0167$ & $0.0796$ & $0.1129$ & $0.1478$\\
       \hline
    \end{tabular}
    \renewcommand{\arraystretch}{1.0}
    \end{table*}

\section{Secular resonant model}\label{sec:Conservative_Model}

In this section, we extend the model presented in \citet{Gomes_Correia_2023} by developing a secular two-satellite model for a system involved in a second order $(p+q)/p$~MMR (hence $q=2$) with small eccentricities and small inclinations.
We have $p=3$ for the 5/3~MMR, but our model is still valid for any $p$ value.

We consider an oblate central body of mass $m_0$ (Uranus) surrounded by two point-mass bodies $m_1$, $m_2 \ll m_0$ (satellites), where the subscript 1 refers to the inner orbit (Ariel) and the subscript 2 refers to the outer orbit (Umbriel). 
The potential energy of the system is given by \citep[eg.][]{Smart_1953}
\begin{equation}\label{eq:general_conservative_gravitational_potential}
    \begin{split}
    U=-\sum_{k=1}^2\frac{\Grav m_0m_k}{r_k}\left[1+ J_2\left(\frac{\Ru}{r_k}\right)^2P_2(\mathbf{\hat{r}}_k\cdot \vv{s})\right]
    -\frac{\mathcal{G}m_1m_2}{|\vr_2-\vr_1|} \ ,
    \end{split}
\end{equation}
where $\mathcal{G}$ is the gravitational constant, $J_2$, $\Ru$ and $\vv{s}$ are the second order gravity field coefficient, the radius, and the spin unit vector of the central body, respectively, $\vv{r}_k$ is the position vector of $m_k$ with respect to the centre-of-mass of $m_0$ (astrocentric coordinates), $r_k = |\vr_k|$ is the norm, $\ur_k = \vr_k / r_k $ is the unit vector, and $P_2(x)=(3x^2-1)/2$ is the Legendre polynomial of degree two. 
We neglect terms in $(\Ru/r_k)^3$ (quadrupolar approximation for the oblateness).

The coefficient $J_2$ mainly depends on Uranus' rotational flattening, and thus on its rotation rate, $\angrot_0$. 
We have \citep[eg.][]{Correia_Rodriguez_2013}
\begin{equation}\label{eq:J20}
 J_{2} = k_{{\rm f},0} \frac{\angrot_0^2 R_0^3}{3\mathcal{G}m_0} \ ,
\end{equation}
where $k_{{\rm f},0}$ is the fluid second Love number for potential.
Using the currently observed $J_2$ and rotation rate of Uranus (Table~\ref{table:physical_orbital_parameters}), we compute $k_{{\rm f},0}= 0.356$.

\subsection{Expansion in elliptical elements}\label{sec:eee}

The Hamiltonian of the problem, $\Energinion$, is obtained by adding the orbital and rotational kinetic energies to Eq.\,(\ref{eq:general_conservative_gravitational_potential}).
We can then expand the Hamiltonian in elliptical elements. 
To the first order in the mass ratios, $m_k/m_0$, and second order in the eccentricities, $e_k$, and the inclinations, $\inc_k$ (with respect to the equatorial plane of the central body), we have \citep[eg.][]{Murray_Dermott_1999}
\begin{equation}\label{genHam}
    \Energinion = \Energinion_K + \Energinion_O + \Energinion_S + \Energinion_R + \Energinion_F + \Energinion_\Theta\ ,
\end{equation}
where
\begin{equation}\label{Hkep}
    \Energinion_K=-\sum_{k=1}^2\frac{\mathcal{G}m_0m_k}{2a_k}
\end{equation}
is the Keplerian part and $a_k$ are the semi-major axes,
\begin{equation}\label{Hobl}
    \begin{split}
        \Energinion_O=-\sum_{k=1}^2\frac{\Grav m_0m_k}{2a_k}J_2\left(\frac{\Ru}{a_k}\right)^2&\Bigg[1+3e_k\cos{(\lambda_k-\varpi_k)}\\
        &+\frac{3}{2}e_k^2\Big(1+3\cos{(2\lambda_k-2\varpi_k)}\Big)\\
        &-\frac{3}{2}\inc_k^2\Big(1-\cos{(2\lambda_k-2\Omega_k)}\Big)\Bigg] 
    \end{split}
\end{equation}
is the contribution from the oblateness of the planet, $\lambda_k$ are the mean longitudes, $\varpi_k$ are the longitudes of the pericentre, and $\Omega_k$ are the longitudes of the nodes,
 \begin{equation}\label{Hsec}
    \begin{split}
        \Energinion_S=-\frac{\mathcal{G}m_1m_2}{8 a_2} \,& \Bigg[  4 b_{1/2}^{(0)}(\alpha) + \Big(e_1^2+e_2^2\Big) \left(2 \alpha \mathcal{D}+\alpha^2 \mathcal{D}^2 \right) \, b_{1/2}^{(0)}(\alpha)  \\
        &+ 2 e_1 e_2 \cos{(\varpi_2-\varpi_1)}\, \Big(2 - 2\alpha - \alpha^2\Big) \, b_{1/2}^{(1)}(\alpha)\\
        &- \,  \, \Big( \inc_1^2 +  \inc_2^2 - 2 \inc_1 \inc_2 \cos{(\Omega_2-\Omega_1)}\Big) \, \alpha \, b^{(1)}_{3/2}(\alpha) \Bigg] \,
    \end{split} 
\end{equation}
is the secular part, $b^{(j)}_s$ are Laplace coefficients, $\alpha=a_1/a_2$ is the semi-major axes ratio, ${\mathcal{D}=\frac{\partial}{\partial \alpha}}$, and
\begin{equation}\label{Hres}
\begin{split}
    \Energinion_R= -\frac{\mathcal{G}m_1m_2}{8a_2} \, & \Bigg[ 
    e_1^2\left(4p^2+11p+6+(4p+6)\alpha\mathcal{D}+\alpha^2\mathcal{D}^2\right)b_{1/2}^{(p+2)}(\alpha)\, \cos{((p+2)\lambdaj-p\lambdai-2\varpi_1)} \\ 
    &- e_2^2 \left(27\alpha+3 \alpha^{-2}\right)  \cos{(3\lambdaj-\lambdai-2\varpi_2)} \\ 
    &+e_2^2\left(4p^2+9p+4+(4p+6)\alpha\mathcal{D}+\alpha^2\mathcal{D}^2\right)b_{1/2}^{(p)}(\alpha)  \cos{\big((p+2)\lambdaj-p\lambdai-2\varpi_2\big)}\\
    &-2 e_1e_2\left(4p^2+10p+6+(4p+6)\alpha\mathcal{D}+\alpha^2\mathcal{D}^2\right)b_{1/2}^{(p+1)}(\alpha)  \cos{\big((p+2)\lambdaj-p\lambdai-\varpi_1-\varpi_2\big)}\\
       & + \inc_1^2 \, \alpha b^{(p+1)}_{3/2}(\alpha) \cos{\big((p+2)\lambdaj-p\lambdai-2\Omega_1\big)} \\
       & + \inc_2^2 \, \alpha b^{(p+1)}_{3/2}(\alpha) \cos{\big((p+2)\lambdaj-p\lambdai-2\Omega_2\big)} \\
       &  -  2 \inc_1 \inc_2 \, \alpha b^{(p+1)}_{3/2}(\alpha) \cos{\big((p+2)\lambdaj-p\lambdai-\Omega_2-\Omega_1\big)}    \Bigg] \, ,
\end{split}
\end{equation}
is the contribution from the second order resonant terms ($q=2$).
The term in $\Energinion_F$ corresponds to the remaining terms of the disturbing function that depend on other combinations of the angles $\lambdai$, $\lambdaj$, $\varpi_1$, $\varpi_2$, $\Omega_1$, and $\Omega_2$ that do not appear in the expressions of $\Energinion_S$ or $\Energinion_R$.
Finally, the last term in the Hamiltonian (Eq.\,(\ref{genHam})), corresponds to the total rotational kinetic energy.
We assume that all bodies rotate about the axis of maximal inertia (gyroscopic approximation), and thus \citep[eg.][]{Goldstein_1950}
\begin{equation}\label{Hrot}
    \Energinion_\Theta=\sum_{k=0}^2\frac{\Theta_k^2}{2 C_k} \ ,
\end{equation}
where $C_k = \zeta_k m_k R_k^2$ is the principal moment of inertia, $\zeta_k$ is a structure constant (Table~\ref{table:physical_orbital_parameters}), 
\begin{equation}\label{Lrot}
 \Theta_k = C_k \, \angrot_k 
\end{equation}
is the rotational angular momentum, $\angrot_k = \dot \theta_k$ is the rotational angular velocity, and $\theta_k$ is the rotation angle.

\subsection{Action-angle resonant variables}
We can rewrite the Hamiltonian (Eq.\,(\ref{genHam})) using a set of canonical action-angle variables. 
For that purpose, we adopt Andoyer variables for the rotation ($\Theta_k, \theta_k$), and Poincar\'e variables for the orbits $(\Lambda_k,\lambda_k;\Sigmaa_k,-\varpi_k;\Psia_k,-\Omega_k)$, with
\begin{equation}\label{Lambdak}
    \Lambda_k = \beta_k\sqrt{\mu_k a_k} \ ,
\end{equation}
\begin{equation}\label{Sigmak}
    \Sigmaa_k=\Lambda_k\left(1-\sqrt{1-e_k^2}\right) \approx \frac{1}{2}\Lambda_k e_k^2 \, ,
\end{equation}
\begin{equation}\label{Psik}
    \Psia_k = \left(\Lambda_k-\Sigma_k\right) (1-\cos \inc_k) \approx \frac{1}{2} \Lambda_k \inc_k^2 \ ,
\end{equation}
 where $\beta_k=m_0m_k/(m_0+m_k)$ and $\mu_k=\mathcal{G}(m_0+m_k)$.
 Since we aim to study the dynamics of the system near the $(p+2)/p$~MMR, we introduce the near resonant angle
\begin{equation}\label{eq:resonace_argument}
    \sigma = \left(1+\frac{p}{2}\right)\lambda_2-\frac{p}{2}\lambda_1 \ ,
\end{equation}
which is present in all terms of the resonant Hamiltonian (Eq.\,(\ref{Hres})).
Each term corresponds to a resonant combination:
\begin{equation}\label{psiiang}
\begin{split}
&\sigmai = \sigma-\varpi_1 \ ,   &\sigmaj  = \sigma-\varpi_2 \ , \quad &\sigma_3 = (\sigmai + \sigmaj)/2 \ ,  \\
&\psii = \sigma - \Omega_1 \ ,   &\psij  = \sigma - \Omega_2 \ , \quad &\psia_3 = (\psii + \psij)/2 \ .
\end{split}
\end{equation}
We additionally introduce the angle
\begin{equation}\label{addangles}
\gamma =\frac{p}{2} \left( \lambda_1 - \lambda_2 \right) =  \lambda_2 - \sigma \ ,
\end{equation}
and modify the rotation variables as
\begin{equation}\label{rotation_angles}
    \vartheta_0 = \theta_0 - \sigma \ ,  \quad \vartheta_1 = \theta_1 - \sigma \quad \mathrm{and} \quad \vartheta_2 = \theta_2 - \sigma \ ,
\end{equation}
such that
\begin{equation}\label{transform_angles}
\left[\begin{array}{c} 
\sigma \\  \gamma \\ \sigmai \\ \sigmaj \\ \psii \\ \psij \\ \vartheta_0 \\ \vartheta_1 \\ \vartheta_2
\end{array}\right] 
\equiv {\cal S} \,
\left[\begin{array}{c} 
\lambda_1 \\ \lambda_2 \\ -\varpi_1 \\ -\varpi_2 \\ -\Omega_1 \\ -\Omega_2 \\ \theta_0 \\ \theta_1 \\ \theta_2
\end{array}\right] \ ,
\end{equation}
with
\begin{equation}
{\cal S} =  \left[\begin{array}{ccccccccc} 
-p/2 & 1 + p/2  & 0 & 0 & 0 & 0 & 0 & 0 & 0 \\
 p/2 & -p/2     & 0 & 0 & 0 & 0 & 0 & 0 & 0 \\
-p/2 & 1 + p/2  & 1 & 0 & 0 & 0 & 0 & 0 & 0 \\
-p/2 & 1 + p/2  & 0 & 1 & 0 & 0 & 0 & 0 & 0 \\
-p/2 & 1 + p/2 & 0 & 0 & 1 & 0 & 0 & 0 & 0 \\
-p/2 & 1 + p/2 & 0 & 0 & 0 & 1 & 0 & 0 & 0 \\
 p/2 & -1 - p/2 & 0 & 0 & 0 & 0 & 1 & 0 & 0 \\
 p/2 & -1 - p/2 & 0 & 0 & 0 & 0 & 0 & 1 & 0 \\
 p/2 & -1 - p/2 & 0 & 0 & 0 & 0 & 0 & 0 & 1 
\end{array}\right]  \ .
\end{equation}

We then obtain for the canonical conjugated actions \citep[eg.,][]{Goldstein_1950}
\begin{equation} \label{canonic_var}
\left[\begin{array}{c} 
\Sigma \\ \Gamma \\ \tilde \Sigmaa_1 \\ \tilde \Sigmaa_2 \\ \tilde \Psia_1 \\ \tilde \Psia_2 \\ \tilde \Theta_0 \\ \tilde \Theta_1 \\ \tilde \Theta_2
\end{array}\right] 
= ( {\cal S}^{-1} )^T \,
\left[\begin{array}{c} 
\Lambdai \\ \Lambdaj \\ \Sigmaa_1 \\ \Sigmaa_2 \\ \Psii \\ \Psij \\ \Theta_0 \\ \Theta_1 \\ \Theta_2
\end{array}\right] \ ,
\end{equation}
where
\begin{equation}\label{canonic_var_Sigma}
\Sigma = \Lambdai + \Lambdaj - \Sigmai - \Sigmaj - \Psii - \Psij + \Theta_0+\Theta_1+\Theta_2 \ ,
\end{equation}
\begin{equation}\label{canonic_var_Gamma}
\Gamma = \left(1+\frac{2}{p}\right)\Lambdai+\Lambdaj \ ,
\end{equation}
\begin{equation}\label{canonic_var_other}
\tilde \Sigmaa_k = \Sigmaa_k \ , \quad  \tilde \Psia_k = \Psia_k \ , \quad \mathrm{and} \quad \tilde \Theta_k = \Theta_k \ .
\end{equation}

\subsubsection{Conserved quantities}\label{consqtt}

We can rewrite the Hamiltonian (Eq.\,(\ref{genHam})) using the resonant canonical variables (Eq.\,(\ref{canonic_var})).
For the actions, we can replace the semi-major axes, the eccentricities, and the inclinations using the relations (\ref{Lambdak}), (\ref{Sigmak}) and (\ref{Psik}). 
We obtain
\begin{equation}\label{smak_ecckL_inckL}
 a_k = \frac{\Lambda_k^2}{\beta_k^2 \mu_k} \ , \quad
 e_k \approx \sqrt{\frac{2\Sigmaa_k}{\Lambda_k}} \ , \quad
\mathrm{and} \quad
 \inc_k \approx \sqrt{\frac{2 \Psia_k}{\Lambda_k}}  \ .
\end{equation}
Since $\Lambda_k$ are no longer actions of the resonant variables, they must be obtained from the canonical actions (Eq.\,(\ref{canonic_var})) as
\begin{equation}\label{GammaOne}
    \Lambdai = \Gamma_1-\frac{p}{2}(\Sigmai+\Sigmaj+\Psii+\Psij) \ ,\\
\end{equation}
\begin{equation}\label{GammaTwo}
    \Lambdaj = \Gamma_2+\left(1+\frac{p}{2}\right)(\Sigmai+\Sigmaj+\Psii+\Psij) \ ,
\end{equation}
where
\begin{equation}\label{Gb1_Gb2}
\Gamma_1  = \frac{p}{2} \Gamma \left( 1 - \Delta \right) 
\quad \mathrm{and} \quad
\Gamma_2 = -\frac{p}{2} \Gamma \left(1 -\left(1+\frac{2}{p}\right) \Delta \right)  \ ,
\end{equation}
with
\begin{equation}\label{DeltaRef}
\Delta = \left(\Sigma - \Theta\right) / \Gamma
\quad \mathrm{and} \quad
\Theta=\Theta_0+\Theta_1+\Theta_2  \ .
\end{equation}

In the approximation of small eccentricities and inclinations, we have $\Sigmaa_k \ll \Lambda_k$ and $\Psia_k \ll \Lambda_k$ (Eq.\,(\ref{smak_ecckL_inckL})), and so we also have $\Sigmaa_k \ll \Gamma_k$ and  $\Psia_k \ll \Gamma_k$, which allows us to write 
\begin{align}
    \Lambdai^\alpha \approx \  & \Gamma_1^\alpha\left[1-\alpha\,\frac{p}{2}\frac{\Sigmai+\Sigmaj+\Psii+\Psij}{\Gamma_1}\right] \ ,\label{eq:Lambda_1_alpha}\\
    \Lambdaj^\alpha \approx \ & \Gamma_2^\alpha\left[1+\alpha \,\left(1+\frac{p}{2}\right)\frac{\Sigmai+\Sigmaj+\Psii+\Psij}{\Gamma_2}\right] \ , \label{eq:Lambda_2_alpha}
\end{align}
\begin{equation}\label{ek_inck}
    e_k \approx \sqrt{\frac{2\Sigmaa_k}{\Gamma_k}} \ ,
    \quad \mathrm{and} \quad 
    \inc_k \approx \sqrt{\frac{2\Psia_k}{\Gamma_k}} \ .
\end{equation}

The replacement of the resonant angles is straightforward (Eq.\,(\ref{transform_angles})).
The rotation angles $\theta_k$ do not appear in the expression of the Hamiltonian (Eq.\,(\ref{genHam})). Then, the canonical angles $\vartheta_k$ also do not appear, and so their conjugated actions, $\Theta_k$, are constants of motion, and so is $\Theta$ (Eq.\,(\ref{DeltaRef})).
The remaining angles are combined as (Eqs.\,(\ref{Hobl})$-$(\ref{Hres}))
\begin{equation}
        \cos\Big(k_1 \lambdai + k_2 \lambdaj + k_3 \varpi_1 + k_4 \varpi_2 + k_5 \Omega_1 + k_6 \Omega_2 \Big) \ , \quad k_j \in \mathbb{Z} \ ,
\end{equation}
with $ \sum_j k_j=0$, 
which corresponds to the d'Alembert rule (conservation of the angular momentum).
Then, using the resonant angles (Eq.\,(\ref{transform_angles})), we obtain for all terms
\begin{equation}
        \cos \left( \Big(k_1+\frac{2}{p}k_1+k_2\Big)\gamma-k_3\sigmai-k_4\sigmaj-k_5\psii - k_6 \psij \right) \ .
\end{equation}
The angle $\sigma$ does not appear in the Hamiltonian (Eq.\,(\ref{genHam})), and so its conjugated action, $\Sigma$ (Eq.\,(\ref{canonic_var_Sigma})), is also a constant of motion. 

\subsubsection{Average}\label{cqaa}

Near the MMR, the resonant angles $\sigmaa_k$ and $\psia_k$ vary much slower than $\gamma$, that is, $\dot{\sigmaa_k} , \dot{\psia_k} \ll \dot{\gamma}$. Therefore, to the first order in $m_k/m_0$, we can construct the resonant secular normal form of the Hamiltonian (Eq.\,(\ref{genHam})) by simply averaging over $\gamma$:
\begin{equation}\label{averHam}
\MedH = \langle \Energinion \rangle_\gamma = \frac{1}{2 p \pi} \int_0^{2 p \pi} \Energinion \, d \gamma \ . 
\end{equation}
As a result, $\langle \Energinion_F \rangle_\gamma = 0$, and since $\gamma$ no longer appears in the expression of the averaged Hamiltonian, the conjugated variable, $\Gamma$ (Eq.\,(\ref{canonic_var})), also becomes a constant of motion.
We thus reduce a problem with initially nine degrees-of-freedom to a problem with four degrees-of-freedom, ($\Sigmai,\sigmai; \Sigmaj, \sigmaj; \Psii, \psii; \Psij, \psij$), and five parameters, $\Sigma$, $\Gamma$, $\Theta_0$, $\Theta_1$ and $\Theta_2$.
The auxiliary quantities $\Gamma_1$, $\Gamma_2$ (Eq.\,(\ref{Gb1_Gb2})), and $\Delta$ (Eq.\,(\ref{DeltaRef})) are also constant.

The resonant secular Hamiltonian then finally reads
\begin{equation}
\label{resHam}
    \begin{split}
       \MedH &=(\Ka+\Sa)(\Sigmai+\Sigmaj+\Psii+\Psij)
        +\Kb(\Sigmai+\Sigmaj+\Psii+\Psij)^2    \\
        &+(\Oa+\Sb)\Sigmai+(\Ob+\Sc)\Sigmaj     
        +\Sd\sqrt{\Sigmai}\sqrt{\Sigmaj}\cos{(\sigmai-\sigmaj)}\\
        &+(\Oc+\Se)\Psii+(\Od+\Sf)\Psij                     
        +\Sg\sqrt{\Psii}\sqrt{\Psij}\cos{(\psii-\psij)}    \\
        &+\Ra\Sigmai\cos{(2\sigmai)}+\Rb\Sigmaj\cos{(2\sigmaj)}     
        +\Rc\sqrt{\Sigmai}\sqrt{\Sigmaj}\cos{(\sigmai+\sigmaj)}    \\
        &+\Rd\Psii\cos{(2\psii)}+\Re\Psij\cos{(2\psij)}     
        +\Rf\sqrt{\Psii}\sqrt{\Psij}\cos{(\psii+\psij)} \ ,
    \end{split} 
\end{equation}
where $\mathcal{K}$ stands for the Keplerian coefficients (Eq.\,(\ref{Hkep})), $\mathcal{O}$ for the oblateness coefficients (Eq.\,(\ref{Hobl})), $\mathcal{S}$ for secular coefficients (Eq.\,(\ref{Hsec})) and $\mathcal{R}$ for resonant coefficients (Eq.\,(\ref{Hres})).
The Keplerian part needs to be expanded to the second order in $\Sigmaa_k$ and $\Psia_k$ (fourth order in the eccentricities and the inclinations)
because $\Kb$ is much larger than the remaining coefficients. 
The explicit expression of all these coefficients is given in Appendix~\ref{sec:Conservative_Hamiltonian_terms}.

\subsection{Complex rectangular coordinates}\label{crcoord}

When expressed in the resonant variables $(\Sigma_k, \sigma_k; \Psia_k, \psia_k)$, the equations of motion may experience some singularities when $\Sigma_k=0$ or $\Psia_k=0$, due to the terms in $\Sd, \ \Sg, \ \Rc$, and $\Rf$ (Eq.\,(\ref{resHam})).
We thus perform a second canonical change of variables to complex rectangular coordinates 
$(\Sigmaa_k, \sigmaa_k; \Psia_k, \psia_k) \rightarrow (\cartxk, \ii \cartxbk; \cartyk, \ii \cartybk)$, where
\begin{equation}\label{eq:complex_cartesian_coordinates}
    \cartxk = \sqrt{\Sigmaa_k}\ep^{\ii\sigmaa_k} \quad \text{and} \quad \cartyk = \sqrt{\Psia_k}\ep^{\ii\psia_k} \ ,
\end{equation}
and $\cartxbk$ and $\cartybk$ are the complex conjugate of $\cartxk$ and $\cartyk$, respectively.
From Eq.\,(\ref{ek_inck}) we have 
\begin{equation}\label{yinc}
\cartxk \approx e_k \sqrt{\frac{\Gamma_k}{2}} \ep^{\ii\sigma_k} \quad \text{and} \quad 
\cartyk \approx \inc_k \sqrt{\frac{\Gamma_k}{2}} \ep^{\ii\psia_k}  \ ,
\end{equation}
and so these variables are proportional to the eccentricities and the inclinations, respectively.
In this new set of canonical variables, the resonant secular Hamiltonian (Eq.\,(\ref{resHam})) becomes
\begin{equation}\label{resHamcomplex}
    \begin{split}
       \MedH &=(\Ka+\Sa)\left(\cartxi\cartxbi+\cartxj\cartxbj+\cartyi\cartybi+\cartyj\cartybj\right)
        +\Kb\left(\cartxi\cartxbi+\cartxj\cartxbj+\cartyi\cartybi+\cartyj\cartybj\right)^2    \\
        &+(\Oa+\Sb)\cartxi\cartxbi+(\Ob+\Sc)\cartxj\cartxbj+\frac{\Sd}{2}\left(\cartxi\cartxbj+\cartxbi\cartxj\right)\\
        &+(\Oc+\Se)\cartyi\cartybi+(\Od+\Sf)\cartyj\cartybj+\frac{\Sg}{2}\left(\cartyi\cartybj+\cartybi\cartyj\right)    \\
        &+\frac{\Ra}{2}\left(\cartxi^2+\cartxbi^2\right)+\frac{\Rb}{2}\left(\cartxj^2+\cartxbj^2\right)+\frac{\Rc}{2}\left(\cartxi\cartxj+\cartxbi\cartxbj\right)    \\
        &+\frac{\Rd}{2}\left(\cartyi^2+\cartybi^2\right)+\frac{\Re}{2}\left(\cartyj^2+\cartybj^2\right)+\frac{\Rf}{2}\left(\cartyi\cartyj+\cartybi\cartybj\right) \ .
    \end{split} 
\end{equation}
The equations of motion are simply obtained from Eq.\,(\ref{resHamcomplex}) using the Hamilton equations, as
\begin{equation}\label{eq:gen_equations}
    \dot x_k = \ii \frac{\partial \MedH}{\partial \cartxbk} \quad \text{and} \quad
    \dot y_k = \ii \frac{\partial \MedH}{\partial \cartybk} \ ,
\end{equation}
which yields, for the conservative resonant dynamics,
\begin{equation}\label{eq:conservative_motion_equations_x1}
    \dot{x}_{1}=\ii \, \Bigg[\left(\Ka+\Sa\right)\cartxi+2\Kb\left(\cartxi\cartxbi+\cartxj\cartxbj+\cartyi\cartybi+\cartyj\cartybj\right)\cartxi
    +\left(\Oa+\Sb\right)\cartxi+\frac{\Sd}{2}\cartxj+\Ra\cartxbi+\frac{\Rc}{2}\cartxbj\Bigg] \ ,
\end{equation}
\begin{equation}\label{eq:conservative_motion_equations_x2}
    \dot{x}_{2}=\ii \, \Bigg[\left(\Ka+\Sa\right)\cartxj+2\Kb\left(\cartxi\cartxbi+\cartxj\cartxbj+\cartyi\cartybi+\cartyj\cartybj\right)\cartxj
    +\left(\Ob+\Sc\right)\cartxj+\frac{\Sd}{2}\cartxi+\Rb\cartxbj+\frac{\Rc}{2}\cartxbi \Bigg] \ ,
\end{equation}
\begin{equation}\label{eq:conservative_motion_equations_y1}
    \dot{y}_{1}=\ii \, \Bigg[\left(\Ka+\Sa\right)\cartyi+2\Kb\left(\cartxi\cartxbi+\cartxj\cartxbj+\cartyi\cartybi+\cartyj\cartybj\right)\cartyi
    +\left(\Oc+\Se\right)\cartyi+\frac{\Sg}{2}\cartyj+\Rd\cartybi+\frac{\Rf}{2}\cartybj\Bigg] \ ,
\end{equation}   
and 
\begin{equation}\label{eq:conservative_motion_equations_y2}
    \dot{y}_{2}=\ii \, \Bigg[\left(\Ka+\Sa\right)\cartyj+2\Kb\left(\cartxi\cartxbi+\cartxj\cartxbj+\cartyi\cartybi+\cartyj\cartybj\right)\cartyj
    +\left(\Od+\Sf\right)\cartyj+\frac{\Sg}{2}\cartyi+\Re\cartybj+\frac{\Rf}{2}\cartybi \Bigg] \ .
\end{equation}

\section{Tidal evolution}\label{sec:tidal_evol}

Up to this point, we only considered the resonant dynamics (Sect.~\ref{sec:Conservative_Model}), which is conservative and therefore the average semi-major axes remain constant. 
However, dissipative tidal interactions are expected to induce an orbital evolution  of the Uranian satellites. 
The tidal contributions can be obtained by considering an additional tidal potential \cite[][]{Darwin_1880, Kaula_1964, Mignard_1979}.

\subsection{Tidal potential energy}

Tides arise from differential and inelastic deformations of an extended body owing to the gravitational effect of a perturber.
The resulting distortion gives rise to a tidal bulge, which modifies the gravitational potential of the extended body. 
As the perturber interacts with the additional potential field, the amount of tidal potential energy is given by \citep[eg.][]{Lambeck_1980} 
\begin{equation}\label{eq:tidal_potential}
    U_\ij= - k_{2,\elli} \frac{\mathcal{G}m_i^2}{R_\elli} \left(\frac{R_\elli}{r_\ij}\right)^3 \left(\frac{R_\elli}{r_\ij'}\right)^3 P_2 (\ur_\ij \cdot \ur_\ij')  \ ,
\end{equation}
where $R_\elli$ and $k_{2,\elli}$ are the radius and the elastic second Love number for potential of the extended body, respectively, and $\vr_\ij $ is the position vector between the centre-of-mass  of the extended body and that of the perturber with mass $m_i$.
The tidal friction within the extended body introduces a time delay, $\dt_\elli$, between the maximal deformation, at $\vr_\ij = \vr_\ij (t)$, and the initial perturbation, at $\vr_\ij ' = \vr_ \ij (t-\dt_\elli)$.
Although tidal effects do not preserve the total energy, it is possible to extend the Hamiltonian formalism from Sect.~\ref{sec:Conservative_Model}, by considering the primed quantities, $\vr_\ij'$, as parameters \citep{Mignard_1979}.
The tidal Hamiltonian then reads (Eqs.\,(\ref{genHam}) and (\ref{eq:tidal_potential}))
\begin{equation}
\label{tidalH}
    \Energinion_t=\Energinion + U_{01} + U_{10} + U_{02} + U_{20} \ .
\end{equation}
We neglect the satellite-satellite interaction terms, $U_{12}$ and $U_{21}$.
For the considered terms, we note that $\vr_{0 k} = - \vr_{k 0} = \vr_k$. 
Therefore, in the following expressions of $U_\ij$ we assume the subscript $k=\mathrm{max}(i,j)$.
As in Sect.~\ref{sec:eee}, we first expand $U_\ij$ in elliptical elements. 
To the first order in the mass ratios, and second order in the eccentricities and in the inclinations, we have
\begin{equation}
    \begin{split}
    U_\ij & =
     -  k_{2,\elli} \frac{\mathcal{G}  m_i^2 R_\elli^5 }{4 a_k^3 a_k'^3}\Bigg[1+3\cos (2 \lambda_k-2 \lambda_k'-2 \thetid+2 \thetid') \\
     & +\frac{3}{2} e_k \bigg( 2\cos{\left(\lambda_k-\varpi_k\right)}-\cos{\left(\lambda_k-2\lambda_k'+\varpi_k-2 \thetid+2 \thetid'\right)}
         +7 \cos{\left(3\lambda_k-2\lambda_k'-\varpi_k-2 \thetid+2 \thetid'\right)} \bigg)\\
     & +\frac{3}{2} e_k' \bigg( 2\cos{\left(\lambda_k'-\varpi_k'\right)}-\cos{\left(2\lambda_k-   \lambda_k'-\varpi_k'-2 \thetid+2 \thetid'\right)}
        +7 \cos{\left(2\lambda_k-3\lambda_k'+\varpi_k'-2 \thetid+2 \thetid'\right)} \bigg)\\
     & +\frac{3}{2}e_k^2 \bigg( 1+3\cos{\left(2\lambda_k-2\varpi_k\right)} 
        - 5 \cos{\left(2\lambda_k-2\lambda_k'-2\thetid+2\thetid'\right)} 
        +17 \cos{\left(4\lambda_k-2\lambda_k'-2\varpi_k-2\thetid+2\thetid'\right)}\bigg)\\
     & +\frac{3}{2}e_k'^2 \bigg( 1+3\cos{\left(2\lambda_k'-2\varpi_k'\right)} 
        - 5 \cos{\left(2\lambda_k-2\lambda_k'-2\thetid+2\thetid'\right)} 
        +17 \cos{\left(2\lambda_k-4\lambda_k'+2\varpi_k'-2\thetid+2\thetid'\right)}\bigg)\\
     & \begin{split} +\frac{3}{4}e_k e_k' \bigg(& 6\cos{\left(\lambda_k-\lambda_k'-\varpi_k+\varpi_k'\right)}
        +6\cos{\left(\lambda_k+\lambda_k'-\varpi_k-\varpi_k' \right)} \\ 
        &+\cos{\left(\lambda_l-\lambda_k'+\varpi_k-\varpi_k'-2\thetid+2\thetid' \right)}\\ 
        &-9\cos{\left(\lambda_l-3\lambda_k'+\varpi_k+\varpi_k'-2\thetid+2\thetid' \right)} 
        -9\cos{\left(3\lambda_l-\lambda_k'-\varpi_k-\varpi_k'-2\thetid+2\thetid' \right)}\\ 
        &+47\cos{\left(3\lambda_l-3\lambda_k'-\varpi_k+\varpi_k'-2\thetid+2\thetid' \right)} \bigg)\end{split}\\
     & -\frac{3}{2} \inc_k^2 \, \bigg( 1-\cos \left(2 \lambda_k-2 \Omega_k \right)  
          +\cos \left(2 \lambda_k-2 \lambda_k' - 2 \thetid+2 \thetid' \right)  
          -\cos \left(2 \lambda_k'-2 \Omega_k+2 \thetid-2 \thetid' \right) \bigg)\\
     & \begin{split}-\frac{3}{2} \inc_k'^2 \, \bigg(& 1-\cos \left(2 \lambda_k'-2 \Omega_k' \right)  
          +\cos \left(2 \lambda_k-2 \lambda_k'-2 \thetid+2 \thetid' \right)  
          -\cos \left(2 \lambda_k-2 \Omega_k'-2 \thetid+2 \thetid' \right)\bigg) \end{split}\\
     & \begin{split}+3 \inc_k \inc_k' \, \bigg(& \cos \left(2 \lambda_k-2 \lambda_k'-\Omega_k+\Omega_k'-\thetid+\thetid' \right)
          +\cos \left(\Omega_k-\Omega_k'-\thetid+\thetid' \right) \phantom{\bigg)} \\
         & - \cos \left(2 \lambda_k-\Omega_k-\Omega_k'-\thetid+\thetid' \right) \phantom{\bigg)} 
         - \cos \left(2 \lambda_k'-\Omega_k-\Omega_k'+\thetid-\thetid' \right)\bigg)\Bigg]\ .\end{split} 
    \end{split} 
\end{equation}

We first perform the canonical change of variables that uses the resonant angles (Eq.\,(\ref{transform_angles})) and then change to the complex Cartesian coordinates (Eq.\,(\ref{eq:complex_cartesian_coordinates})). 
We get
\begin{equation*}
    \begin{split}
    U_\ij=- & k_{2,\elli} \frac{\Grav m_i^2 R_\elli^5 \beta_k^{12} \mu_k^6}{16 \Gamma_k^7 \Gamma_k'^7}\Bigg[4\Gamma_k \Gamma_k' 
    +12\Gamma_k' \, \cartxk \cartxbk +12\Gamma_k \, \cartxk' \cartxbk' -12\Gamma_k' \, \cartyk \cartybk-12\Gamma_k \, \cartyk' \cartybk'\\
    &- 24 p_k \, \bigg(\Gamma_k' (\cartxi \cartxbi +\cartxj \cartxbj +\cartyi \cartybi +\cartyj \cartybj\bigg)\\
    &- 24 p_k \, \bigg(\Gamma_k (\cartxi' \cartxbi' +\cartxj' \cartxbj' +\cartyi' \cartybi' +\cartyj' \cartybj'\bigg)\\
    &+12 \, \bigg( \Gamma_k \Gamma_k'-\Gamma_k'(5 \cartxk \cartxbk - \cartyk \cartybk)-\Gamma_k (5 \cartxk' \cartxbk' - \cartyk' \cartybk')\\
        &\quad\begin{split}-6 p_k \ \Big(&\Gamma_k' (\cartxi \cartxbi +\cartxj \cartxbj + \cartyi \cartybi + \cartyj \cartybj) +\Gamma_k (\cartxi' \cartxbi' +\cartxj' \cartxbj' + \cartyi' \cartybi' + \cartyj' \cartybj')\Big) \bigg)\cos \left(2 (\vthetid-\vthetid'-q_k (\gamma -\gamma' ))\right) \end{split}\\
    &+ 6\sqrt{2}\, (\cartxk+\cartxbk) \sqrt{\Gamma_k}\Gamma_k'\cos{(q_k\gamma)}\\
    &+ 6\sqrt{2}\, (\cartxk'+\cartxbk') \Gamma_k \sqrt{\Gamma_k'}\cos{(q_k\gamma')}\\
    &- 3\sqrt{2}\, (\cartxk+\cartxbk) \sqrt{\Gamma_k}\Gamma_k'\cos{\left(q_k(\gamma-2\gamma')-2\vthetid+2\vthetid'\right)}\\
    &- 3\sqrt{2}\, (\cartxk'+\cartxbk') \Gamma_k\sqrt{\Gamma_k'}\cos{\left(q_k(2\gamma-\gamma')-2\vthetid+2\vthetid'\right)}\\
    &+ 21\sqrt{2}\, (\cartxk+\cartxbk) \sqrt{\Gamma_k}\Gamma_k'\cos{\left(q_k(3\gamma-2\gamma')-2\vthetid+2\vthetid'\right)}\\
    &+ 21\sqrt{2}\, (\cartxk'+\cartxbk') \Gamma_k\sqrt{\Gamma_k'}\cos{\left(q_k(2\gamma-3\gamma')-2\vthetid+2\vthetid'\right)}\\
   &+6\, \Gamma_k \, \bigg(3\cartxk'^2+3\cartxbk'^2+\cartyk'^2+\cartybk'^2\bigg) \cos (2 q_k \gamma')\\
   &+6\, \Gamma_k \, \bigg(3\cartxk'^2+3\cartxbk'^2+\cartyk'^2+\cartybk'^2\bigg) \cos (2 q_k \gamma')\\
    &+18\,\left(\cartxk\cartxbk'+\cartxk'\cartxbk\right)\sqrt{\Gamma_k \Gamma_k'}\cos{\left(q_k(\gamma-\gamma')\right)}\\
    &+18\,\left(\cartxk\cartxk'+\cartxbk\cartxbk'\right)\sqrt{\Gamma_k \Gamma_k'}\cos{\left(q_k(\gamma+\gamma')\right)}\\
    &+102\,\Gamma_k'\left(\cartxk^2+\cartxbk^2\right)\cos{\left(2(2 q_k \gamma-q_k \gamma' -\vthetid+\vthetid')\right)}\\
    &+102\,\Gamma_k\left(\cartxk'^2+\cartxbk'^2\right)\cos{\left(2(q_k \gamma-2 q_k \gamma' -\vthetid+\vthetid')\right)}\\
    &+3\, \sqrt{\Gamma_k \Gamma_k'}\left(\cartxk \cartxbk'+\cartxk'\cartxbk\right) \cos{\left(q_k(\gamma-\gamma')-2\vthetid+2\vthetid'\right)}\\
    &+147\, \sqrt{\Gamma_k \Gamma_k'}\left(\cartxk \cartxbk'+\cartxk'\cartxbk\right) \cos{\left(3 q_k(\gamma-\gamma')-2\vthetid+2\vthetid'\right)}\\
    &-21\, \sqrt{\Gamma_k \Gamma_k'}\left(\cartxk \cartxk'+\cartxbk \cartxbk'\right) \cos{\left(q_k(\gamma-3\gamma')-2\vthetid+2\vthetid'\right)}\\
    &-21\, \sqrt{\Gamma_k \Gamma_k'}\left(\cartxk \cartxk'+\cartxbk \cartxbk'\right) \cos{\left(q_k(3\gamma-\gamma')-2\vthetid+2\vthetid'\right)}\\
    &+6\,\Gamma_k \, \bigg(\cartyk'^2+\cartybk'^2\bigg) \cos \left(2 (q_k \gamma-\vthetid+\vthetid')\right) 
    +6\,\Gamma_k' \, \bigg(\cartyk^2+\cartybk^2\bigg) \cos \left(2 (q_k \gamma'+\vthetid-\vthetid')\right) \\
    &+12 \sqrt{\Gamma_k \Gamma_k'} \, \bigg(\cartyk \cartybk'+\cartyk' \cartybk\bigg)\cos \Big(\vthetid-\vthetid'\Big)\\
    &+12 \sqrt{\Gamma_k \Gamma_k'} \, \bigg(\cartyk \cartybk'+\cartyk' \cartybk\bigg) \cos \left(2 q_k (\gamma -\gamma' )-\vthetid+\vthetid'\right)\\
    \end{split}
    \end{equation*}
    \begin{equation}\label{sec:tidal_hamiltonian}
        \begin{split}
    &-12 \sqrt{\Gamma_k \Gamma_k'} \, \bigg(\cartyk \cartyk'+\cartybk \cartybk'\bigg) \cos (2 q_k \gamma-\vthetid+\vthetid')\\
    &-12 \sqrt{\Gamma_k \Gamma_k'} \, \bigg(\cartyk \cartyk'+\cartybk \cartybk'\bigg) \cos (2 q_k \gamma'+\vthetid-\vthetid')\\
    &+ 6\sqrt{2}\, \ii \,(\cartxk-\cartxbk) \sqrt{\Gamma_k}\Gamma_k'\sin{(q_k\gamma)}
    + 6\sqrt{2}\, \ii \,(\cartxk'-\cartxbk') \Gamma_k \sqrt{\Gamma_k'}\sin{(q_k\gamma')}\\
    &+ 3\sqrt{2}\, \ii \, (\cartxk-\cartxbk) \sqrt{\Gamma_k}\Gamma_k'\sin{\left(q_k(\gamma-2\gamma')-2\vthetid+2\vthetid'\right)}\\
    &- 3\sqrt{2}\, \ii \,(\cartxk'-\cartxbk') \Gamma_k\sqrt{\Gamma_k'}\sin{\left(q_k(2\gamma-\gamma')-2\vthetid+2\vthetid'\right)}\\
    &+ 21\sqrt{2}\, \ii \,(\cartxk-\cartxbk) \sqrt{\Gamma_k}\Gamma_k'\sin{\left(q_k(3\gamma-2\gamma')-2\vthetid+2\vthetid'\right)}\\
    &- 21\sqrt{2}\, \ii \,(\cartxk'-\cartxbk') \Gamma_k\sqrt{\Gamma_k'}\sin{\left(q_k(2\gamma-3\gamma')-2\vthetid+2\vthetid'\right)}\\
    &+102\, \ii \,\Gamma_k' \left(\cartxk^2-\cartxbk^2\right)\sin{\left(2(2 q_k \gamma-q_k \gamma' -\vthetid+\vthetid')\right)}\\
    &-102\, \ii \,\Gamma_k  \left(\cartxk'^2-\cartxbk'^2\right)\sin{\left(2(q_k \gamma-2 q_k \gamma' -\vthetid+\vthetid')\right)}\\
    &+18\, \ii \, \left(\cartxk\cartxbk'+\cartxk'\cartxbk\right)\sqrt{\Gamma_k \Gamma_k'}\sin{\left(q_k(\gamma-\gamma')\right)}\\
    &+18\, \ii \, \left(\cartxk\cartxk'+\cartxbk\cartxbk'\right)\sqrt{\Gamma_k \Gamma_k'}\sin{\left(q_k(\gamma+\gamma')\right)}\\
    &-3\, \ii \, \sqrt{\Gamma_k \Gamma_k'}\left(\cartxk \cartxbk'-\cartxk'\cartxbk\right) \sin{\left(q_k(\gamma-\gamma')-2\vthetid+2\vthetid'\right)}\\
    &+147\, \ii \, \sqrt{\Gamma_k \Gamma_k'}\left(\cartxk \cartxbk'-\cartxk'\cartxbk\right) \sin{\left(3 q_k(\gamma-\gamma')-2\vthetid+2\vthetid'\right)}\\
    &+21\, \ii \, \sqrt{\Gamma_k \Gamma_k'}\left(\cartxk \cartxk'-\cartxbk \cartxbk'\right) \sin{\left(q_k(\gamma-3\gamma')-2\vthetid+2\vthetid'\right)}\\
    &-21 \, \ii \, \sqrt{\Gamma_k \Gamma_k'}\left(\cartxk \cartxk'-\cartxbk \cartxbk'\right) \sin{\left(q_k(3\gamma-\gamma')-2\vthetid+2\vthetid'\right)}\\
    &+6 \, \ii \,\Gamma_k' \, \bigg(3\cartxk^2-3\cartxbk^2+\cartyk^2-\cartybk^2\bigg) \sin \Big(2 q_k \gamma\Big)\\
    &+6 \, \ii\, \Gamma_k \, \bigg(3\cartxk'^2-3\cartxbk'^2+\cartyk'^2-\cartybk'^2\bigg) \sin \Big(2 q_k \gamma'\Big)\\
    &+6 \, \ii \, \Gamma_k' \, \bigg(\cartyk^2-\cartybk^2 \bigg) \sin \left(2 (q_k \gamma'+\vthetid-\vthetid')\right)\\
    &+6 \, \ii \, \Gamma_k \, \bigg(\cartyk'^2-\cartybk'^2 \bigg) \sin \left(2 (q_k \gamma-\vthetid+\vthetid')\right)\\
    &+12 \, \ii \, \sqrt{\Gamma_k \Gamma_k'} \, \bigg(\cartyk \cartybk'-\cartyk' \cartybk\bigg)\sin \Big(\vthetid-\vthetid'\Big)\\
    &+12 \, \ii \, \sqrt{\Gamma_k \Gamma_k'} \, \bigg(\cartyk \cartybk'-\cartyk' \cartybk\bigg) \sin \left(2 q_k (\gamma -\gamma' )-\vthetid+\vthetid'\right)\\
    &-12 \, \ii \, \sqrt{\Gamma_k \Gamma_k'} \, \bigg(\cartyk \cartyk'-\cartybk \cartybk'\bigg) \sin (2 q_k \gamma-\vthetid+\vthetid')
    -12 \, \ii \, \sqrt{\Gamma_k \Gamma_k'} \, \bigg(\cartyk \cartyk'-\cartybk \cartybk'\bigg) \sin (2 q_k \gamma'+\vthetid-\vthetid') \Bigg]\ ,
    \end{split}
\end{equation}
where $p_1=-p/2$, $p_2=1+p/2$, $q_1=1+2/p$ and $q_2=1$.

We note again that $\sigma$ does not appear in the expression of $U_\ij$ (Eq.\,(\ref{sec:tidal_hamiltonian})). 
As a result, in the presence of tides, the parameter $\Sigma$ (Eq.\,(\ref{canonic_var_Sigma})) remains conserved. 
On the other hand, the fast angle $\gamma$ is still present.
Yet, at this stage, we cannot perform an average as in Sect.~\ref{cqaa}, because $\gamma'$ is considered as a parameter that can later cancel with $\gamma$ (see Eq.\,(\ref{eq:gamma'_expansion})).

\subsection{Secular equations of motion}

The equations of motion are obtained from Eq.\,(\ref{tidalH}) using the Hamilton equations.
The additional contributions from tides derive only from the $U_\ij$ (Eq.\,(\ref{sec:tidal_hamiltonian})), and are given by ($k=1,2$)
\begin{equation}\label{tidal:eom}
\begin{split}
\dot x_k &=  \ii \frac{\partial U_{0k}}{\partial \cartxbk} + \ii \frac{\partial U_{k0}}{\partial \cartxbk} \ , \\
\dot y_k &=\ii \frac{\partial U_{0k}}{\partial \cartybk} + \ii \frac{\partial U_{k0}}{\partial \cartybk} \ ,  \\
\dot \Theta_0 &= - \frac{\partial U_{10}}{\partial\vartheta_0} - \frac{\partial U_{20}}{\partial\vartheta_0} \ , \\
\dot \Theta_k &= - \frac{\partial U_{0k}}{\partial\vartheta_k} \ , \\
\dot \Gamma &= - \frac{\partial U_{01}}{\partial\gamma} - \frac{\partial U_{10}}{\partial\gamma} - \frac{\partial U_{02}}{\partial\gamma} - \frac{\partial U_{20}}{\partial\gamma}  \ .
\end{split}
\end{equation}
We should also write the equations for $\dot \gamma$ and $\dot \vartheta_k$. 
However, these angles disappear from the equations of motion with some of the following simplifications, and so we do not need them to get a closed set for the secular evolution of the system.

To handle the expression of the primed quantities, we need to use a tidal model.
For simplicity, we adopt here the weak friction model \citep[eg.][]{Singer_1968, Alexander_1973}, which assumes a constant and small time delay between the tidal potential and the perturbing mass, $\dt_\elli$.
This model is widely used and provides simple expressions for the tidal interactions, because it can be made linear \citep[eg.][]{Mignard_1979},
\begin{equation}\label{linear_model}
\lambda_k' \approx \lambda_k - n_k \dt_\elli \ , \quad \mathrm{and} \quad
\thetid' \approx \thetid - \angrot_\elli \dt_\elli \ .
\end{equation}
The remaining primed quantities follow as,
\begin{equation}\label{eq:x_k'_expansion}
    \cartxk' \approx \cartxk - \ii \cartxk \left( p_2 \mmj + p_1 \mmi\right)\dt_\elli \ ,
\end{equation}
\begin{equation}\label{eq:y_k'_expansion}
    \cartyk' \approx \cartyk - \ii \cartyk \left( p_2 \mmj + p_1 \mmi\right)\dt_\elli \ ,
\end{equation}
\begin{equation}\label{eq:gamma'_expansion}
    \gamma' \approx \gamma - p_1 \left(\mmj-\mmi\right)\dt_\elli \ , 
\end{equation} 
\begin{equation}\label{eq:vartheta'_expansion}
    \vartheta_\elli' \approx \vartheta_\elli + \left( p_2 \mmj + p_1 \mmi - \angrot_\elli \right)\dt_\elli \ ,
\end{equation}
with
\begin{equation}
\overline{n}_k = \beta_k^3 \mu_k^2 / \Gamma_k^{3} 
\ .
\end{equation}

We then replace expressions (\ref{eq:y_k'_expansion}) to (\ref{eq:vartheta'_expansion}) into the equations of motion (\ref{tidal:eom}) and average over the fast angle $\gamma$ (as in Eq.\,(\ref{averHam})) to finally get the secular equations for the tidal evolution
\begin{equation}\label{eq:tidal_x1_equation_of_motion}
    \begin{split}
    \dot{x}_{1}=&-\frac{3}{2}\frac{\mathcal{D}_{1 0}}{\Gamma_1^{13}}\Big(\ii(2p+5)+(19\mmi-12\angrot_0) \dt_0\Big) \cartxi 
    -\frac{3}{2}\frac{\mathcal{D}_{0 1}}{\Gamma_1^{13}}\Big(\ii(2p+5)+(19\mmi-12\angrot_1) \dt_1\Big) \cartxi \\
    &+3\ii (p+2)\frac{\mathcal{D}_{2 0}}{\Gamma_2^{13}}\,\cartxi +3\ii (p+2)\frac{\mathcal{D}_{0 2}}{\Gamma_2^{13}}\,\cartxi \ ,
    \end{split}
\end{equation}
\begin{equation}\label{eq:tidal_x2_equation_of_motion}
   \begin{split}
    \dot{x}_{2}=&-3\ii p\frac{\mathcal{D}_{1 0}}{\Gamma_1^{13}}\,\cartxj-\frac{3}{2}\frac{\mathcal{D}_{2 0}}{\Gamma_2^{13}}\Big(\ii(1-2p)+(19\mmj-12\angrot_0)\dt_0\big)\cartxj\\
    &-3\ii p\frac{\mathcal{D}_{0 1}}{\Gamma_1^{13}}\,\cartxj-\frac{3}{2}\frac{\mathcal{D}_{0 2}}{\Gamma_2^{13}}\Big(\ii(1-2p)+(19\mmj-12\angrot_2)\dt_2\Big)\cartxj \ ,
   \end{split}
\end{equation}
\begin{equation}\label{eq:tidal_y1_equation_of_motion}
    \begin{split}
    \dot{y}_{1}=&-\frac{3}{2}\frac{\mathcal{D}_{1 0}}{\Gamma_1^{13}}\left(2\ii p+\mmi \dt_0\right) \cartyi +3\ii (p+2)\frac{\mathcal{D}_{2 0}}{\Gamma_2^{13}}\,\cartyi
    -\frac{3}{2}\frac{\mathcal{D}_{0 1}}{\Gamma_1^{13}}\left(2\ii p+\mmi \dt_1\right) \cartyi +3\ii (p+2)\frac{\mathcal{D}_{0 2}}{\Gamma_2^{13}}\,\cartyi \ ,
    \end{split}
\end{equation}
\begin{equation}\label{eq:tidal_y2_equation_of_motion}
   \begin{split}
    \dot{y}_{2}=&-3\ii p\frac{\mathcal{D}_{1 0}}{\Gamma_1^{13}}\,\cartyj+\frac{3}{2}\frac{\mathcal{D}_{2 0}}{\Gamma_2^{13}}\big(2\ii(p+2)-\mmj\dt_0\big)\cartyj
    -3\ii p\frac{\mathcal{D}_{0 1}}{\Gamma_1^{13}}\,\cartyj+\frac{3}{2}\frac{\mathcal{D}_{0 2}}{\Gamma_2^{13}}\big(2\ii(p+2)-\mmj\dt_2\big)\cartyj \ ,
   \end{split}
\end{equation}
\begin{equation}\label{eq:tidal_Gamma_equation_of_motion}
     \begin{split}
     \dot{\Gamma}=3\frac{\mathcal{D}_{1 0}}{\Gamma_1^{13}}\left(1+\frac{2}{p}\right)&\bigg[\left(27\cartxi\cartxbi-\cartyi\cartybi+6p(\cartxi\cartxbi+\cartxj\cartxbj+\cartyi\cartybi+\cartyj\cartybj)+\Gamma_1\right)\angrot_0\\
     &-\left(46\cartxi\cartxbi+6p(\cartxi\cartxbi+\cartxj\cartxbj+\cartyi\cartybi+\cartyj\cartybj)+\Gamma_1\right)\mmi\bigg]\dt_0\\
     +3\frac{\mathcal{D}_{0 1}}{\Gamma_1^{13}}\left(1+\frac{2}{p}\right)&\bigg[\left(27\cartxi\cartxbi-\cartyi\cartybi+6p(\cartxi\cartxbi+\cartxj\cartxbj+\cartyi\cartybi+\cartyj\cartybj)+\Gamma_1\right)\angrot_1\\
     &-\left(46\cartxi\cartxbi+6p(\cartxi\cartxbi+\cartxj\cartxbj+\cartyi\cartybi+\cartyj\cartybj)+\Gamma_1\right)\mmi\bigg]\dt_1\\
     +3\frac{\mathcal{D}_{2 0}}{\Gamma_2^{13}}\left(1+\frac{2}{p}\right)&\left[\bigg(27\cartxj\cartxbj-\cartyj\cartybj-6(p+2)(\cartxi\cartxbi+\cartxj\cartxbj+\cartyi\cartybi+\cartyj\cartybj)+\Gamma_2\right)\angrot_0\\
     &-\left(46\cartxj\cartxbj-6(p+2)(\cartxi\cartxbi+\cartxj\cartxbj+\cartyi\cartybi+\cartyj\cartybj)+\Gamma_1\right)\mmj\bigg]\dt_0\\
     +3\frac{\mathcal{D}_{0 2}}{\Gamma_2^{13}}\left(1+\frac{2}{p}\right)&\left[\bigg(27\cartxj\cartxbj-\cartyj\cartybj-6(p+2)(\cartxi\cartxbi+\cartxj\cartxbj+\cartyi\cartybi+\cartyj\cartybj)+\Gamma_2\right)\angrot_2\\
     &-\left(46\cartxj\cartxbj-6(p+2)(\cartxi\cartxbi+\cartxj\cartxbj+\cartyi\cartybi+\cartyj\cartybj)+\Gamma_1\right)\mmj\bigg]\dt_2 \ ,
     \end{split}
\end{equation}
\begin{equation}\label{eq:tidal_Theta0_equation_of_motion}
\begin{split}
     \dot{\Theta}_0=-3\frac{\mathcal{D}_{1 0}}{\Gamma_1^{13}}&\bigg[\left(15\cartxi\cartxbi-\cartyi\cartybi+6p(\cartxi\cartxbi+\cartxj\cartxbj+\cartyi\cartybi+\cartyj\cartybj)+\Gamma_1\right)\angrot_0\\
     &-\left(27\cartxi\cartxbi-\cartyi\cartybi+6p(\cartxi\cartxbi+\cartxj\cartxbj+\cartyi\cartybi+\cartyj\cartybj)+\Gamma_1\right)\mmi\bigg]\dt_0\\
     -3\frac{\mathcal{D}_{2 0}}{\Gamma_2^{13}}&\bigg[\left(15\cartxj\cartxbj-\cartyj\cartybj-6(p+2)(\cartxi\cartxbi+\cartxj\cartxbj+\cartyi\cartybi+\cartyj\cartybj)+\Gamma_2\right)\angrot_0\\
     &-\left(27\cartxj\cartxbj-\cartyj\cartybj-6(p+2)(\cartxi\cartxbi+\cartxj\cartxbj+\cartyi\cartybi+\cartyj\cartybj)+\Gamma_2\right)\mmj\bigg]\dt_0 \ ,
     \end{split}
\end{equation}
\begin{equation}\label{eq:tidal_Theta1_equation_of_motion}
\begin{split}
     \dot{\Theta}_1=-3\frac{\mathcal{D}_{0 1}}{\Gamma_1^{13}}&\bigg[\left(15\cartxi\cartxbi-\cartyi\cartybi+6p(\cartxi\cartxbi+\cartxj\cartxbj+\cartyi\cartybi+\cartyj\cartybj)+\Gamma_1\right)\angrot_1\\
     &-\left(27\cartxi\cartxbi-\cartyi\cartybi+6p(\cartxi\cartxbi+\cartxj\cartxbj+\cartyi\cartybi+\cartyj\cartybj)+\Gamma_1\right)\mmi\bigg]\dt_1 \ ,
     \end{split}
\end{equation}
\begin{equation}\label{eq:tidal_Theta2_equation_of_motion}
\begin{split}
     \dot{\Theta}_2=-3\frac{\mathcal{D}_{0 2}}{\Gamma_2^{13}}&\bigg[\left(15\cartxj\cartxbj-\cartyj\cartybj-6(p+2)(\cartxi\cartxbi+\cartxj\cartxbj+\cartyi\cartybi+\cartyj\cartybj)+\Gamma_2\right)\angrot_2\\
     &-\left(27\cartxj\cartxbj-\cartyj\cartybj-6(p+2)(\cartxi\cartxbi+\cartxj\cartxbj+\cartyi\cartybi+\cartyj\cartybj)+\Gamma_2\right)\mmj\bigg]\dt_2 \, ,
     \end{split} 
\end{equation}  
where
\begin{equation}
    \mathcal{D}_\ij=k_{2,\elli}\mathcal{G}m_i^2\beta_k^{12}\mu_k^6 R_\elli^5 
    \quad \mathrm{with} \quad k = \mathrm{max}(i,j) \ .
\end{equation}

We note that in the expressions of $\dot{x}_k$ (Eqs.\,(\ref{eq:tidal_x1_equation_of_motion}) and (\ref{eq:tidal_x2_equation_of_motion})) and $\dot{y}_k$ (Eq.\,(\ref{eq:tidal_y1_equation_of_motion}) and ({\ref{eq:tidal_y2_equation_of_motion})), we have a conservative contribution (imaginary terms) and a dissipative contribution (real terms in $\dt_\elli$).
The conservative contributions result from a permanent tidal deformation and only slightly modify the fundamental frequencies of the system, while the dissipative contributions modify the secular evolution.

Tidal dissipation induce variations in the parameters $\Gamma$ (Eq.\,(\ref{eq:tidal_Gamma_equation_of_motion})) and $\Theta_k$ (Eqs.\,(\ref{eq:tidal_Theta0_equation_of_motion})$-$(\ref{eq:tidal_Theta2_equation_of_motion})).
Then, the coefficients $\mathcal{K}$, $\mathcal{O}$, $\mathcal{S}$ and $\mathcal{R}$ appearing in the Hamiltonian (\ref{resHamcomplex}) slowly change in time (Eqs.\,(\ref{Gb1_Gb2})$-$(\ref{DeltaRef})), which translates into a secular evolution of the system.
We also note that for the oblateness terms (Eqs.\,(\ref{Oaterm})$-$(\ref{Odterm})), changes are observed not only due to $\Gamma$, but also in $J_2$ because of the evolution of Uranus' rotation rate, $\angrot_0 =  \Theta_0 / C_0 $ (Eq.\,(\ref{eq:J20})).

\section{Planar dynamics}
\label{plandyn}

\citet{Tittemore_Wisdom_1988} have studied the passage through the 5/3~MMR between Ariel and Umbriel using a secular planar two-satellite model, which has only two degrees-of-freedom.
To better compare with our results, in this section we also restrict our model to the planar case ($\inc_k = 0$).

\subsection{Conservative motion}

The planar case is equivalent to set $\cartyk=0$ in the resonant secular Hamiltonian (Eq.\,(\ref{resHamcomplex})), which suppresses the terms in $\Oc$, $\Od$, $\Se$, $\Sf$, $\Sg$, $\Rd$, $\Re$, and $\Rf$, and further simplifies the terms in $\Ka$, $\Kb$, and $\Sa$, yielding to
\begin{equation}\label{resplanarHamcomplex}
    \begin{split}
       \MedH &=(\Ka+\Sa)\left(\cartxi\cartxbi+\cartxj\cartxbj\right)+\Kb\left(\cartxi\cartxbi+\cartxj\cartxbj\right)^2    \\
        &+(\Oa+\Sb)\cartxi\cartxbi+(\Ob+\Sc)\cartxj\cartxbj+\frac{\Sd}{2}\left(\cartxi\cartxbj+\cartxbi\cartxj\right)\\
        &+\frac{\Ra}{2}\left(\cartxi^2+\cartxbi^2\right)+\frac{\Rb}{2}\left(\cartxj^2+\cartxbj^2\right)+\frac{\Rc}{2}\left(\cartxi\cartxj+\cartxbi\cartxbj\right) \ .
    \end{split} 
\end{equation}

The values of $m_k$, $J_2$, and $\Ru$ are relatively well determined for the Uranian system (Table~\ref{table:physical_orbital_parameters}).
Therefore, to compute the remaining coefficients $\mathcal{K}$, $\mathcal{O}$, $\mathcal{S}$, and $\mathcal{R}$ appearing in the Hamiltonian (Eq.\,(\ref{resHamcomplex})), we only need to know the values of the parameters $\Gamma_1$ and $\Gamma_2$ (see Appendix~\ref{sec:Conservative_Hamiltonian_terms}), which in turn depend on the parameters $\Gamma$ and $\Delta$ (Eq.\,(\ref{Gb1_Gb2})).

In Paper~I, we performed a backwards tidal evolution of the semi-major axes of Ariel and Umbriel, and estimated that the 5/3~MMR between Ariel and Umbriel was crossed about 640~Myr ago, with 
\begin{equation}\label{nominal:sma}
    a_1/\Ru=7.39054 \quad \mathrm{and} \quad a_2/\Ru=10.38909 \,.
\end{equation}
Adopting these semi-major axes and considering for simplicity $e_k=\inc_k=0$, we obtain (Eq.\,(\ref{canonic_var_Gamma}))
\begin{equation}\label{Gammafix}
{\Gamma=\num{2.647289e-12}  \ \mathrm{M_\odot \, au^2 \, yr^{-1}} } \ .
\end{equation}
The conservative dynamics is not very sensitive to the $\Gamma$ parameter \citep[eg.][]{Tittemore_Wisdom_1988, Tittemore_Wisdom_1989}, and so we fix it at the reference value estimated for the near resonance encounter (Eq.\,(\ref{Gammafix})).

The dynamics of the 5/3~MMR thus mainly depends on the $\Delta$-parameter (Eq.\,(\ref{DeltaRef})), which measures the proximity to the resonance.
Following \citet{Delisle_etal_2012} and \citet{Gomes_Correia_2023}, we write
\begin{equation}\label{delta:equation}
\delta = \frac{\Delta}{\Delta_r} - 1 \ ,
\end{equation}
where $\Delta_r$ is the value of $\Delta$ at the circular planar ($e_k=\inc_k=0$) nominal resonance, that is (Eq.\,(\ref{canonic_var}) and (\ref{DeltaRef})), 
\begin{equation}\label{Deltari}
\Delta_r = \left(\Lambda_{1,r} + \Lambda_{2,r}\right) / \Gamma_r \ ,
\end{equation}
when $n_1/n_2=5/3$, where $n_k$ is the mean motion of the satellite with mass $m_k$.
At the nominal resonance, using Kepler's third law, we have 
\citep{Gomes_Correia_2023}
\begin{equation}\label{Deltarf}
\Delta_r = \left( 1 + \epsilon \left(\tfrac53\right)^{1/3} \right) \left( \tfrac53 + \epsilon \left(\tfrac53\right)^{1/3} \right)^{-1}   \ , \quad \mathrm{with} \quad \epsilon \approx \frac{m_2}{m_1} \ .
\end{equation}

\subsubsection{Equilibrium points}
The equilibrium points correspond to stationary solutions of the Hamiltonian. 
They can be found by solving 
\begin{equation}\label{eq:equilibrium_points_condition_cartesian}
    \frac{\partial\MedH}{\partial\cartxi}=0 \quad \mathrm{and} \quad \frac{\partial\MedH}{\partial\cartxj}=0 \ ,
\end{equation}
which correspond to the roots of equations (\ref{eq:conservative_motion_equations_x1}) and (\ref{eq:conservative_motion_equations_x2}) with $\cartyk=0$.
Splitting these equations in their real and imaginary parts, ${\cartxk = x_{k,r} + \ii x_{k,i}}$, and following a similar procedure as in \citet{Gomes_Correia_2023}, we find that stable equilibria can only occur when the real roots are null, i.e., $x_{1,r} = x_{2,r} = 0$.
We then focus on the imaginary roots to determine the exact position of the stable equilibria. 
Since ${x_{1,r}=x_{2,r}=0}$, we have (Eq.\,(\ref{eq:complex_cartesian_coordinates}))
\begin{equation}\label{imag_roots_x}
x_{1,i} = \pm \sqrt{\Sigmai }\quad \mathrm{and} \quad x_{2,i}= \pm \sqrt{\Sigmaj} \ ,
\end{equation}
with
\begin{equation}\label{root0_Sigma}
\Sigmai=\Sigmaj=0 \ ,
\end{equation}
or
\begin{equation}\label{root1_Sigma}
    \Sigmai = \frac{(\Rc-\Sd)\epsx-2(\Ka+\Oa+\Sa+\Sb-\Ra)}{4 \Kb (1+\epsx^2)} \ ,
\end{equation}
\begin{equation}\label{root2_Sigma}
\Sigmaj = \epsx^2 \, \Sigmai \ ,
\end{equation}
where 
\begin{equation}\label{root_split}
\begin{split}
    \epsx=&\frac{\Oa - \Ob - \Ra + \Rb + \Sb - \Sc}{\Rc - \Sd}
   \pm \frac{ \sqrt{(\Oa - \Ob - \Ra + \Rb + \Sb - \Sc)^2 + (\Rc - \Sd)^2}}{\Rc - \Sd} \ .
   \end{split}
\end{equation}

The equilibrium point at $\Sigmai=\Sigmaj=0$ (Eq.\,(\ref{root0_Sigma})) is always present, although it can be stable or unstable.
The remaining equilibria (Eqs.\,(\ref{root1_Sigma})$-$(\ref{root_split})) only exist for some $\delta$-values.
In Fig.~\ref{Fig:equilibrium_points}, we show the evolution of the equilibrium points as a function of $\delta$. We rescale $x_{k,i}$ by $\sqrt{\Gamma_k/2}$, such that we can translate the different equilibria in terms of eccentricities (Eq.\,({\ref{yinc})).

\begin{figure}
    \centering
    \includegraphics[width=0.6\linewidth]{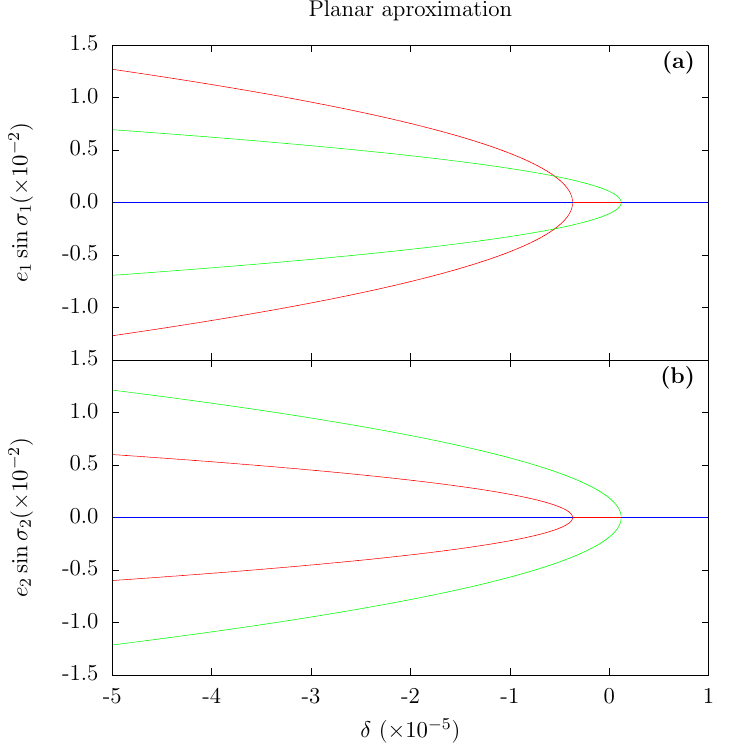}
    \caption{Evolution of the equilibrium points as function of $\delta$. The green lines represent stable points inside the resonance (in a libration region), the red lines represent hyperbolic points (unstable), and the blue lines represent stable fixed points (in a circulation region).}
    \label{Fig:equilibrium_points}
\end{figure}

For positive $\delta$ values far from zero, there is only one equilibrium point at $e_k=0$, which is stable (in blue colour).
For $\delta = \num{1.2e-6}$, there is a first bifurcation in the equilibria: two new stable equilibrium appear at non-zero eccentricity (in green colour), while the point at $e_k=0$ becomes unstable. 
For $\delta = \num{-3.7e-6}$, which is close to the resonance nominal value $\delta=0$ (Eq.\,(\ref{Deltari})), a second bifurcation arises: two additional unstable equilibrium points appear at non-zero eccentricity (in red colour), while the point at $e_k=0$ becomes stable again.
This geometry of the eccentricity equilibrium points in the planar case is similar to the one obtained for the inclination equilibrium points in the circular approximation \citep[see Fig.~1 in][]{Gomes_Correia_2023}.

\subsubsection{Energy levels}
\label{sec:energylev}

For a deeper understanding of the dynamics, we can examine the energy levels of the resonant planar Hamiltonian (Eq.\,(\ref{resplanarHamcomplex})) for various values of $\delta$ (Eq.\,(\ref{Deltarf})).
Given that our problem involves two degrees of freedom, thus four dimensions, we need to depict these levels on sectional planes. 
To ensure that all stable equilibria are visible, we choose the plane ($x_{1,i}, x_{2,i}$) with $x_{1,r}=x_{2,r}=0$ (Eq.\,(\ref{imag_roots_x})).
In Fig.~\ref{fig:level_curves}, we show the energy levels for three distinct values of $\delta$, each representative of the three equilibrium possibilities depicted in Fig.~\ref{Fig:equilibrium_points}.
Once more, we rescale the $x_{k,i}$ by $\sqrt{\Gamma_k/2}$, and so we actually show the energy levels in the plane ($e_1 \sin \sigmai, e_2 \sin \sigmaj $) with $\cos \sigmai = \cos \sigmaj = 0$.
As for the equilibrium points, we observe that the typology of the energy level curves in the eccentric planar case is also similar to that of the energy level curves obtained for the inclinations in the circular case \citep[see Fig.~2 in][]{Gomes_Correia_2023}, although here the symmetry axis of the figure is more tilted.

\begin{figure}
    \centering
    \includegraphics[width=0.4\linewidth]{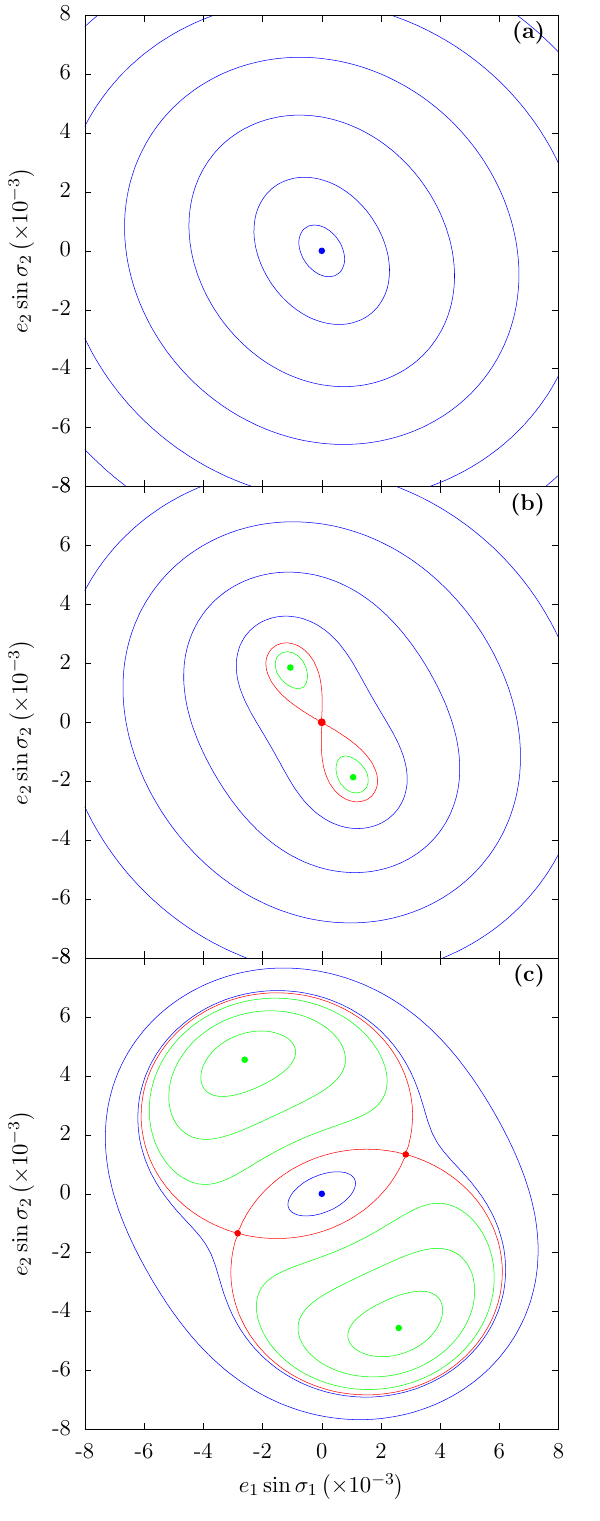}
    \caption{Energy level curves in the plane ($e_1 \sin \sigmai, e_2 \sin \sigmaj$) with $\cos \sigmai = \cos \sigmaj =0$, for $\delta=\num{5e-6}$ (top), $\delta=0$ (middle), and $\delta=\num{-5e-6}$ (bottom). Stable equilibria are coloured in green (libration) and blue (circulation), while unstable equilibria are coloured in red, as well as the level curves that correspond to the separatrix.}
    \label{fig:level_curves}
\end{figure}

For $\delta = \num{5e-6} > 0 $ (Fig.~\ref{fig:level_curves}\,a) there is a single equilibrium point $(x_{1,i}=0, x_{2,i}=0)$ at the centre (in blue). 
It corresponds to a fix point of the planar Hamiltonian (Eq.\,(\ref{resplanarHamcomplex})), encircled by a circulation region. 
Therefore, in this case (and for higher $\delta$-values), all trajectories lie outside the 5/3~MMR

For $\delta=0$ (Fig.~\ref{fig:level_curves}\,b), the system is at the nominal resonance (Eq.\,(\ref{delta:equation})).
Here, the equilibrium point at the centre  $(x_{1,i}=0, x_{2,i}=0)$ is still present (in red), but it becomes unstable.
Notably, a separatrix in a tilted 8-shape emerges from this point, encompassing two additional stable equilibrium points (depicted in green).
Trajectories inside the separatrix that encircle the stable points are in libration and correspond to orbits inside the 5/3~MMR.
Trajectories outside the separatrix are in circulation.

Finally, for $\delta=\num{-5e-6} < 0$ (Fig.~\ref{fig:level_curves}\,c), five equilibrium points exist.
Two hyperbolic points (depicted in red) give rise to a separatrix with two 'banana' shapes. This separatrix delineates the phase space into regions of libration and circulation.
There are two stable points (in green), one inside each banana island.
Trajectories revolving around these points undergo libration and correspond to orbits within the 5/3~MMR.
The point at the origin $(x_{1,i}=0, x_{2,i}=0)$ (in blue) is again stable and inside a small circulation region.
Trajectories outside the separatrix are also in circulation.
This phase space configuration persists for smaller $\delta$-values, but the central circulation region becomes larger, while the resonant islands become smaller.

\subsubsection{Capture probabilities}
\label{sec:cap_prob}

Understanding the system's behaviour upon crossing the 5/3~MMR is not straightforward due to the problem's two degrees of freedom. 
To gain insight into the critical eccentricities that either trap the system in resonance or allow to skip it, we can construct a simplified one-degree-of-freedom model and estimate the capture probability for various eccentricities, following \citet{Tittemore_Wisdom_1988}.

Using the same method described in \citet{Gomes_Correia_2023} to study the effect of inclination on the capture probability of the 5/3 MMR between Ariel and Umbriel, we obtain the capture probability as a function of the eccentricity,
\begin{equation}\label{eq:capture_probility_equation}
    P_\mathrm{cap} = 1 + \frac{\partial \Jk^-}{\partial\Gamma} \bigg/ \, \frac{\partial \Jk^+}{\partial\Gamma} \ ,
\end{equation}
where,
\begin{equation}
    \begin{split}
        \frac{\partial \Jk^-}{\partial\Gamma}&=\frac{1}{\Bk^2}\left(\arcsin{\left(\sqrt{-\frac{\Ck}{\Ak}}\right)}+\frac{\pi }{2}\right)(\Bk \Ak'-\Ak \Bk')
        +\frac{1}{\Bk^2}\sqrt{-\frac{\Ck+\Ak}{\Ck}} (\Bk \Ck'-\Ck \Bk')   \ ,
    \end{split}
\end{equation}

\begin{equation}
    \begin{split}
        \frac{\partial \Jk^+}{\partial\Gamma}&=\frac{1}{\Bk^2}\left(\arcsin{\left(\sqrt{-\frac{\Ck}{\Ak}}\right)}-\frac{\pi }{2}\right)(\Bk \Ak' - \Ak \Bk')
        +\frac{1}{\Bk^2}\sqrt{-\frac{\Ck+\Ak}{\Ck}} (\Bk \Ck'-\Ck \Bk') \ , 
    \end{split}
\end{equation}
and $\Ak' = {\partial \Ak}/{\partial\Gamma}$, $\Bk' = {\partial \Bk}/{\partial\Gamma}$, and $\Ck' = {\partial \Ck}/{\partial\Gamma}$. Note that 
since $\Gamma$ is the only time dependent quantity in the expressions of $\Ak$, $\Bk$, and $\Ck$ (appendix~\ref{sec:Conservative_Hamiltonian_terms}), $\dot{\Jk}= \dot{\Gamma} \, {\partial\Jk} / {\partial\Gamma}$ (we neglect the small changes in $\Theta$ from the oblateness terms, ${\cal O}_k$).
For the eccentricity, the single resonance terms are
\begin{equation}\label{simpleHam}
    \MedH_k=\Ak \cartxk \cartxbk + \Bk \left( \cartxk \cartxbk \right)^2+\Ck/2\left(\cartxk^2+\cartxbk^2\right)
\end{equation}
with 
\begin{equation}
    \mathcal{A}_1=\Ka+\Oa+\Sa+\Sb \ , \quad \mathcal{B}_1 = \Kb \ , \quad \mathcal{C}_1 = \Ra \ ,
\end{equation}
\begin{equation}
    \mathcal{A}_2=\Ka+\Ob+\Sa+\Sc \ , \quad \mathcal{B}_2 = \Kb \ , \quad \mathcal{C}_2 = \Rb \ .
\end{equation}

\begin{figure}
    \centering
    \includegraphics[width=0.45\linewidth]{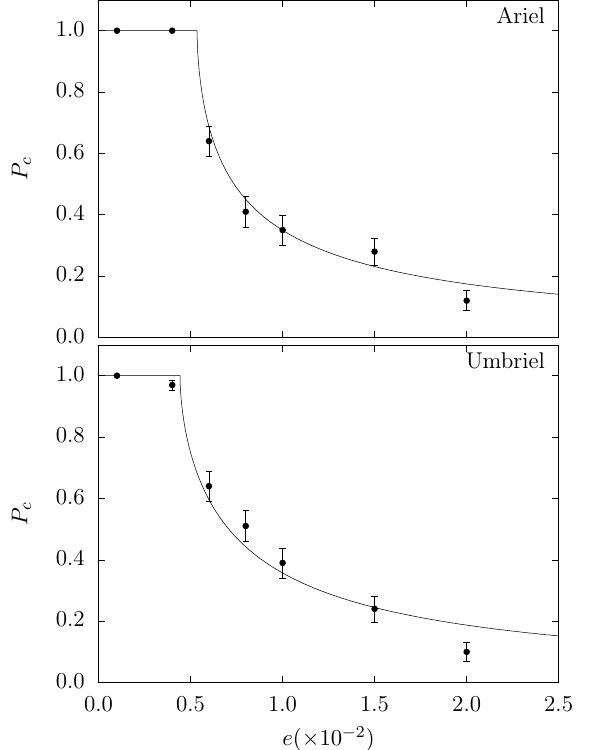}
    \caption{Capture probabilities in the $\sigmai$ (top) and $\sigmaj$ (bottom) resonances. The solid line gives the theoretical approximation given by Eq.\,(\ref{eq:capture_probility_equation}), while the dots give the results of numerical simulations. We ran 100 initial conditions with $\sigmaa_k$ differing by $1.8\degree$.} 
   \label{fig:capture_probability_eAU}
\end{figure}

In Fig.~\ref{fig:capture_probability_eAU}, we show the probability of capture in the $\sigmai$ and $\sigmaj$ resonances obtained with expression (\ref{eq:capture_probility_equation}).
For some eccentricity values,   numerical integrations of the equations of motion derived from the simplified Hamiltonian (Eq.\,(\ref{simpleHam})) together with the secular tidal equations (Eqs.\,(\ref{eq:tidal_x1_equation_of_motion}), (\ref{eq:tidal_x2_equation_of_motion}), and (\ref{eq:tidal_Theta0_equation_of_motion})) are also presented.
For each initial eccentricity, we ran 100 simulations where the initial angle $\sigmaa_k$ is uniformly sampled.
The amount of simulations captured in resonance are marked with a dot.
The statistical fluctuation, represented as error bars, were estimated using binomial statistics.
We observe a close correspondence between the theoretical curve (Eq.\,(\ref{eq:capture_probility_equation})) and the output of the numerical simulations, i.e, the adiabatic approximation holds.
These results align with those presented in Fig.~31 in \citet{Tittemore_Wisdom_1988}.

In Fig.~\ref{fig:capture_probability_eAU}, we observe that for initial eccentricities lower than $~0.005$ the system is consistently captured in resonance. 
However,  as the initial eccentricity increases, the capture probability quickly decreases, it becomes $\sim 40\%$ for $e_k \sim 0.01$.
This results suggest that a system with nearly circular orbits is unlikely to escape the 5/3~MMR. Conversely, for eccentricities higher than about $0.01$, it may be able to evade it.

We cannot entirely rely on the conclusions drawn from the simplified Hamiltonian, primarily for two reasons. 
Firstly, the complete planar Hamiltonian (Eq.\,(\ref{resplanarHamcomplex})) depends on the eccentricity of the other body.
When we simplified the Hamiltonian (Eq.\,(\ref{simpleHam})) for $\cartxi$, we omitted all terms involving $\cartxj$ (and vice versa) by setting $\cartxj=0$.
However, if we adopt $\cartxj \ne 0$, additional terms emerge in the Hamiltonian, resulting in a distinct distribution of capture probabilities.
Secondly, the complete Hamiltonian has two degrees-of-freedom, and thus, for certain combinations of eccentricity values, the system can exhibit chaotic behaviour (see next section).

\subsubsection{Chaotic diffusion}
\label{sec:stabmaps}

The energy levels obtained in Sect.~\ref{sec:energylev} allow us to identify the different regions of the phase space (Fig.~\ref{fig:level_curves}), but a priori they do not correspond to trajectories followed by the system.
Indeed, since the planar problem has two degrees-of-freedom, hence four dimensions, the energy levels show the trajectories when they cross the section plane with $x_{1,r}=x_{2,r}=0$, which only remain constant for the equilibrium points.
Therefore, to study the global dynamics, in this section we adopt the frequency analysis method \citep{Laskar_1990, Laskar_1993PD} to map the diffusion of the orbits, as explained in \citet{Gomes_Correia_2024proc}.

In Fig.~\ref{fig:level_curves}, we observe that the more diverse dynamics occurs for $\delta=-5 \times 10^{-6}$.
We then use this value to construct the diffusion maps.
For each map we fix an energy value and build a grid of $200 \times 200$ equally distributed initial conditions in the plane ($x_{1,i},x_{1,r}$).
We fix $x_{2,r}=0$ for all initial conditions and compute $x_{2,i}$ from the total energy (Eq.\,(\ref{resplanarHamcomplex})).
Since the planar Hamiltonian is a fourth-degree function of $\cartxk$, the intersection of the constant energy manifold with a plane can yield up to four roots (families). Each family corresponds to a different dynamical behavior, requiring individual plotting. However, due to their symmetry, we only need to show two of them. We chose to represent the families with the positive roots, labelled 1 and 2.
We then numerically integrate the planar equations of motion (\ref{eq:conservative_motion_equations_x1}) and (\ref{eq:conservative_motion_equations_x2}), with $\cartyk=0$, for a time $T$.
Finally, we perform a frequency analysis of $\cartxi$, using the software TRIP \citep{Gastineau_Laskar_2011} over the time intervals $[0,T/2]$ and $[T/2,T]$, and determine the main frequency in each interval, $f_{\text{in}}$ and $f_{\text{out}}$, respectively.
The stability of the orbit is measured by the index
\begin{equation}
D \equiv \left\lvert1 - \frac{f_{\text{out}}}{f_{\text{in}}}\right\rvert \ ,
\label{deltaindex}
\end{equation}
which estimates the stability of the orbital long-distance diffusion \citep{Dumas_Laskar_1993}.
The larger $D$, the more orbital diffusion exists.
For stable motion, we have $D \sim 0$, while $D \ll 1$ if the motion is weakly perturbed, and $D \sim 1$ when the motion is irregular.
It is difficult to determine the precise value of $D$ for which the motion is stable or unstable, but a threshold of stability $D_s$ can be estimated such that most of the trajectories with $D < D_s$ are stable \citep[for more details see][]{Couetdic_etal_2010}.
The diffusion index $D$ is represented by a logarithmic colour scale calibrated such that blue and green correspond to quasi-periodic trajectories ($D \ll D_s$), while orange and red correspond to chaotic motion ($D \gg D_s$).

We show the diffusion maps of Ariel for family~1 in Fig.~\ref{fig:stabmap_ariel_fam1} and for family~2 in Fig.~\ref{fig:stabmap_ariel_fam2}.
We rescaled $\cartxk$ again by $\sqrt{\Gamma_k/2}$, and so we actually show the surface sections in the plane ($e_1 \sin \sigmai, e_1 \cos \sigmai $) with $\cos \sigmaj = 0$.
Each panel corresponds to a different energy value $\MedH/\Energinion_0$ (Eq.\,(\ref{resplanarHamcomplex})), corresponding to the levels shown in Fig.~\ref{fig:level_curves}\,c, where $\Energinion_0=\num{5.789e-20}$~$M_\odot \, \mathrm{au}^2 \, \mathrm{yr}^{-2}$ is the energy of the of the separatrix.
The lowest energies occur in the circulation regions, $\MedH<\SecH$, while the largest energies occur in the libration region, $\MedH>\SecH$.
The inner circulation region is delimited by $0<\MedH<\SecH$, where $\MedH=0$ corresponds to the energy of the equilibrium point with $x_1=x_2=0$.
For this energy range, there are four families, while for the remaining energies only two families exist.

\begin{figure*}
    \centering
   \includegraphics[width=0.9\textwidth]{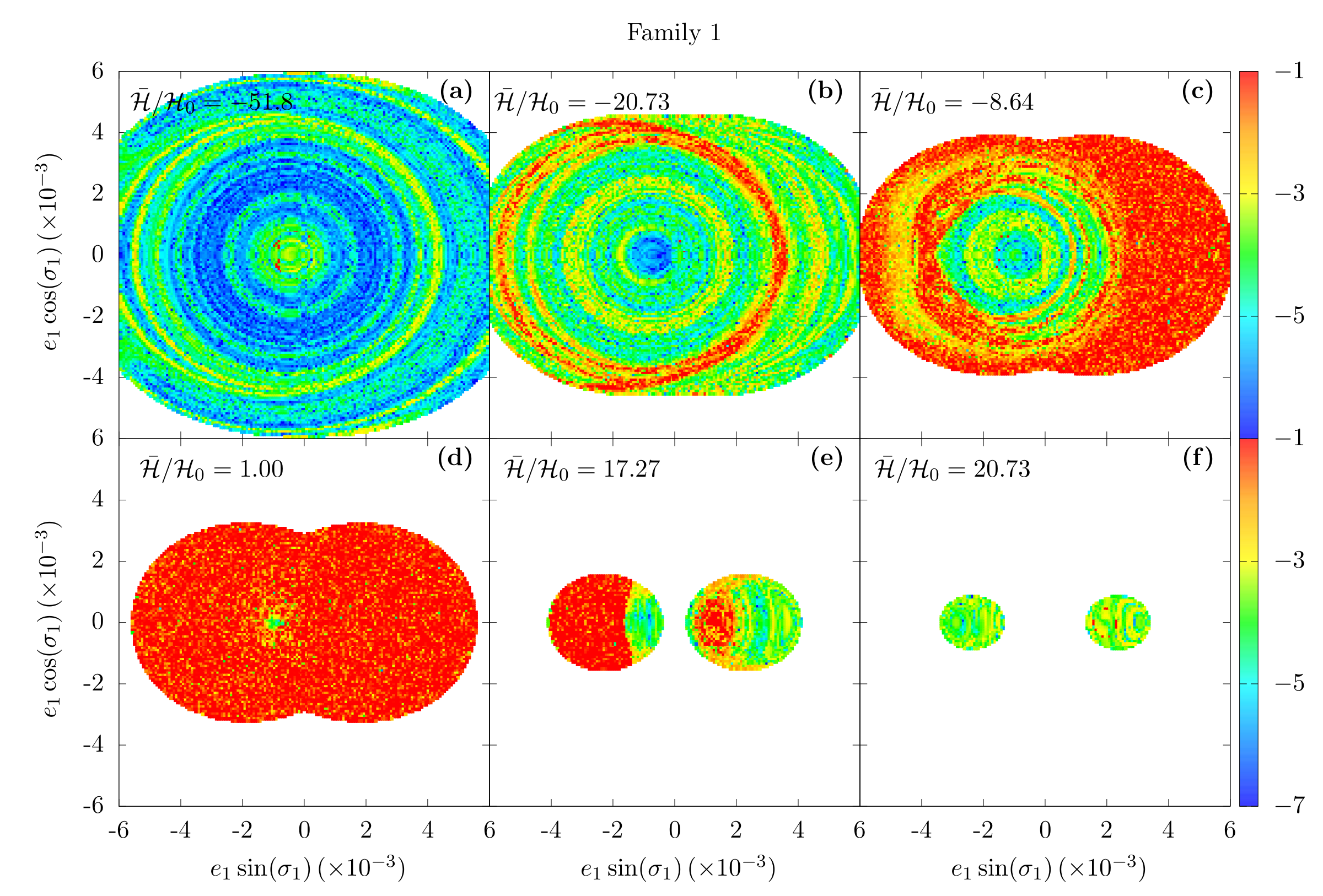}
    \caption{Diffusion maps for Ariel in the plane ($e_1 \sin \sigmai, e_1 \cos \sigmai $) with $\cos \sigmaj=0$ and $\delta=-5 \times 10^{-6}$, for trajectories corresponding to family 1. Each panel was obtained with a different energy value, and $\Energinion_0=\num{5.789e-20}$~$\rm{M_\odot \, au^2 \, yr^{-2}}$. The colour scale corresponds to the relative frequency diffusion index in logarithmic scale (Eq.\,(\ref{deltaindex})). More negative values correspond to stable orbits, while larger values correspond to more chaotic orbits. }
    \label{fig:stabmap_ariel_fam1}
\end{figure*}

\begin{figure}
    \centering
   \includegraphics[width=0.4\linewidth]{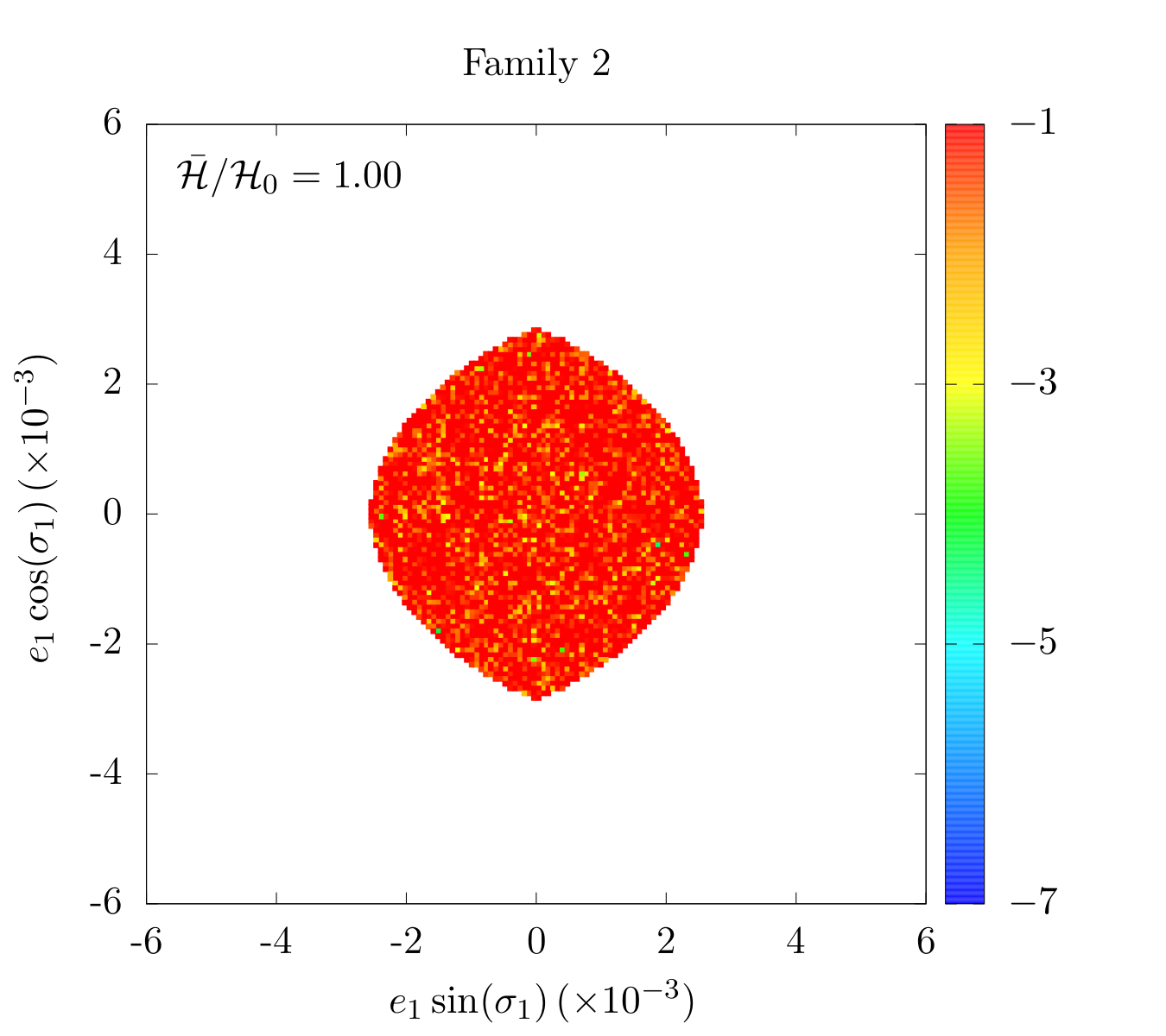}
    \caption{Diffusion map for Ariel in the plane ($e_1 \sin \sigmai, e_1 \cos \sigmai $) with $\cos \sigmaj=0$ and $\delta=-5 \times 10^{-6}$, for trajectories corresponding to family 2 and $\MedH = \Energinion_0=\num{5.789e-20}$~$\rm{M_\odot \, au^2 \, yr^{-2}}$. The colour scale is the same from Fig.~\ref{fig:stabmap_ariel_fam1}.}
    \label{fig:stabmap_ariel_fam2}
\end{figure}

For $\MedH \ll 0$ (Fig.~\ref{fig:stabmap_ariel_fam1}\,a), only family 1 exists, and we observe that the system is always quasi-periodic, corresponding to trajectories in the outer circulation region (blue line trajectories in Fig.~\ref{fig:level_curves}\,c).
As the energy increases, concentric chaotic regions start to emerge outside the separatrix (Fig.~\ref{fig:stabmap_ariel_fam1}\,b) and become progressively dominanting (Fig.~\ref{fig:stabmap_ariel_fam1}\,c). 

As the energy becomes positive, $\MedH>0$, the stable internal quasi-periodic region becomes increasingly smaller, up to $\MedH=\SecH$ (red line trajectories of Fig.~\ref{fig:level_curves}\,c), were the phase space is completely dominated by chaotic trajectories (Fig.~\ref{fig:stabmap_ariel_fam1}\,d).
For $0<\MedH<\SecH$, family 2 also exists (Fig.~\ref{fig:stabmap_ariel_fam2}). 
Yet, it maintains the regime observed in the separatrix, with chaos dominating all trajectories.

Finally, for $\MedH\gg\SecH$ (green resonance regions of Fig.~\ref{fig:level_curves}\,c), two resonant islands emerge. 
Here, the phase-space can be stable again (Fig.~\ref{fig:stabmap_ariel_fam1}\,e), and confined to small eccentricity ranges. 
For $\MedH$ close to the libration centers, the trajectories are fully stable and quasi-periodic (Fig.~\ref{fig:stabmap_ariel_fam1}\,f).

From the analysis of the planar diffusion maps, we conclude that, as for the circular dynamics \citep{Gomes_Correia_2023,Gomes_Correia_2024proc},  the planar dynamics of the 5/3~MMR between Ariel and Umbriel is highly intricate and contingent upon the energy of the system.
In fact, the energy depends on the value of the eccentricities (Eq.\,(\ref{resplanarHamcomplex})), given by the variables $x_1$ and $x_2$ (Eq.\,(\ref{yinc})).
Therefore, the value of the eccentricity of Ariel and Umbriel when the system encounters the resonance can trigger completely different behaviours.
For $\MedH \ll \SecH$, the motion is quasi-periodic. 
Still, for a wide range of energies close to the separatrix, and mainly for $\MedH \sim \SecH$, the motion is mostly chaotic.
Only for $\MedH \gg \SecH$, the motion is again quasi-periodic, but only possible in libration with a small amplitude around the high eccentricity stable equilibrium points (Fig.~\ref{Fig:equilibrium_points}).

\subsection{Numerical simulations}
\label{sec:Numerical_integration_planar}

Owing to the chaotic diffusion, the passage of Ariel and Umbriel through the 5/3 MMR is a stochastic problem.
Depending on the initial eccentricity of these satellites, the system may experience quite different evolutions (Sect.~\ref{sec:stabmaps}).
For the trajectories crossing the chaotic regions, the final outcome is unpredictable and can only be accessed using a large number of numerical simulations to have an accurate statistics of all possible scenarios.
Therefore, in this section we integrate the conservative differential equations (\ref{eq:conservative_motion_equations_x1}) and (\ref{eq:conservative_motion_equations_x2}) together with the dissipative tidal equations (\ref{eq:tidal_x1_equation_of_motion}), (\ref{eq:tidal_x2_equation_of_motion}), and (\ref{eq:tidal_Gamma_equation_of_motion}) to (\ref{eq:tidal_Theta2_equation_of_motion}) with $\cartyk=0$ (planar motion).

\subsubsection{Setup}\label{sec:num_setup}

When the 5/3 MMR is crossed, we cannot perform a backwards integration. 
We need to place the system slightly before the resonance encounter and then integrate it forwards.
It is not possible to determine the exact semi-major axes prior to resonance crossing. 
Nonetheless, if the system does not spend much time in resonance, the semi-major axes should not differ much from the nominal values (Eq.\,(\ref{nominal:sma})).
We still need to slightly decrease $a_1$ (or increase $a_2$) to move the system out of the nominal resonance. 
Since tides are stronger in Ariel, we opted to shift $a_1$ and keep $a_2$ constant:
\begin{equation}\label{preres:sma}
    a_1/\Ru=7.3868 \, , \quad a_2/\Ru=10.3891 \, . 
\end{equation}
These values of the semi-major axes allow us to compute the initial $\Gamma$ parameter (Eq.\,(\ref{canonic_var})). 
For a given set of initial eccentricity values, we then also compute the initial value of $\Delta$ (Eq.\,(\ref{DeltaRef})), which translates into an initial $\delta > 0$ (Eq.\,(\ref{delta:equation})).

The physical properties of Uranus and its satellites can be found in Table~\ref{table:physical_orbital_parameters}, while the constant $\Sigma$ parameter is conserved and can also be obtained from the present system (Eq.\,(\ref{canonic_var_Sigma})),
\begin{equation}\label{SigmaTOT}
\Sigma = \num{9.367247e-10} \ \mathrm{M_\odot \, au^2 \, yr^{-1}} \ .
\end{equation}
The initial rotation rate of Uranus 
is obtained from the conservation of $\Sigma$ (Eq.\,(\ref{canonic_var_Sigma})), together with Eqs.\,(\ref{GammaOne}) and (\ref{GammaTwo}),
\begin{equation}\label{eq:angrot_correction}
    \angrot_0=\frac{1}{C_0}\left(\Sigma-\Gamma_1-\Gamma_2-\Theta_1-\Theta_2\right) \ ,
\end{equation}
where the $\Gamma_k$ (Eq.\,(\ref{Lambdak})) are obtained with the pre-resonance semi-major axes (Eq.\,(\ref{preres:sma})), and the rotation rate of the satellites is assumed to be synchronous with the orbital period, that is, $\Theta_k = C_k n_k $ (Eq.\,(\ref{Lrot})). 

Finally, for the tidal dissipation of Uranus, we adopt $\klU = 0.104$ \citep{Gavrilov_Zharkov_1977} and $\dt_0=0.617~\mathrm{s}$ (corresponding to $Q_0 = 8\,000$, see Sect.~3.1 in Paper~I for details), which translates into
\begin{equation}\label{k2sQ}
    \klU \, \dt_0 = 0.064 \, \mathrm{s} \ .
\end{equation}
As for the satellites, we adopt ${\kli=\num{1.02e-2}}$ and ${\klj=\num{7.35e-2}}$ \citep{Chen_etal_2014}, together with $\dt_1=69.3$~s and $\dt_2=113.9$~s (corresponding to $Q_k = 500$, see Sect.~3.2 in Paper~I for details), yielding to 
\begin{equation}
\kli \, \dt_1 = 0.707 \, \mathrm{s} \ , \quad \mathrm{and} \quad \klj \, \dt_2 = 0.837\,  \mathrm{s} \,.
\end{equation}

\subsubsection{Comparison with analytical estimations}

Tidal effects are usually weak and introduce only a small drift in the phase space of the resonant dynamics.
To verify that the adiabatic approximation holds in the planar case, we ran two numerical simulations with different initial conditions and then superimposed the output in the equilibria map as a function of $\delta$ (Fig.~\ref{Fig:equilibrium_points}).
The results are shown in Fig.~\ref{fig:overlap}.
Since $\angrot_0 / n_k > 1$, tidal effects are expected to increase the value of $\Gamma$ (Eq.\,(\ref{eq:tidal_Gamma_equation_of_motion})) and thus decrease the value of $\delta$ (Eq.\,(\ref{delta:equation})).
Therefore, the results as a function of time must be read from the right to the left.

\begin{figure*}
    \centering
    \includegraphics[width=\textwidth]{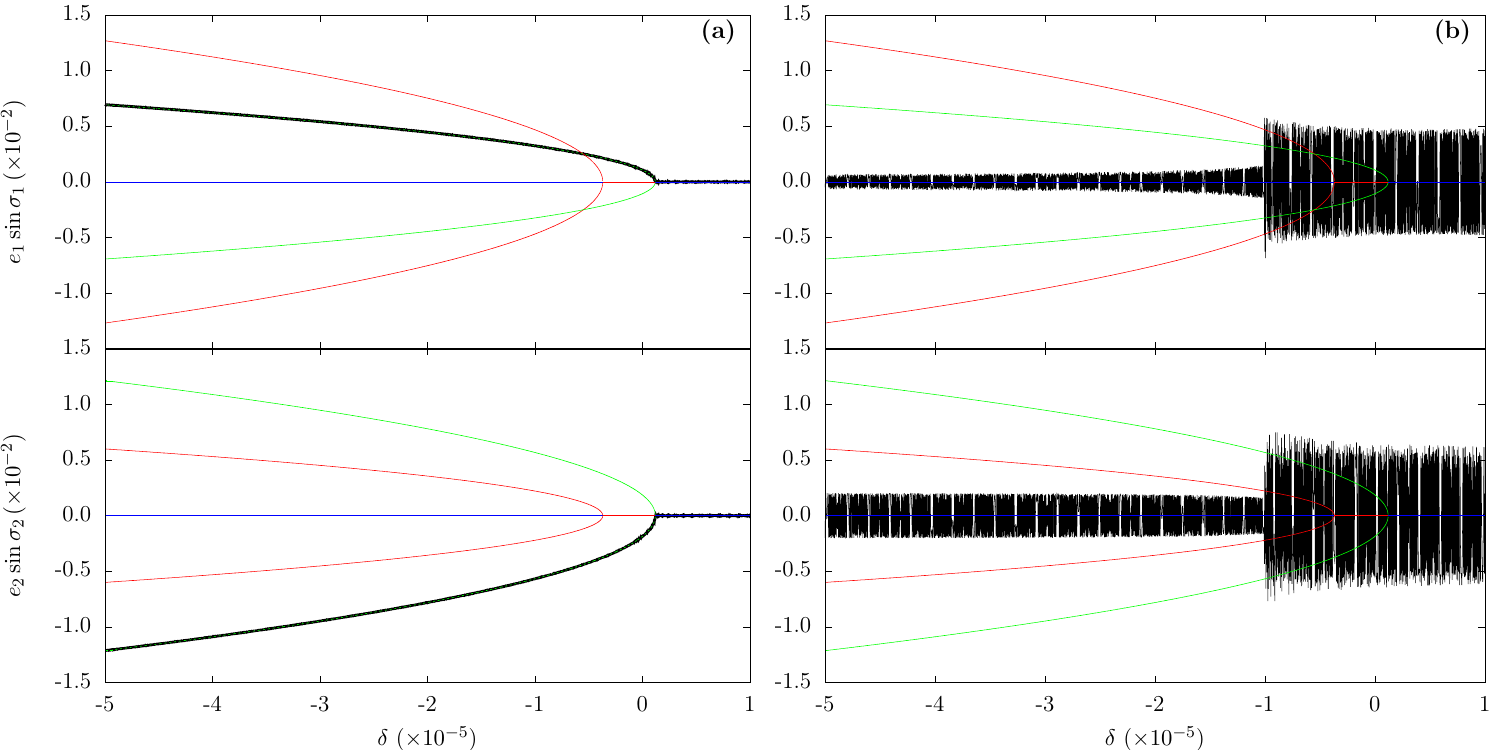}
    \caption{Two examples of tidal evolution of the system as a function of $\delta$ for small initial  $e_1=e_2=10^{-5}$ (a) and higher initial $e_1=e_2=0.005$ (b). The results of the numerical simulations (in black) are superimposed in the equilibria map for the eccentricity (Fig.~\ref{Fig:equilibrium_points}). We show the evolution for the resonant angle $\sigmai$ (top) and $\sigmaj$ (bottom). The results as a function of time must be read from the right to the left.   \label{fig:overlap}}
\end{figure*}

In Fig.~\ref{fig:overlap}\,a, we adopted very small initial eccentricities for both satellites, $e_1 = e_2 = 10^{-5}$.
At the beginning of the simulation, when $\delta > 0$, the system is in circulation with a small amplitude around the equilibrium point at zero $(\cartxi=0, \cartxj=0)$.
For $\delta \approx 0$, the system encounters the 5/3~MMR and two stable equilibrium points emerge, while the equilibrium point at zero becomes unstable.
Because the amplitude of oscillation is small, the system is forced to follow one of the two resonance branches. 
Therefore, as the system evolves with $\delta < 0$, the eccentricities increase.
For initially near circular orbits, it is then impossible to avoid capture in the 5/3~MMR (see also Sect.~\ref{sec:cap_prob}).

In Fig.~\ref{fig:overlap}\,b, we adopted higher initial eccentricities for both satellites, $e_1 = e_2 = 0.005$.
The initial evolution for $\delta > 0$ is similar to the case with lower initial eccentricities, except that the amplitude of oscillation is 500 times larger in this case.
Therefore, as the system encounters the resonance at $\delta \approx 0$, it is not able to follow one of the resonant branches and it remains in a chaotic region around the separatrix (see Sect.~\ref{sec:stabmaps}).
After some time  in the chaotic region with $\delta < 0$, the inner circulation region around the equilibrium point at zero becomes again stable, and the system finds a way there.
We thus confirm that for orbits with some initial eccentricity, the system can experience a chaotic regime for some time, after which it can escape the 5/3~MMR \citep{Tittemore_Wisdom_1988, Cuk_etal_2020}.

\subsubsection{Results}
\label{planar_results_sect}

We explored a mesh of initial eccentricities ranging between $10^{-5}$ up to $0.020$, with a stepsize of $0.005$. For each pair $(e_1,e_2)$, we integrated a set of $1\,000$ simulations evenly sampled over the resonance angle $\sigma$ (Eq.\,(\ref{eq:resonace_argument})), over 100~Myr, in a total of 25\,000 experiments.

The outcome of the resonance crossing is similar to the one for the circular approximation \citep{Gomes_Correia_2023}, although now it takes place whenever at least one resonant angles, $\sigmak$ (Eq.\,(\ref{canonic_var})), switches from circulation to libration.
We observe that the trajectories can: a) be permanently captured for 100~Myr; b) be captured but escape in less than 100~Myr; or c) quickly skip the resonance.
In Fig.~\ref{fig:simulations_planar_case}, we show one example of each kind (for initial $e_1=0.015$ and $e_2=0.005$).

\begin{figure*}
    \centering
    \includegraphics[width=\textwidth]{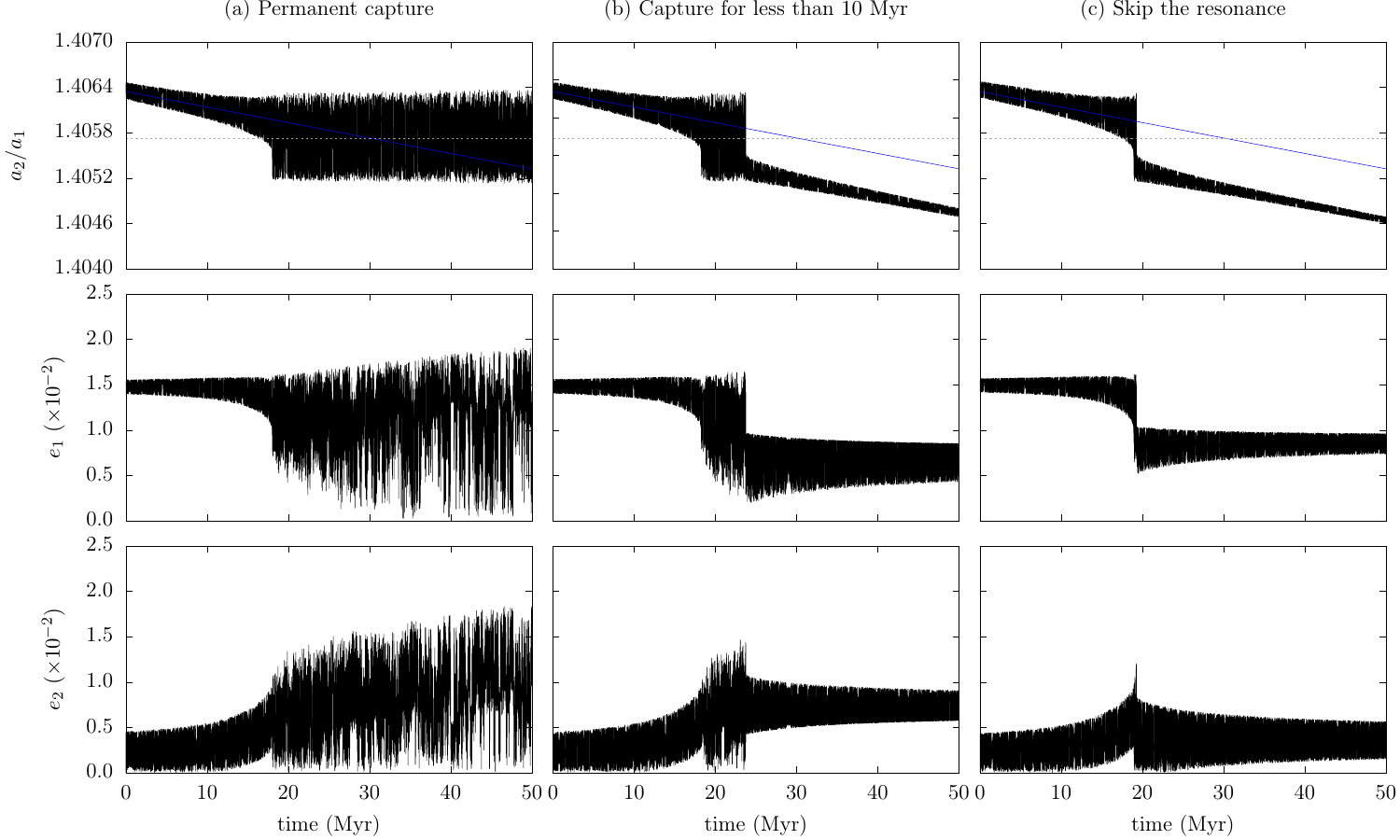}
    \caption{Three examples of a system crossing the 5/3~MMR with initial $e_1 = 0.015$ and $e_2=0.005$. We plot the ratio between the semi-major axes (top), the eccentricity of Ariel (middle), and the eccentricity of Umbriel (bottom) as a function of time. Each column corresponds to a different simulation. We show an example of a system that is permanently caught in resonance (a), one that is captured but evades the resonance in less than 10~Myr (b), and another that skips the resonance without capture (c). The blue line gives the asymptotic evolution (see Sect.~3.1 in Paper~I), while the dashed line gives the position of the nominal resonance (Eq.\,(\ref{nominal:sma})). \label{fig:simulations_planar_case}}
\end{figure*}

For each simulation, we evaluated whether capture in resonance occurred by analysing if the semi-major axis ratio $a_2/a_1$ becomes constant or close to the nominal resonance value (Eq.\,(\ref{nominal:sma})).
We consider that a trajectory escapes the resonance when it skips the resonance, or it is captured for less than 10~Myr \citep[for more details see][]{Gomes_Correia_2023}.
In Table~\ref{tab:statistics_planar_case}, we show a summary of the outcome of the resonance crossing for all the initial conditions explored.

We observe that for initial eccentricities of Ariel, $e_1$, smaller than 0.005, the satellites are always captured in the 5/3~MMR, preventing a future evolution to the currently observed system.
These results confirm those obtained by \citet{Tittemore_Wisdom_1988}.
They are also in agreement with the estimations of the capture probability computed with a one degree-of-freedom simplified model (Fig.~\ref{fig:capture_probability_eAU}).
On the other hand, contrarily to previous analysis, we note that for $e_1 \lesssim 0.005$, the initial eccentricity of Umbriel does not seem to have a significant impact in the number of resonance crossings.

\begin{table}
    \caption{Escape probability from the $5/3$~MMR between Ariel-Umbriel for different initial eccentricities $(e_1,e_2)$ assuming a planar model.  \label{tab:statistics_planar_case} }
    \centering
    \begin{tabular}{|c|*{5}{c|}}
    \hline
     \backslashbox{$e_1$}{$e_2$} & \, $10^{-5}$ \, & 0.005 & 0.010 & 0.015 & 0.020 \\
     \hline
   $10^{-5}$ & 0.0  & 0.0   & 0.0   & 0.0   & 2.1 \\
    0.005       & 0.0  & 0.1   & 0.5   & 0.2   & 0.1 \\
    0.010       & 21.9 & 31.8  & 25.5  & 15.8  & 8.5 \\
    0.015       & 56.2 & 63.6  & 53.8  & 39.7  & 29.7 \\
    0.020       & 77.6 & 80.8  & 68.5  & 57.9  & 43.2 \\
    \hline
    \end{tabular}
\end{table}

For $e_1 \gtrsim 0.01$, we find that the escape probability increases with the initial eccentricity of Ariel.
The number of escapes can attain values as high as $80\%$ for $e_1= 0.02$.
Nevertheless, these numbers are only maximised for small initial eccentricities of Umbriel ($e_2 \lesssim 0.005$), because the escape probability decreases for higher initial eccentricities of this satellite.
In brief, in the frame of a planar model, we conclude that the eccentricity of Ariel plays a crucial role in the passage through the 5/3~MMR.

\section{Complete secular dynamics}\label{sec:Numerical_integration}

Adopting a secular circular model, \citet{Gomes_Correia_2023} have shown that non-zero inclinations of both satellites can also be important in the 5/3~MMR crossing.
In addition, the currently observed inclinations are expected to correspond to the inclination values just after escaping the resonance (see Paper~I), and can thus be used to put constraints on the pre-resonance configuration.
Therefore, we now consider the complete secular problem presented in Sects.~\ref{sec:Conservative_Model} and \ref{sec:tidal_evol}, that simultaneously takes into account non-zero eccentricities and non-zero inclinations.

The full secular problem has four degrees of freedom, and so it can only be accessed through numerical simulations and statistical analysis.
In this section, we integrate the complete set of conservative differential equations (\ref{eq:conservative_motion_equations_x1}) to (\ref{eq:conservative_motion_equations_y2}) together with the complete set of dissipative tidal equations (\ref{eq:tidal_x1_equation_of_motion}) to (\ref{eq:tidal_Theta2_equation_of_motion}).
We adopt exactly the same numerical setup from Sect.~\ref{sec:num_setup}.

\subsection{Resonance crossing}

The exact location of the resonant islands is given by the resonant angles $\sigmak$ and $\psik$ (Eq.\,(\ref{psiiang})), that is, 
\begin{equation}
\dot \sigmaa_k \approx 5 n_2 - 3 n_1 - 2 g_k = 0 \ ,
\end{equation}
or
\begin{equation}
\dot \psia_k \approx 5 n_2 - 3 n_1 - 2 s_k = 0 \ ,
\end{equation}
where $g_k\simeq d\varpi_k/dt$ and $s_k \simeq d\Omega_k/dt$ are the frequencies of the secular modes (see Sect.~2.3 in Paper~I).
The exact mean motion ratio for each resonance island is then given by
\begin{equation}\label{resonant_ratio}
    \frac{n_1}{n_2}=\frac{5}{3}-\frac{2}{3}\frac{g_k}{n_2} \ ,
         \quad \mathrm{or} \quad 
    \frac{n_1}{n_2}=\frac{5}{3}-\frac{2}{3}\frac{s_k}{n_2} \ .
\end{equation}
We have $g_1>g_2>s_2>s_1$ for the satellites of Uranus \citep[eg.][Paper~I]{Laskar_1986}.
Since $d (n_1/n_2) / dt <0$, the system first encounters the inclination resonances at $\psik$.
Therefore, we initially expect some perturbation on the inclinations \citep{Gomes_Correia_2023}, followed by some perturbations on the eccentricities (Sect.~\ref{plandyn}). 

\begin{figure*}
    \centering
    \includegraphics[height=0.34\textheight]{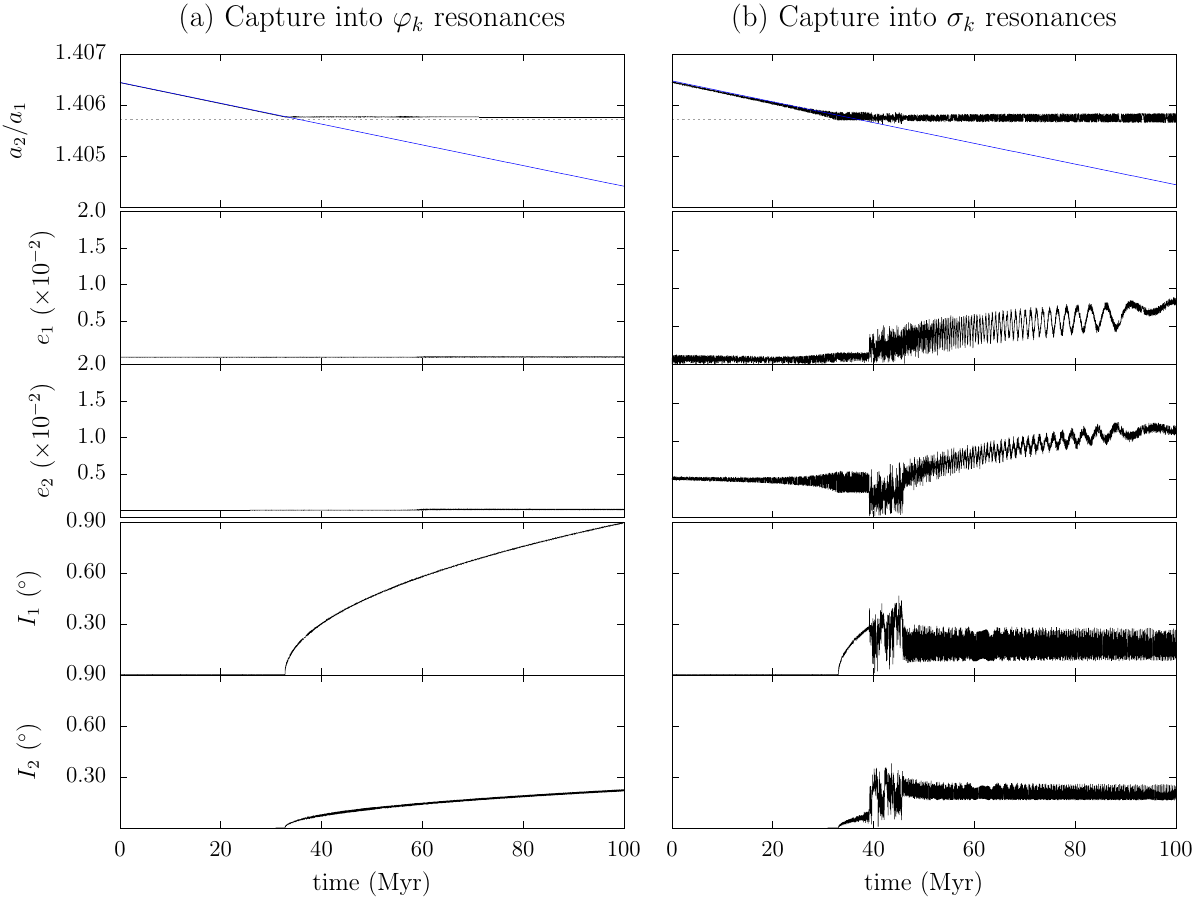}\\
    \vspace{0.5cm}    \includegraphics[height=0.34\textheight]{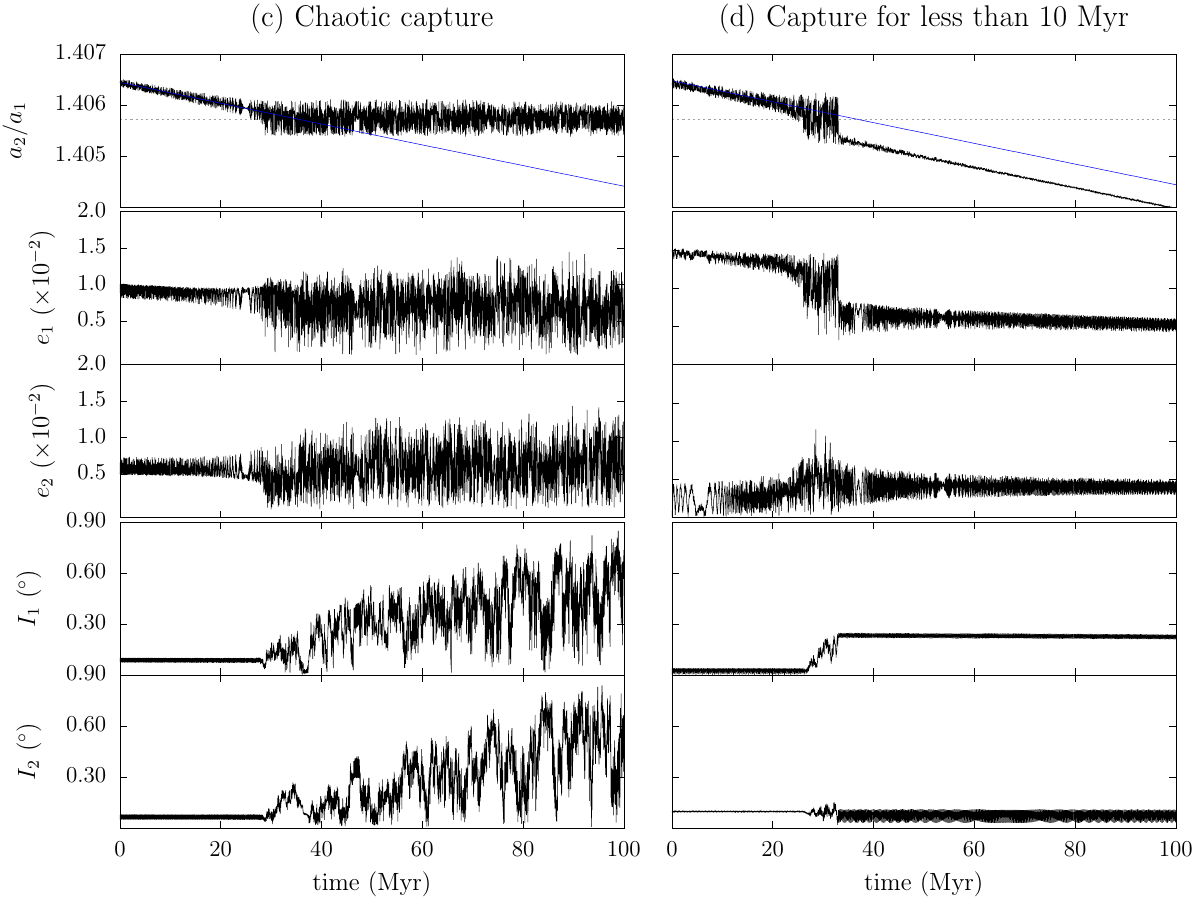}
    \caption{Four examples of a system crossing the 5/3~MMR with the secular model and initial
    $e_1=e_2=10^{-5}$ and $\inc_1=\inc_2=0.001\degree$ (a); 
    $e_1=10^{-5}$, $e_2=\num{5.0e-3}$ and $\inc_1=\inc_2=0.001\degree$ (b);  
    $e_1=10^{-2}$, $e_2=\num{5.0e-3}$, $\inc_1=0.1\degree$, and $\inc_2=0.05\degree$ (c); 
    $e_1=\num{1.5e-2}$, $e_2=10^{-5}$, $\inc_1=0.001\degree$, and $\inc_2=0.10\degree$ (d).
    They correspond respectively to
    permanent capture in the resonant arguments $\psik$ (a),
    permanent capture in the resonant arguments $\sigmak$ (b),
    permanent capture in the chaotic region (c), 
    and to a system that escapes the resonance in less than 10~Myr (d).
    From the top to the bottom, we show the semi-major axes ratio $a_2/a_1$; the eccentricities of Ariel and Umbriel, respectively; and the inclinations of Ariel and Umbriel, respectively. The blue line gives the asymptotic evolution (see Sect.~3.1 in Paper~I), while the dashed line gives the position of the nominal resonance (Eq.\,(\ref{nominal:sma})).}
    \label{fig:eccentric_inclined}
\end{figure*}

In Fig.~\ref{fig:eccentric_inclined}, we provide four examples of the typical behaviours during the resonance crossing.
Before the resonance encounter, the semi-major axes always follow the asymptotic evolution of an isolated two-body system (see Sect.~3.1 in Paper~I), and the resonant arguments $\sigmak$ and $\psik$ circulate.
In Fig. \ref{fig:eccentric_inclined}\,a, we show an example of capture solely in the angles $\psik$.
At $t \approx 30$~Myr, the semi-major axes ratio locks with a value slightly higher than $(5/3)^{2/3} \approx 1.4$, and the resonance angles $\psik$ start to librate, while the angles $\sigmak$ continue to circulate. 
Just after the capture into resonance, the inclinations of both satellites grow steadily.
In Fig. \ref{fig:eccentric_inclined}\,b, we show an example of capture in the angles $\sigmak$. 
Until $t\approx40$~Myr, the evolution is similar to the one previously observed in Fig.~\ref{fig:eccentric_inclined}\,a, where capture in $\psik$ occurs at $t\approx 32$~Myr. 
However, at $t\approx 40$~Myr, the system quits the libration region, and both the eccentricities and the inclinations start to evolve chaotically. 
After about 7~Myr in this regime, the resonant angles $\sigmak$ start to librate, while the angles $\psik$ circulate. 
From that point on, the eccentricities grow steadily on average, while the inclinations become constant on average.
We also note that, between $t\approx40$~Myr and $47$~Myr, there is a slight decrease in the mean motion ratio constant value, resulting from the switch between the resonant arguments, from $\psik$ to $\sigmak$ (Eq.\,(\ref{resonant_ratio})).
In Fig. \ref{fig:eccentric_inclined}\,c, we show an example of permanent capture in the chaotic regime.
The system is directly captured in the chaotic zone at $t\approx32$ Myr. 
Then, the eccentricities and inclinations evolve chaotically and grow on average.
Finally, in Fig. \ref{fig:eccentric_inclined}\,d, we show an example of a short term capture inside the 5/3~MMR.
The system is again directly captured in the chaotic region, which excites both eccentricities and inclinations, but manages to evade it in less than 10~Myr. 
After the resonance is crossed, the semi-major axes ratio returns to the predicted asymptotic evolution.

\subsection{Escape probability}
\label{escape_prob_full}

In order to get a global view of the possible outcomes of the resonance crossing, we explored a mesh of initial eccentricities ranging between $10^{-5}$ and $0.02$, with a stepsize of $0.005$, combined with a mesh of initial inclinations ranging between $0.001 \degree$ and $0.2\degree$, with a stepsize of $0.05\degree$, totalling 625 different initial combinations of $e_1$, $e_2$, $\inc_1$, and $\inc_2$. 
For each initial combination, we performed 1\,000 simulations for 100~Myr, evenly sampled over the resonant angle $\sigma$ (Eq.\,(\ref{eq:resonace_argument})). 
For each run, we adopt the same numerical setup from Sect.\,\ref{sec:num_setup}}, and determined that the system is captured in resonance if the semi-major axes ratio $a_2/a_1$ is still constant after 10~Myr inside the resonance as in the planar case (Sect.~\ref{planar_results_sect}).
In Appendix \ref{app:escape_probabilities} (Table \ref{tab:eccentric_inclined_escape_probability}), we list the results that we obtained for the complete set of initial conditions.

We observe that for initial eccentricities smaller than about $0.005$, long term capture is certain regardless of the initial inclinations. 
Actually, this very reduced dependency on the initial inclinations is observed throughout the whole mesh of initial conditions. 
That is, for the same pair of initial eccentricities ($e_1$, $e_2$), the escape probability is not very sensitive to changes in the initial pair of inclinations ($\inc_1$, $\inc_2$).
Therefore, for a better analysis of the impact of the eccentricities on the escape probability of the 5/3~MMR, for each initial pair of ($e_1$,$e_2$), we can combine the results from the 25 combinations of initial pairs ($\inc_1$,$\inc_2$).
The condensed results are shown in Table~\ref{tab:condensed_statistics}.

\begin{table}
    \caption{Escape probability from the 5/3~MMR between Ariel-Umbriel for different initial eccentricities $(e_1,e_2)$, combining all sets of initial inclination ($\inc_1,\inc_2$), computed from Table~\ref{tab:eccentric_inclined_escape_probability}.} 
    \label{tab:condensed_statistics}    \centering
    \begin{tabular}{|c|*{5}{c|}}
    \hline
     \backslashbox{$e_1$}{$e_2$} & $10^{-5}$ & 0.005 & 0.010 & 0.015 & 0.020 \\
     \hline
    $10^{-5}$ & 0.0  & 0.0   & 5.1   & 14.3   & 14.7 \\
    0.005       & 0.0   & 0.1   & 11.2  & 15.4   & 12.5 \\
    0.010       & 17.4  & 31.1  & 33.7  & 23.9  & 13.8 \\
    0.015       & 59.9  & 64.9  & 48.3  & 28.2  & 17.7 \\
    0.020       & 75.9  & 78.8  & 51.7  & 38.5  & 25.2 \\
    \hline
    \end{tabular}
\end{table}

By comparing Table~\ref{tab:statistics_planar_case}, obtained for the planar approximation (Sect.~\ref{plandyn}), with Table~\ref{tab:condensed_statistics}, obtained for the full eccentric and inclined problem, we observe that there is a good agreement between the two results, in particular for initial $e_2 \le 0.005$.
As in the planar approximation, the escape probability is more sensitive to variations in $e_1$ than in $e_2$. Furthermore, the escape probability is maximised for initial high values of $e_1$ and $e_2\le0.005$ (about $80\%$ of escapes).
However, for initial $e_1 \ge  0.01$ and $e_2 > 0.005$, the escape probability decreases more significantly in the general case than in the planar approximation.
Inversely, for initial $e_1 \le  0.005$ and $e_2 > 0.005$, the escape probability is almost null in the planar case, while it becomes non-negligible and increases with $e_2$ in the general case.

\subsection{Monte Carlo simulations}\label{sec:monte_carlo}

By adopting an initial discrete distribution on the eccentricities and inclinations, we were able to quantify the escape probability for each set of initial conditions.
As a result, we have seen that the escape probability from the 5/3~MMR between Ariel and Umbriel is mostly dominated by the initial eccentricities and enhanced for initial $e_1=0.02$ and $e_2\le0.005$.
However, the final distribution of the inclinations is also very important, because it allows us to exclude all sets of initial conditions that fail to reproduce the current system \citep{Gomes_Correia_2023}.
Indeed, tidal dissipation is very inefficient to damp the inclinations (see Sect.~3.3 in Paper~I), and so we expect that the inclination values ($\inc_1$,$\inc_2$) just after the resonance crossing match the currently observed mean values (Table~\ref{table:physical_orbital_parameters}).

The discrete nature of the distribution adopted in Sect.~\ref{escape_prob_full} may conceal some of the initial conditions that can reproduce the current configuration of the system.  
To ensure that all potential combinations of initial conditions are thoroughly examined, we employ the Monte Carlo method to encompass the entire range of initial combinations.
We conducted one million simulations in which, for each run, we selected random initial eccentricities within the range of $0$ and $0.02$, random initial inclinations within the range of 0 and $0.2\degree$, and random resonant angles $\sigmak$ and $\psik$ between $0^\circ$ and $360^\circ$. 

Overall, we find that $30\%$ of the simulations successfully avoided capture.
It is a non-negligible number, but it strongly depends on the initial conditions (Sect.~\ref{escape_prob_full}).
Therefore, we first look at the distribution of the initial eccentricities and initial inclinations that avoided capture into resonance.
In Fig. \ref{fig:colormaps}, we divided the range of initial eccentricities $(0.00,0.02)$ and initial inclinations  $(0.00\degree,0.2\degree)$ into a grid of $50\times50$ equally spaced intervals.
We then counted the number of initial conditions that escape capture in less than 10~Myr and divided by the total number of simulations in each bin.
This provides a more in-depth examination of the escape probability distribution.
Rather than relying on a fixed discrete set of initial conditions, we can now analyse a more continuous dataset, and pinpoint the values where behavioural changes occur in the initial conditions.

\begin{figure*}
    \centering    \includegraphics[width=\textwidth]{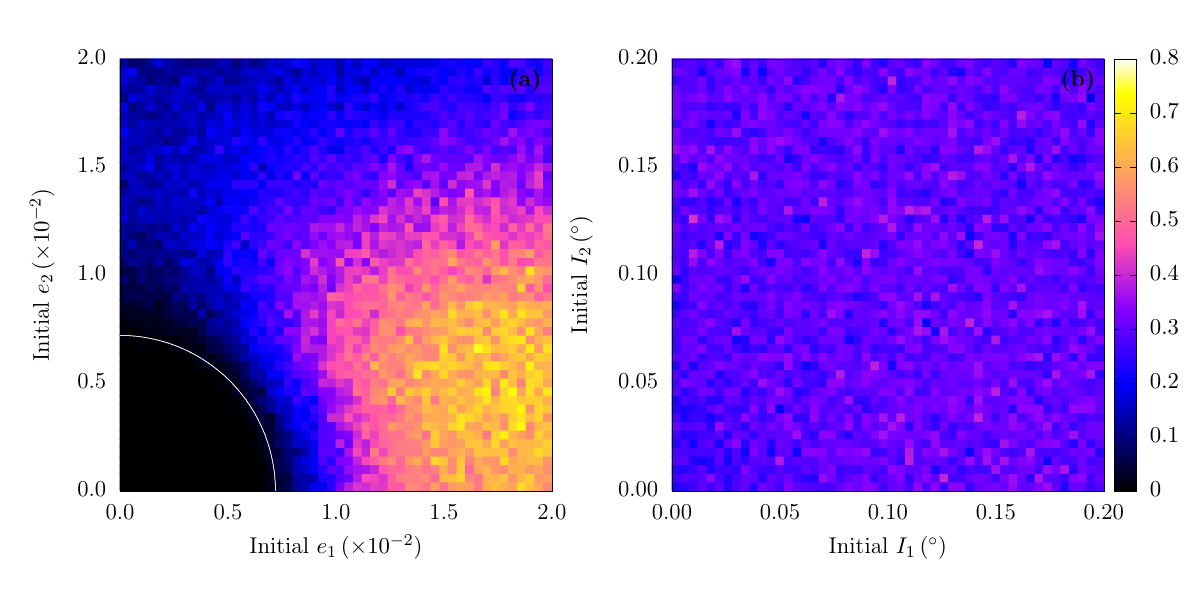}
    \caption{Escape probability from the 5/3 MMR between Ariel and Umbriel for a grid of $50\times50$ intervals, equally spaced between 0 and $0.02$ for the initial eccentricities (a), and between 0 and $0.2\degree$ for the initial inclinations (b). The white circle corresponds to $\left(e_1^2+e_2^2\right)^{1/2}=0.0072$.}
    \label{fig:colormaps}
\end{figure*}

The analysis of the initial eccentricity distribution (Fig.~\ref{fig:colormaps}\,a), shows two distinct features: i) for initial ${(e_1^2+e_2^2)^{1/2} \lesssim 0.007}$, all simulations were captured into resonance; ii) Ariel's high initial eccentricities facilitate the evasion, for initial $e_1 \gtrsim 0.01$ and $e_2 \lesssim 0.01$, the escape probability reaches $\sim 80 \%$, whereas for the inverse case, that is, for initial $e_1 \lesssim 0.01$ and $e_2 \gtrsim 0.01$, the escape probability is around $\sim 30\%$.
From the analysis of the initial inclinations distribution (Fig.~\ref{fig:colormaps}\,b), we observe a uniform spread across the phase space, devoid of any discernible trends, with escapes probabilities around $\sim 35 \%$.
We conclude that the results obtained with the Monte Carlo simulations are consistent with the results obtained with the discrete sampling (Sect.~\ref{escape_prob_full}) and can thus be used for a deeper statistical analysis of the resonance crossing problem.

In Fig.~\ref{fig:initial_inc}, we show the distribution of the final eccentricities and final inclinations.
We observe that both distributions are delimited by two straight lines crossing at the origin, that is, at ${(e_1,e_2)=(0,0)}$ and ${(\inc_1,\inc_2)=(0\degree,0\degree)}$, and widens as the values of eccentricity and inclination increase.

\begin{figure*}
    \centering
    \includegraphics[width=\textwidth]{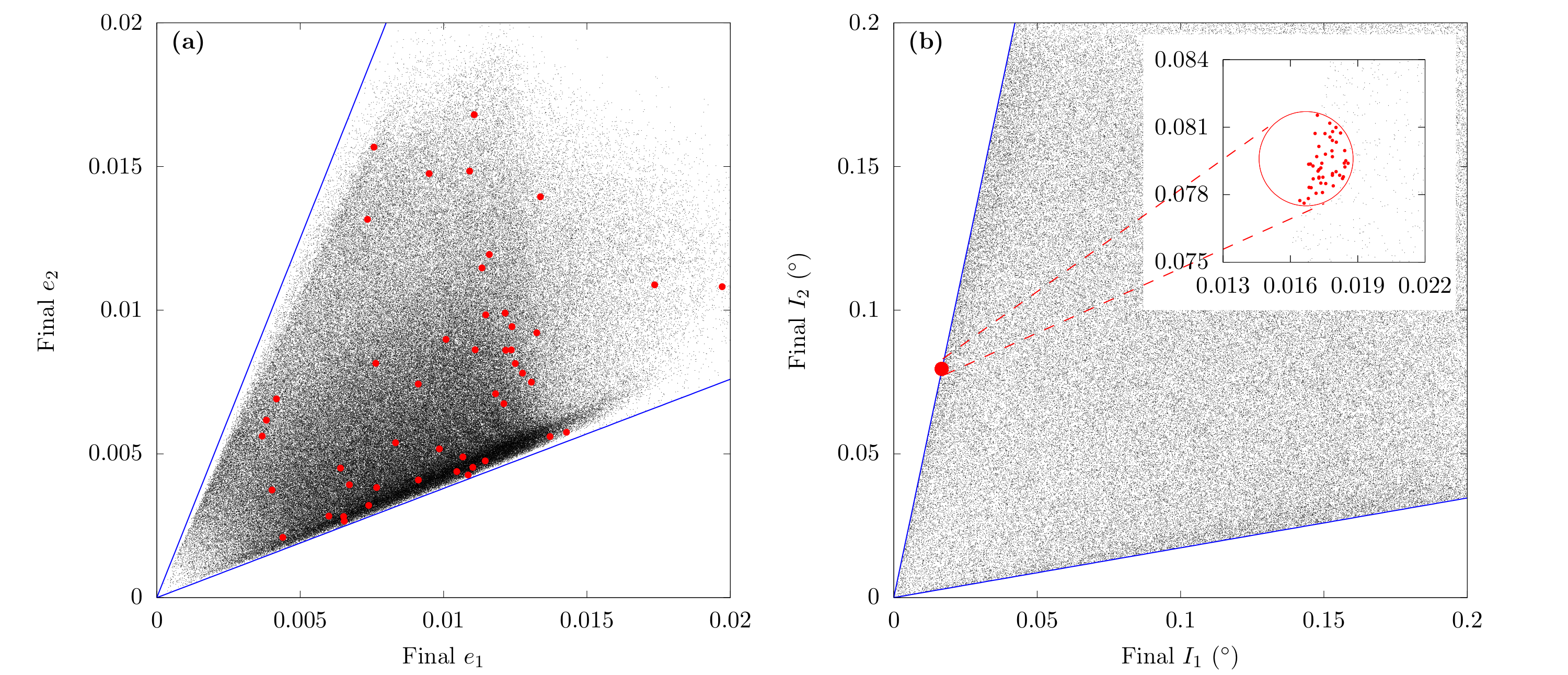}
    \caption{Distribution of the final eccentricities (a) and final inclinations (b) after the 5/3~MMR passage. The figures encompass $295\,953$ runs with random initial ($e_1,e_2,\inc_1,\inc_2$) that evade the resonance capture. From (b), we selected the points (in red) that lay inside a circle centred at the current mean inclinations of Ariel and Umbriel and a radius of $0.002\degree$ (Table~\ref{table:physical_orbital_parameters}). \label{fig:initial_inc} }
\end{figure*}

For the final eccentricity distribution (Fig.~\ref{fig:initial_inc}\,a), we note that there is a denser concentration of points at the lower edge of the confinement region, that is, for higher final $e_1$ and lower final $e_2$. 
In addition, there is a clear void of results for final eccentricities below $0.001$. 
Both observations are related with the results shown in Fig.~\ref{fig:colormaps}\,a for the initial eccentricity distribution. 
Indeed, in order to escape capture, the initial eccentricity must be higher than $\sim 0.007$, and so it is more challenging to obtain near zero eccentricities after the system escapes the resonance. 
Similarly, the clustering of points in the region where $e_1>e_2$ can be attributed to the higher probability of escape associated with higher initial eccentricity of Ariel ($e_1>0.01$) combined with lower initial eccentricity of Umbriel ($e_2<0.015$).

From the distribution of the final inclinations (Fig.~\ref{fig:initial_inc}\,b), we can easily identify the simulations that closely match the presently observed mean values of $\inc_1$ and $\inc_2$ (Table~\ref{table:physical_orbital_parameters}).
For that purpose, we  established a circular region around the current mean values, with a radius of $0.002\degree$ (corresponding to $\sim10\%$ error in the current inclination of Ariel). 
These points were coloured in red.
We observe that the final eccentricities of these points are scattered (Fig.~\ref{fig:initial_inc}\,a), lacking any noticeable pattern. 

Using solely the points that match the current mean inclinations (Table~\ref{table:physical_orbital_parameters}) after the 5/3~MMR passage, in Fig.~\ref{fig:initial_inc2}, we show the distribution of the corresponding initial eccentricities and initial inclinations before the resonant encounter.
For the distribution of the initial eccentricities (Fig.~\ref{fig:initial_inc2}\,a), we observe that the selected points are evenly distributed over the interval, provided that $e_1>0.005$. 
On the other hand, we observe that the initial inclinations of the selected points (Fig.~\ref{fig:initial_inc2}\,b) are well constrained within the interval $0.01\degree < \inc_1 < 0.05\degree$ and $0.06\degree < \inc_2 < 0.09\degree$.
We hence conclude that this set of orbital elements prior to the resonant encounter best explains the currently observed system.
We also remind that the final eccentricity values after the resonance passage are not important, because we expect tides to quickly damp them to the current values (see Sect.~3.2 in Paper~I).

\begin{figure*}
    \centering
    \includegraphics[width=\textwidth]{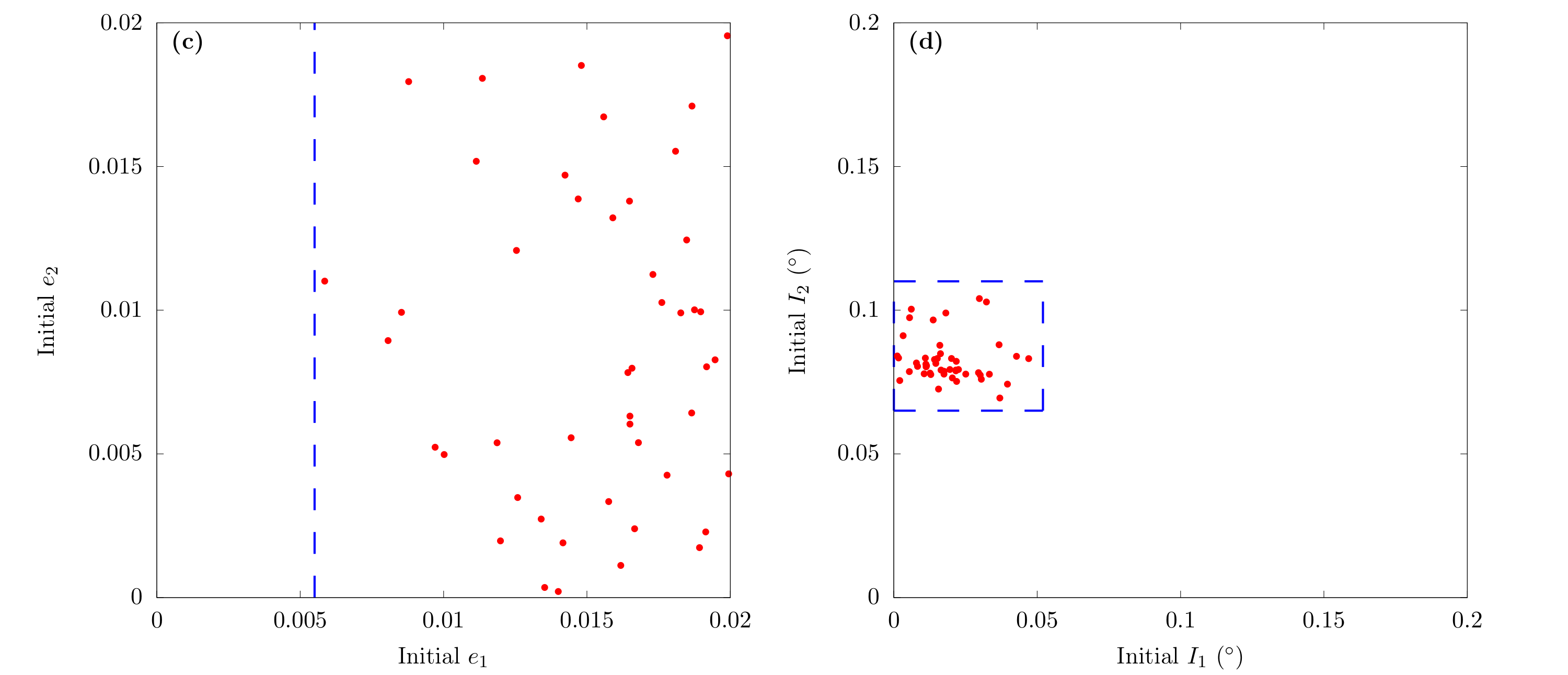}
    \caption{Distribution of the initial eccentricities (a) and initial inclinations (b) that reproduce the current mean inclinations (Table~\ref{table:physical_orbital_parameters}).}
    \label{fig:initial_inc2}
\end{figure*}

\subsection{Final inclination distributions}

In the previous section we have identified the most favourable initial conditions to cross the 5/3~MMR between Ariel and Umbriel.
However, the resonance crossing is a stochastic process, and so identical initial conditions may lead to completely different evolutionary scenarios (Sect.~\ref{sec:stabmaps}).

The average initial inclination values that comply with the present system are given by $\inc_1=0.036\degree$ and $\inc_2=0.082\degree$ (Fig.~\ref{fig:initial_inc2}\,b).
To investigate whether these initial inclinations can effectively replicate the present ones, we employed the same methodology used in Sect.~\ref{escape_prob_full}.
We fixed the initial ($\inc_1,\inc_2$) at the averaged values, and explored a mesh of initial eccentricities ranging between $10^{-5}$ and $0.02$, with a step size of 0.005. 
For each set of initial ($e_1,e_2$), we ran 1\,000 simulations for 100~Myr, equally distributed over the resonance angle $\sigma$, totalizing  25\,000 simulations.

In Fig.~\ref{fig:histogram_102_104}, we show the histograms of the final inclinations distributed over 51 classes ranging between $0.0\degree$ and $0.3\degree$, with a step size of $0.006\degree$. 
We only considered the final inclinations from the simulations that escaped the 5/3~MMR, totalising 6\,685 simulations. 
For each pair of initial ($\inc_1,\inc_2$), we combined the results from the 25 combinations of initial ($e_1,e_2$).
The red dot gives the present mean inclinations of each body (Table~\ref{table:physical_orbital_parameters}).
We observe that the histograms display prominent peaks for the final inclinations of Ariel and Umbriel around a well-defined mean value, which is close to the currently observed one.

\begin{figure*}
    \centering    \includegraphics[width=\textwidth]{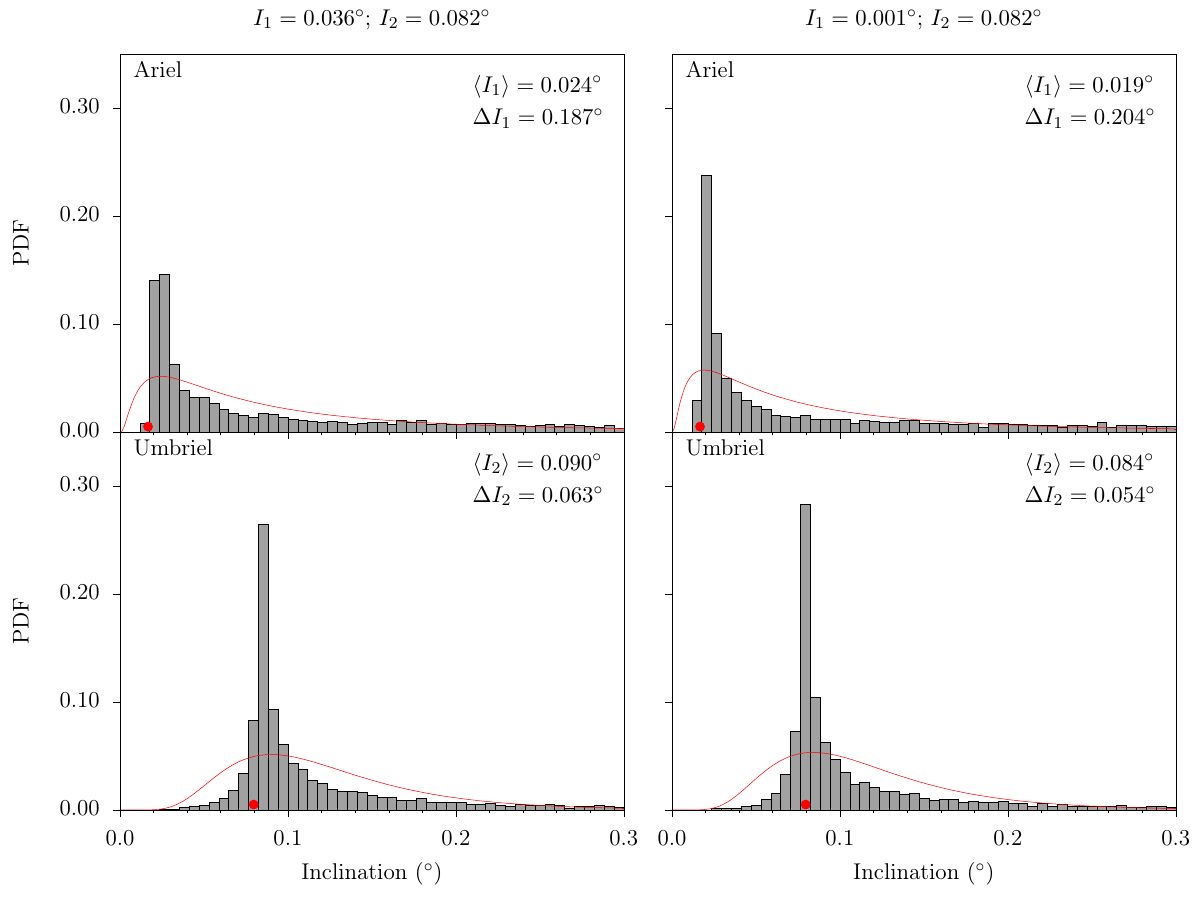}
    \caption{Histograms of the final inclination distributions of Ariel (top) and Umbriel (bottom) for initial inclinations $\inc_1=0.036\degree$ and $\inc_2=0.082\degree$. We show the best fit Lognormal distribution (Eq.\,(\ref{eq:Lognormal_PDF})) to each histogram (red curve) and the corresponding inclination mode, $\langle \inc \rangle$, and standard deviation, $\Delta \inc$. The presently observed inclinations (Table~\ref{table:physical_orbital_parameters}) are marked with red dots. \label{fig:histogram_102_104}}
\end{figure*}

We repeated the experiment for other combinations of the initial inclinations.
To analyse the results systematically, we fitted a Lognormal point distribution function (PDF) to each histogram (with $\xxx = \inc_k$),
\begin{equation}\label{eq:Lognormal_PDF}
    f(\xxx)=\frac{1}{\xxx\,\eta\,\sqrt{2 \pi}}\exp{\left(-\frac{(\ln{\xxx}-\upsilon)^2}{2\,\eta^2}\right)} \, ,
\end{equation}
where $\upsilon$ and $\eta$ are  the parameters that define the distribution. 
The peak of the PDF (mode), is given by
\begin{equation}\label{eq:Lognormal_mode}
    \langle \xxx \rangle=\exp{(\upsilon-\eta^2/2)} \, ,
\end{equation}
and the standard deviation is given by
\begin{equation}\label{eq:Lognormal_variance}
    \Delta \xxx=\sqrt{\left(\exp{(\eta^2)}-1\right)\left(\exp{(2\upsilon+\eta^2)}\right)} \, .
\end{equation}
We verify that the observed mean inclination values are very close to the peak of the distribution. 
For the example shown in Fig.~\ref{fig:histogram_102_104}, we obtain $\inc_1 = 0.024\degree \pm 0.187\degree$ and $\inc_2 = 0.090\degree \pm 0.063\degree$, which compares with the observed mean values $\inc_1 = 0.017\degree$ and $\inc_2 = 0.080\degree$ (Table~\ref{table:physical_orbital_parameters}). 
We hence conclude that the initial conditions determined in Sect.~\ref{sec:monte_carlo} do correspond to reliable representations of the system prior to the resonance crossing.

\section{$N-$body simulations}\label{sec:n_body_simulations}

The results from the previous sections were obtained using a two-satellite secular model. 
In this section, we attempt to validate those results by adopting a more complete non-averaged $N-$body model.
For that purpose, we use the numerical code \texttt{SPINS} \citep{Correia_2018}, that takes into account satellite-satellite interactions, spin dynamics, rotational flattening, and tidal evolution according to the weak friction model\footnote{The numerical results in this section can also be obtained using the open-source code \texttt{TIDYMESS} \citep{Boekholt_Correia_2023}, with option \texttt{tidal\_model = 2}, which corresponds to the weak friction tidal model.}. 
That is, it corresponds to the full non-averaged equations of motion that derive from the complete Hamiltonian given by Eq.\,(\ref{tidalH}).
It is also the same numerical code that we used in Paper~I.

\subsection{Two-satellite simulations}\label{sec:2_body_problem}

We first compare the results of the secular model with the full three-body model (Uranus, Ariel, and Umbriel). 
This allows us to determine whether the secular model provides a good description of the system, or if the high-frequency terms, which were averaged out (Sect.~\ref{cqaa}), introduce some unexpected features.
For a better comparison with the secular model simulations, we adopt here the exact same numerical setup from Sect.~\ref{sec:num_setup}.

In a first experiment, we aim to verify the low impact of the initial inclinations. 
We then fix the initial $\inc_1=0.001\degree$, and vary $\inc_2$ between $0.001\degree$ and $0.20\degree$, with a step size of $0.05\degree$. 
To simultaneously check the role of the initial eccentricity of Ariel into the capture probability, we fix the initial $e_2=10^{-5}$ and vary $e_1$ between $10^{-5}$ and $0.02$, with a step size of $0.005$. 
This totalises 25 initial combinations of $(e_1,\inc_2)$.

The computation time to integrate these initial conditions for 100~Myr with the three-body model is $10^4$ longer than with the secular model. 
Therefore, it is not feasibly to compute 1\,000 runs equally distributed over the resonance angle $\sigma$ for every pair of initial ($e_1,\inc_2$). 
We thus reduced the number of simulations per initial condition by a factor of ten, to 100 runs for each pair, which gives a total number of 2\,500 simulations. 
Then, for each run, we determine again if the system evades the 5/3~MMR between Ariel and Umbriel in less than 10~Myr by evaluating the semi-major axes ratio $a_2/a_1$ (see Sect.~\ref{planar_results_sect}).

In Table~\ref{tab:2_body_problem_statistics}, we show the escape probability for the 25 pairs of initial $(e_1,\inc_2)$ obtained with the three-body model.
For a better comparison, we also show the results obtained with the secular model\footnote{The results from the secular model were obtained in Sect.~\ref{escape_prob_full} and displayed here again for a more convenient comparison between the two sets of results (see also Table~\ref{tab:eccentric_inclined_escape_probability}).}. 
For $e_1\le0.005$, both models show that capture in resonance is certain, independently of the initial $\inc_2$.
For $e_1\ge 0.01$, we observe that the escape probabilities obtained with the two models are in good agreement, despite some small local variations which are most likely due to statistical fluctuations.
In particular, we confirm that outside the secular approximation, there is not much impact from the initial inclination of Umbriel and the initial eccentricity of Ariel remains the key parameter that controls the escape probability.

\begin{table}
    \centering
        \caption{Escape probability from the 5/3~MMR between Ariel and Umbriel for a mesh of 25 initial pairs of $(e_1,\inc_2)$, with $e_2=10^{-5}$ and $\inc_1=0.001\degree$, obtained with three different numerical models. \label{tab:2_body_problem_statistics}} 
    \begin{tabular}{|c|*{6}{c|}} 
    \multicolumn{6}{c}{Secular model (Sect.~\ref{escape_prob_full})}  \\ \hline
    \backslashbox{$\inc_2$}{$e_1$} & $10^{-5}$ & 0.005 & 0.010 & 0.015 & 0.020 \\ \hline
    $0.001\degree$      & 0 & 0 & 22	& 49		& 60 \\
    $0.05\degree$   	& 0 & 0 & 21	& 55   	& 66 \\
    $0.10\degree$   	& 0 & 0 & 19	& 61   	& 72 \\
    $0.15\degree$   	& 0 & 0 & 18	& 63   	& 74  \\
    $0.20\degree$   	& 0 & 0 & 20	& 59   	& 74  \\ \hline
 \multicolumn{6}{c}{}\\ 
     \multicolumn{6}{c}{Two-satellite model (Sect.~\ref{sec:2_body_problem})}\\ \hline
 \backslashbox{$\inc_2$}{$e_1$} & $10^{-5}$ & 0.005 & 0.010 & 0.015 & 0.020\\ \hline
$0.001\degree$	& 0   & 0   & 37  & 46  & 59  \\
$0.05\degree$  	& 0   & 0   & 37  & 54  & 48  \\
$0.10\degree$	& 0   & 0   & 27  & 55  & 54  \\
$0.15\degree$	& 0   & 0   & 16  & 62  & 56  \\
$0.20\degree$	& 0   & 0   & 36  & 67  & 48  \\ \hline
 \multicolumn{6}{c}{}\\ 
     \multicolumn{6}{c}{Five-satellite model (Sect.~\ref{sec:five_satellite simulations})}\\  \hline
     \backslashbox{$\inc_2$}{$e_1$} & $10^{-5}$ & 0.005 & 0.010 & 0.015 & 0.020 \\
     \hline
    $0.001\degree$	& 0 & 0 & 4     & 62    & 82    \\
    $0.05\degree$    	& 0 & 0 & 13    & 72    & 77    \\
    $0.10\degree$    	& 0 & 0 & 7     & 67    & 85    \\
    $0.15\degree$    	& 0 & 0 & 15    & 70    & 86    \\
    $0.20\degree$    	& 0 & 0 & 10    & 65    & 90    \\    \hline
    \end{tabular}
\end{table}

According to the secular model (Sect.~\ref{sec:monte_carlo}), the currently observed system (Table~\ref{table:physical_orbital_parameters}) can be replicated with initial inclinations $(\inc_1,\inc_2)=(0.001\degree,0.082\degree)$.
To ensure a low capture probability in resonance, we chose initial $(e_1,e_2)=(0.015,0.005)$. 
With these initial eccentricity and inclination values, we then integrated one set of 100 simulations evenly distributed across $\sigma$ for 100~Myr using the secular model and another set of 100 simulations using the three-body model.

In Fig.~\ref{fig:histogram_final}, we built a histogram of the final eccentricity and inclination distributions. 
For a better comparison, we overlaid the Lognormal curve obtained with the secular model (Eq.\,(\ref{eq:Lognormal_PDF})).
We observe that the Lognormal curve presents a remarkable adjustment to the results obtained with the three-body model.
For the final eccentricities, the differences can be likely attributed to statistical fluctuations.
For the final inclinations, the distributions also display the same diffusion pattern, although the secular model  exhibits a more pronounced peak.

In order to quantify the agreement between the distributions obtained with both models, we performed the two sample Kolmogorov-Smirnov (KS) test \citep[e.g.][]{Darling_1957}. 
For the eccentricity distributions of Ariel and Umbriel, we obtained KS~$= 0.10$ and KS~$= 0.17$, respectively, while for the inclination distributions of Ariel and Umbriel, we obtained KS~$=0.15$ and KS~$ =0.13$, respectively. 
We thus conclude that the results arising from the secular and the $N-$body models can be derived from similar statistical distributions.

The good agreement observed between the two models demonstrates that the secular model provides a correct description of the two-satellite system composed by Ariel and Umbriel.

\begin{figure*}
    \centering
    \includegraphics[width=0.80\textwidth]{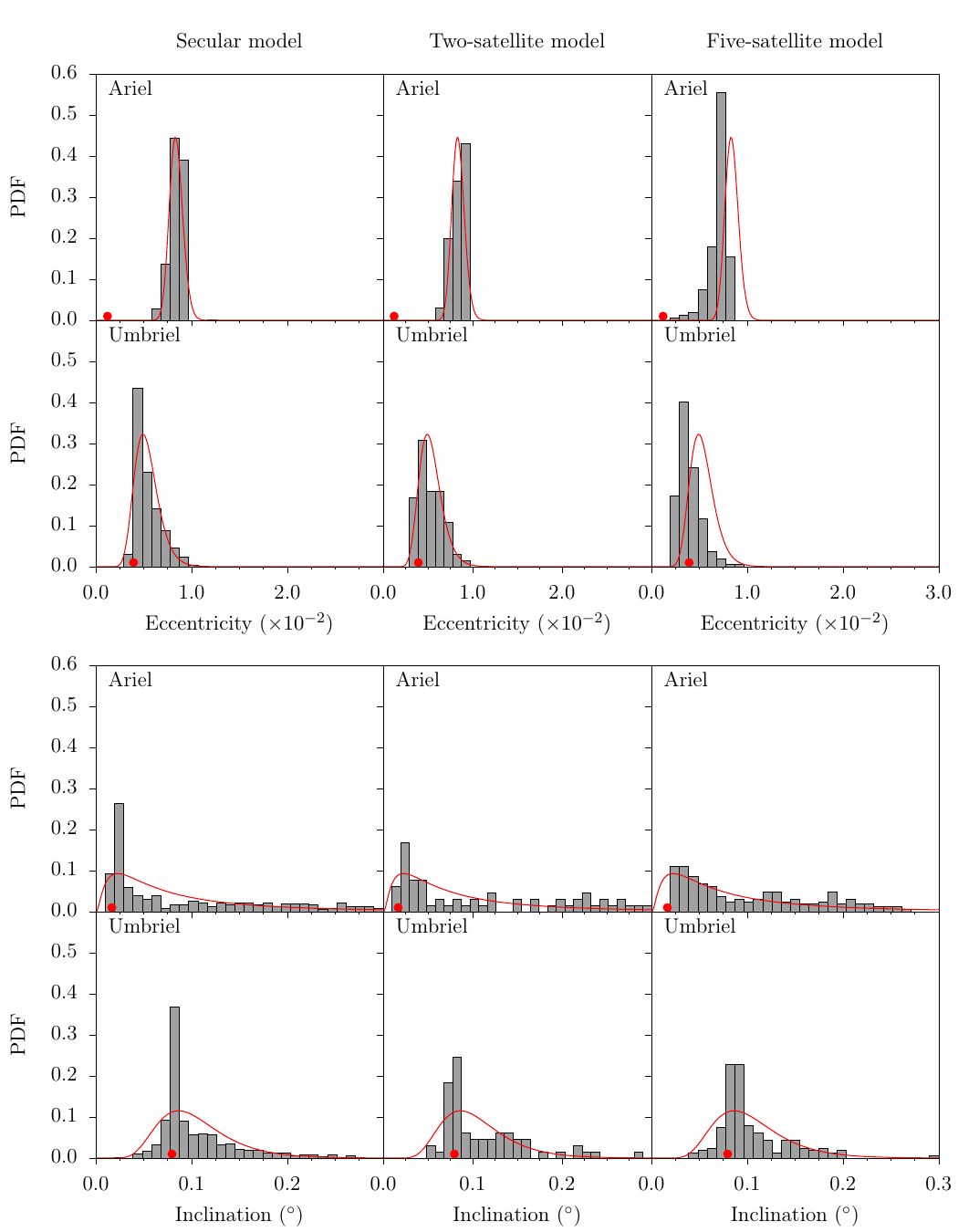}
    \caption{Histograms of the final eccentricity and inclination distributions of Ariel (top) and Umbriel (bottom) for initial $e_1=0.015$, $e_2=0.005$, $\inc_1=0.001\degree$, and $\inc_2=0.082\degree$. We show the results for the secular model (left), the two-satellite model (middle), and the five-satellite model (right). We also show the best fit Lognormal distribution (Eq.\,(\ref{eq:Lognormal_PDF})) to the histogram of the secular model (red curve). The presently observed inclinations (Table~\ref{table:physical_orbital_parameters}) are marked with red dots. \label{fig:histogram_final}}
\end{figure*}

\subsection{Five-satellite simulations}\label{sec:five_satellite simulations}

For simplicity, up to now we have studied the 5/3~MMR passage between Ariel and Umbriel by building models that consider only these two satellites.
However, mutual gravitational interactions with the remaining satellites of Uranus may also influence the architecture of the entire system during the resonance passage \citep{Cuk_etal_2020}.
Therefore, in this section we compare the previous results with the full six-body model (Uranus, Miranda, Ariel, Umbriel, Titania, and Oberon).

The physical properties of the satellites are given in Table \ref{table:physical_orbital_parameters}.
For the Love numbers of Miranda, Titania, and Oberon, we adopt $k_{2,M}=\num{8.84e-4}$, $k_{2,T}=\num{1.99e-2}$, and $k_{2,O}=\num{1.68e-2}$ \citep{Chen_etal_2014}, together with $\dt_{M}=38.9$~s, $\dt_{T}=239.4$~s, and $\dt_{O}=370.2$~s (corresponding to $Q_k = 500$, see Sect.~3.2 in Paper~I for details), yielding to 
\begin{equation}
k_{2,M} \, \dt_{M} = 0.034 \, \mathrm{s} \ , \quad k_{2,T} \, \dt_{T} = 4.764 \, \mathrm{s} \ , \quad \mathrm{and} \quad k_{2,O} \, \dt_{O} = 6.219 \,  \mathrm{s} \ .
\end{equation}
The total angular momentum of the system also needs to be updated (Eq.\,(\ref{SigmaTOT})) to accommodate the five satellites,
\begin{equation}\label{ang_mom_five}
    \Sigma=\num{9.446072e-10} \ \mathrm{M_\odot \, au^2 \, yr^{-1}} \, .
\end{equation}
The semi-major axes of the satellites at the nominal resonance were again calculated as for the two-satellite case (Eq.\,(\ref{nominal:sma})).
The angular velocity of Uranus is corrected according to Eq.\,(\ref{eq:angrot_correction}), such that the current total angular momentum of the system is conserved (Eq.\,(\ref{ang_mom_five})).
Finally, the semi-major axis of Ariel is slightly decreased to move the system from the nominal resonance, leading to
\begin{equation}\label{eq:semi-major_axis_5_body_problem}
    \begin{split}
    a_M/\Ru &= 5.0794 \, , \\
    a_A/\Ru &= 7.3868 \, , \\
    a_U/\Ru &=10.3891 \, , \\
    a_T/\Ru &=17.0671 \, ,\\
    a_O/\Ru &=22.8239\, .
    \end{split}
\end{equation}

We first attempted to reproduce the results in Table~\ref{tab:2_body_problem_statistics}. 
We fixed $e_2=10^{-5}$ and $\inc_1=0.001\degree$ and integrated the whole system over 100~Myr for different initial $(e_1,\inc_2)$ pairs.
For each pair, we ran 100 simulations equally distributed over $\sigma$.
The initial eccentricities and inclinations for Miranda, Titania, and Oberon were kept at the currently observed values (Table \ref{table:physical_orbital_parameters}). 
For each simulation, we again checked if the system evades the 5/3~MMR between Ariel and Umbriel in less than 10~Myr.
When we increase the number of bodies in the system, the computation time increases even more. 
The five-satellite simulations are about four times longer than the two-satellite simulations (Sect.~\ref{sec:2_body_problem}).

By comparing the results between the different models, we verify again that for $e_1 \le 0.01$, it is never possible to escape the 5/3~MMR.
We also confirm that the initial inclination of Umbriel does not play any significant role.
However, some minor differences can be observed with respect to the secular and two-satellite models.
In general, for $e_1 = 0.015$, the five-satellite model provides lower escape probabilities.
On the other hand, for $e_1 \ge 0.015$, it provides higher escape probabilities.
That is, it appears that the perturbations from the remaining satellites initially difficult the resonance crossing, but after some critical value $e_1 \sim 0.01$, these perturbations facilitate the evasion.

In a second experiment, we integrated a set of 100 simulations equally distributed over $\sigma$ with the best suited initial eccentricities and inclinations of Ariel and Umbriel that are able to reproduce the current system.
The initial eccentricities and inclinations for Miranda, Titania, and Oberon were kept at the currently observed values (Table~\ref{table:physical_orbital_parameters}), but for Ariel and Umbriel we adopted $e_1=0.015$, $\inc_1=0.001\degree$, $e_2=0.005$, and $\inc_2=0.082\degree$, respectively, as in Sect.~\ref{sec:2_body_problem}.

In Fig.~\ref{fig:histogram_final}, we built a histogram of the final eccentricity and inclination distributions. 
For a better comparison with the previous models, we overlaid the Lognormal curve obtained with the secular model (Eq.\,(\ref{eq:Lognormal_PDF})).
We observe that there is still a good agreement between the five-satellite model and the secular and two-satellite models.
This is particularly true for the inclination distributions, whose differences with respect to the two-satellite are mostly likely due to statistical fluctuations.
Concerning the eccentricity distributions, there is a slight shift of the peak of maximal eccentricity to lower values.
However, this discrepancy is not really a problem, because in the case of Umbriel this peak is still very close to the currently observed mean value, while in the case of Ariel, any remnant eccentricity is expected to be quickly damped to the currently observed value (see Sect.~3.2 in Paper~I).

\begin{sidewaysfigure*}[htbp]
    \centering
    \includegraphics[width=\textwidth]{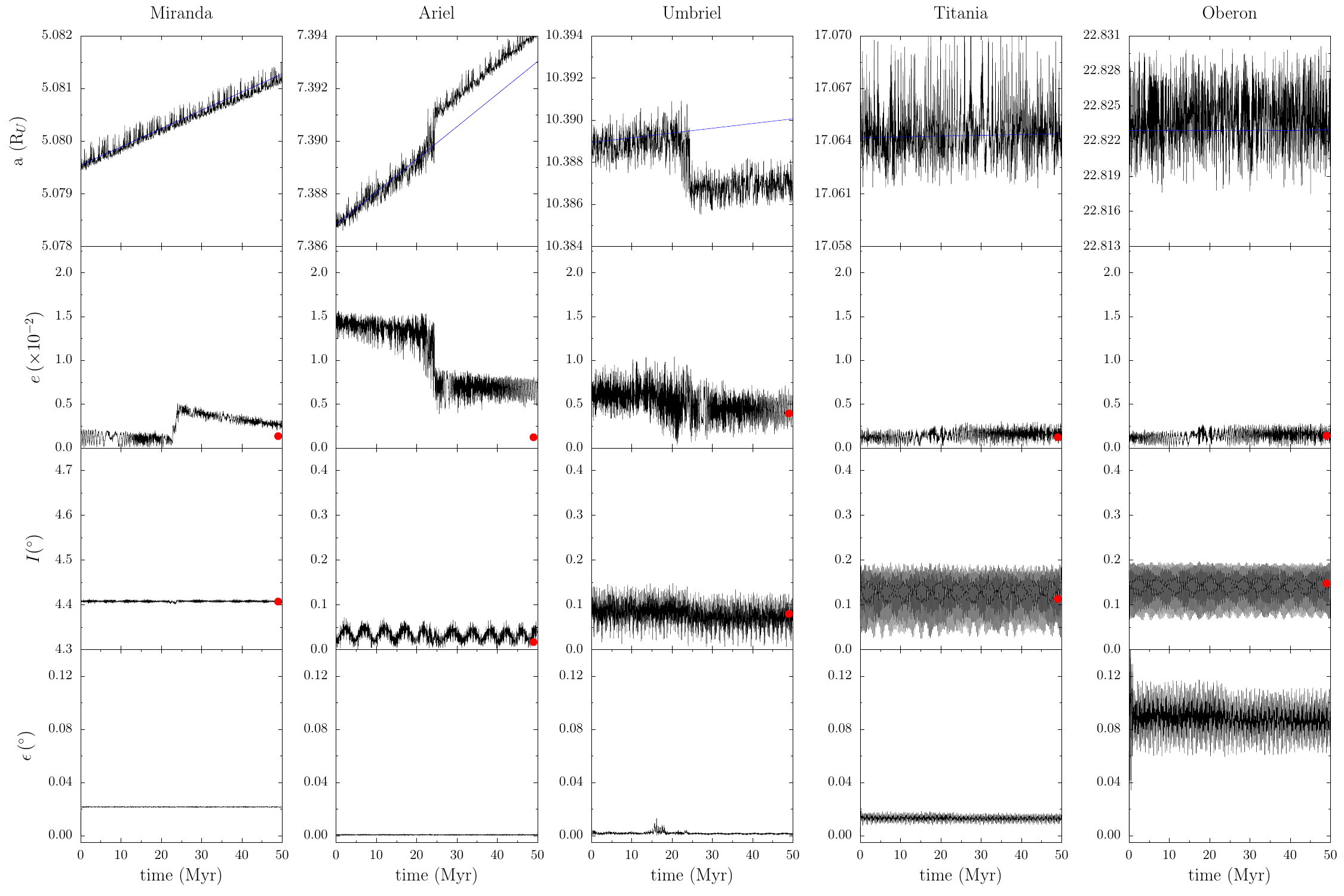}
    \caption{Simulation with the five-satellite model for the passage through the 5/3 MMR between Ariel and Umbriel. From the top to the bottom, we show the semi-major axes, the eccentricities, the inclinations, and the obliquities. The system evolves from initial $e_1=0.015$, $e_2=0.005$, $\inc_1=0.001\degree$, and $\inc_2=0.082\degree$. The initial eccentricities and inclinations of Miranda, Titania, and Oberon correspond to their current mean values (Table~\ref{table:physical_orbital_parameters}). The asymptotic tidal evolution of the semi-major axis (see Sect. 3.1 in Paper I) is represented with a blue line, and the present mean eccentricities and inclinations are marked with red dots. \label{fig:5_body_example_skip}}    
\end{sidewaysfigure*}

In Fig.~\ref{fig:5_body_example_skip}, we plot one example of the 100 simulations with the five-satellite model, where the system skips the 5/3~MMR between Ariel and Umbriel.
We show the evolution of the semi-major axes, eccentricities, inclinations, and obliquities of all satellites.
Prior to the commensurability, the average semi-major axes follow the asymptotic tidal evolution (blue line), while the eccentricities and inclinations oscillate around a mean value close to the initial ones. 
As the system crosses the nominal resonance, the semi-major axes of Ariel and Umbriel quickly shift, placing the mean-motion ratio below the nominal resonance, and thus avoiding entrapment in resonance.
Simultaneously, the eccentricity of Ariel quickly drops, while the eccentricity of Miranda grows. 
The eccentricity of Umbriel, Titania, and Oberon, as well as the inclinations of all the satellites, appear to be unaffected by the resonance crossing.
The subsequent tidal damping on the eccentricity of Miranda and Ariel is expected to drive the system close to the present state after 640~Myr (see Sect.~4 in Paper~I).
This example thus confirms that it is possible to cross the 5/3~MMR between Ariel and Umbriel and recover the currently observed system (Table~\ref{table:physical_orbital_parameters}), as long as the initial eccentricity of Ariel before the resonance is $e_1 \ge 0.015$.
It also shows that the obliquities of all satellites settle into the low obliquity Cassini state~1 (see Paper~I) instead of developing high obliquities \citep{Cuk_etal_2020}.

Furthermore, in simulations where the system was captured for an extended period of time, we observed a significant increase in both the eccentricity and inclination values of Miranda. 
This aligns with the findings of \cite{Cuk_etal_2020}, supporting that, despite not being directly involved in the resonance, due to its relatively small mass, Miranda's orbit is severely affected by a lengthening passage through the 5/3 MMR between Ariel and Umbriel.


\section{Conclusion}

The Uranian satellites Ariel and Umbriel have most likely passed through the 5/3~MMR in the past owing to the tidal evolution of their orbits.
In this paper, we have revisited the crossing of this resonance to ascertain the initial conditions that allow the system to evade capture and evolve into the present configuration.

To address this issue, we developed a secular resonant two-satellite model with low eccentricities and low inclinations, using a Hamiltonian approach similar to \citet{Tittemore_Wisdom_1990}.
However, in our model we introduced the spin evolution of the planet and the satellites, and we adopted the total angular momentum of the system as a canonical variable, which is conserved and naturally removes one degree of freedom from the problem. 
We also developed a Hamiltonian extension to include tides based on the weak friction constant time-lag model \citep[eg.][]{Mignard_1979}, which provides the complete tidal evolution for all variables in the problem.

By applying our model to Ariel and Umbriel, we numerically studied in detail the passage through the 5/3~MMR between the two satellites.
At first, we revisited the problem in the planar approximation to better compare with previous studies \citep{Tittemore_Wisdom_1988}.
The crossing of the 5/3~MMR is a stochastic process, and so we performed a large number of numerical simulations covering many different combinations for the initial eccentricities of Ariel and Umbriel.
We observed that the eccentricity of at least one of the satellites must be higher than $0.007$ to avoid capture in resonance. 
However, we have shown that a high eccentricity of Ariel translates into an increased likelihood of evading the 5/3~MMR. 

We then considered the effects of both eccentricity and inclination in the resonance passage.
Once more, we performed a large number of numerical simulations across a mesh of discrete combinations of initial conditions.
We observed that the inclinations do not impact significantly the escape probability and retrieved similar results as in the planar approximation case.

The discrete nature of the previous analysis compelled us to adopt a Monte Carlo methodology, and perform one million simulations with random initial eccentricities, inclinations, and longitudes.
These results have shown that the optimal configuration to enhance the escape probability is achieved when the initial eccentricities are $e_1 \ge 0.01$ and $e_2 \le 0.015$. 
Following the resonance passage, the eccentricities are usually higher than the currently observed ones (Table \ref{table:physical_orbital_parameters}), but they can be quickly eroded owing to tidal damping, in particular for Ariel.
The Monte Carlo method also allowed us to constrain the initial inclinations of the system, we expect that $\inc_1 \le 0.05\degree$ and $\inc_2 \approx 0.08 \degree$, so that the final inclinations after the resonance passage match the current inclinations of the system (Table \ref{table:physical_orbital_parameters}).

Finally, the results obtained with the secular model were inspected using a complete $N-$body model with two and five-satellites.
We did not observe any significant differences between the three models, which confirms that the long-term evolution through the 5/3~MMR is mainly driven by the secular interactions between Ariel and Umbriel.
Therefore, the results regarding Ariel and Umbriel obtained with the secular model are still valid for the complete, five-satellite case. 
We also observed that the orbits of Titania and Oberon do not show any perceptive modifications when the resonance is shortly skipped.
However, as in \citet{Cuk_etal_2020}, we realize that the orbit of Miranda can be severely excited during the resonance entrapment, increasing both the eccentricity and inclination.
In addiction, the final eccentricity and inclination values observed with the five-satellite simulations closely match with the values estimated in Paper~I as necessary to replicate the current architecture of the system \citep{Gomes_Correia_2024p1}.

To accomplish the relatively high initial eccentricity of Ariel and inclination of Umbriel, some past dynamical events must have occurred that excited both values. 
Since tides are very inefficient to damp the inclination, Umbriel's value can be attributed to the possible passage through the past 3/1~MMR between Miranda and Umbriel \citep{Tittemore_Wisdom_1989}. 
As for the Ariel's initial eccentricity, the strong eccentricity damping requires that the excitation must have occurred shortly before the encounter with the 5/3~MMR.
The best candidates are the third order 7/4~MMR resonance between Ariel and Umbriel (see Fig.~1 in Paper~I) or some first order three-body MMR, which can also excite the eccentricity \citep[eg.][]{Petit_2021}.
Therefore, future work on the past dynamical evolution of the Uranian satellite system should try to address this point.

\section*{Acknowledgements}
This work was supported by COMPETE 2020 and by
FCT - Funda\c{c}\~ao para a Ci\^encia e a Tecnologia, I.P., Portugal, 
through the projects
SFRH/BD/143371/2019,
GRAVITY (PTDC/FIS-AST/7002/2020),
ENGAGE SKA (POCI-01-0145-FEDER-022217), and
CFisUC (UIDB/04564/2020 and UIDP/04564/2020, with DOI identifiers 10.54499/UIDB/04564/2020 and 10.54499/UIDP/04564/2020, respectively).
We acknowledge the Laboratory for Advanced Computing at University of Coimbra (\href{https://www.uc.pt/lca}{https://www.uc.pt/lca}) for providing the resources to perform the numerical simulations.



\bibliographystyle{cas-model2-names}

\bibliography{bibliography.bib}

\begin{appendix}

\section{Conservative Hamiltonian terms}\label{sec:Conservative_Hamiltonian_terms}
We note that,

\begin{equation}
    \Ka= -\frac{p}{2}\frac{\mu_1^2\beta_1^3}{\Gamma_1^3}+\left(1+\frac{p}{2}\right)\frac{\mu_2^2\beta_2^3}{\Gamma_2^3} \, ,
\end{equation}

\begin{equation}
    \Kb= -\frac{3}{2}\left(\frac{p^2}{4}\frac{\mu_1^2\beta_1^3}{\Gamma_1^4}+\left(1+\frac{p}{2}\right)^2\frac{\mu_2^2\beta_2^3}{\Gamma_2^4}\right) \, ,
\end{equation}

\begin{equation}\label{Oaterm}
    \Oa=\frac{3}{2}J_2R^2  \left(-\frac{(2p+q)}{q}\frac{\mu_1^4\beta_1^7 }{\Gamma_1^7}+\frac{2 (p+q)}{q}\frac{\mu_2^4\beta_2^7 }{\Gamma_2^7}\right) \, ,
\end{equation}
\begin{equation}\label{Obterm}
    \Ob=
        \frac{3}{2}J_2 R^2  \left(-\frac{2p}{q}\frac{\mu_1^4\beta_1^7 }{\Gamma_1^7}+\frac{(2p+q)}{q}\frac{\mu_2^4\beta_2^7 }{\Gamma_2^7}\right) \, ,
\end{equation}

\begin{equation}\label{Octerm}
    \Oc=\frac{3}{2}J_2R^2   \left(-(p-1)\frac{\mu_1^4\beta_1^7 }{\Gamma_1^7}+(p+2)\frac{\mu_2^4\beta_2^7}{\Gamma_2^7}\right) \, ,
\end{equation}
\begin{equation}\label{Odterm}
    \Od=\frac{3}{2}J_2R^2    \left(-p\frac{\mu_1^4\beta_1^7 }{\Gamma_1^7}+(p+3)\frac{\mu_2^4\beta_2^7 }{\Gamma_2^7}\right) \, ,
\end{equation}

\begin{equation}
    \Sa =\frac{\mathcal{G}\mu_2m_1m_2\beta_2^2}{\Gamma_2^3} \left(\frac{\mu_2\beta_2^2}{\mu_1\beta_1^2} \left[\left(1+\frac{p}{2}\right)\frac{\Gamma_1^2}{\Gamma_2^2}+\frac{p}{2}\frac{\Gamma_1}{\Gamma_2}\right]D+1+\frac{p}{2}\right)b^{(0)}_{\frac{1}{2}}(\alpha_0) \, ,
\end{equation}
\begin{equation}
    \Sb=-\frac{\mathcal{G}\mu_2m_1m_2\beta_2^2}{2\Gamma_1\Gamma_2^2}\left[\alpha_0D+\frac{1}{2}\alpha_0^2D^2\right]b^{(0)}_{\frac{1}{2}}(\alpha_0) \, ,
\end{equation}
\begin{equation}
    \Sc=-\frac{\mathcal{G}\mu_2m_1m_2\beta_2^2}{2\Gamma_2^3}\left[\alpha_0D+\frac{1}{2}\alpha_0^2D^2\right]b^{(0)}_{\frac{1}{2}}(\alpha_0) \, ,
\end{equation}
\begin{equation}
    \Sd=-\frac{\mathcal{G}\mu_2m_1m_2\beta_2^2}{2\Gamma_2^2\sqrt{\Gamma_1\Gamma_2}}\left[2-2\alpha_0D-\alpha_0^2D^2\right]b^{(1)}_{\frac{1}{2}}(\alpha_0) \, ,
\end{equation}
\begin{equation}
    \Se=\frac{\mathcal{G}\mu_2m_1m_2\beta_2^2}{4\Gamma_1\Gamma_2^2}\alpha_0 b^{(1)}_{\frac{3}{2}}(\alpha_0) \, ,
\end{equation}
\begin{equation}
    \Sf=\frac{\mathcal{G}\mu_2m_1m_2\beta_2^2}{4\Gamma_2^3}\alpha_0 b^{(1)}_{\frac{3}{2}}(\alpha_0) \, ,
\end{equation}
\begin{equation}
    \Sg=-\frac{\mathcal{G}\mu_2m_1m_2\beta_2^2}{2\Gamma_2^2\sqrt{\Gamma_1\Gamma_2}}\alpha_0b^{(1)}_{\frac{3}{2}}(\alpha_0) \, ,
\end{equation}

\begin{equation}
    \Ra=-\frac{\mathcal{G}\mu_2m_1m_2\beta_2^2}{2\Gamma_1\Gamma_2^2}\left(\frac{8p^2+11pq+3q^2}{q^2}+\frac{4p+3q}{q}\alpha_0D\right.
    \left.+\frac{\alpha_0^2}{2}D^2\right)b^{(p+2)}_{\frac{1}{2}}(\alpha_0) \, ,
\end{equation}
\begin{equation}
    \Rb=-\frac{\mathcal{G}\mu_2m_1m_2\beta_2^2}{2\Gamma_2^3}\left(\frac{8p^2+9pq+2q^2}{q^2}+\frac{4p+3q}{q}\alpha_0D\right.
    \left.+\frac{\alpha_0^2}{2}D^2\right)b^{(p)}_{\frac{1}{2}}(\alpha_0) +\Rb^{*} \, ,
\end{equation}
where
\begin{equation}
    \Rb^*=\begin{cases}\frac{9}{4}\frac{\beta_1\beta_2\mu_1\mu_2}{\Gamma_1\Gamma_2^2} \quad \text{if} \quad p=1\\ 0 \quad \text{if} \quad p\neq1 \end{cases} \, ,
\end{equation}
\begin{equation}
    \Rc=-\frac{\mathcal{G}\mu_2m_1m_2\beta_2^2}{\Gamma_2^2\sqrt{\Gamma_1\Gamma_2}}\left(\frac{8p^2+10pq+3q^2}{q^2}+\frac{4p+3q}{q}\alpha_0D\right.
    \left.+\frac{\alpha_0^2}{2}D^2\right)b^{(p+1)}_{\frac{1}{2}}(\alpha_0) \, ,
\end{equation}
\begin{equation}
    \Rd=-\frac{\mathcal{G}\mu_2m_1m_2\beta_2^2}{4 \Gamma_1 \Gamma_2^2} \alpha_0 b^{(p+1)}_{\frac{3}{2}}(\alpha_0) \, ,
\end{equation}
\begin{equation}
    \Re=-\frac{\mathcal{G}\mu_2m_1m_2\beta_2^2}{4 \Gamma_2^3} \alpha_0 b^{(p+1)}_{\frac{3}{2}}(\alpha_0) \, ,
\end{equation}
\begin{equation}
    \Rf=\frac{\Grav \mu_2m_1m_2\beta_2^2}{2\Gamma_2^2\sqrt{\Gamma_1\Gamma_2}}\alpha_0 b^{(p+1)}_{\frac{3}{2}}(\alpha_0)\, ,
\end{equation}
where $D=\frac{\partial}{\partial \alpha_0}$ and $\alpha_0=0.7114$ is equal to $\alpha=a_1/a_2$ at nominal resonance (Eq.\ref{nominal:sma}).

\section{Complete secular dynamics: escape probabilities}
\label{app:escape_probabilities}

\begin{sidewaystable*}[htbp]
	\captionsetup{width=.96\linewidth}
    \caption{Escape probability from the 5/3 Ariel-Umbriel MMR for a mesh of 625 initial $(e_1,e_2,\inc_1,\inc_2)$. Each set encompass a total 1\,000 runs equally distributed over the resonance angle $\sigma$ (Eq.\,(\ref{eq:resonace_argument})). \label{tab:eccentric_inclined_escape_probability}}
    \tiny
    \centering
     \renewcommand{\arraystretch}{1.5} 
    \begin{tabular}{|c|c||c c c c c||c c c c c||c c c c c||c c c c c||c c c c c|}
    \cline{3-27}
    \multicolumn{2}{c|}{} & \multicolumn{5}{c|}{$e_2=\num{1e-5}$} & \multicolumn{5}{c|}{$e_2=\num{5e-3}$} & \multicolumn{5}{c|}{$e_2=\num{1e-2}$} & \multicolumn{5}{c|}{$e_2=\num{1.5e-2}$} & \multicolumn{5}{c|}{$e_2=\num{2e-2}$}\\     
     \cline{2-27}
      \multicolumn{1}{c|}{}& \backslashbox{$\inc_1\,(\degree)$}{$\inc_2\,(\degree)$} & 0.001 & \num{0.05} & \num{0.10} & \num{0.15} & \num{0.20}
                                    & 0.001 & \num{0.05} & \num{0.10} & \num{0.15} & \num{0.20}
                                    & 0.001 & \num{0.05} & \num{0.10} & \num{0.15} & \num{0.20}
                                    & 0.001 & \num{0.05} & \num{0.10} & \num{0.15} & \num{0.20}
                                    & 0.001 & \num{0.05} & \num{0.10} & \num{0.15} & \num{0.20}\\
    \cline{2-27}\noalign{\vskip\doublerulesep\vskip-\arrayrulewidth}\hline
    \multirow{5}{*}{\rotatebox{90}{$e_1=\num{1e-5}$}}&    \num{0.001}    & 0.0 & 0.0 & 0.0 & 0.0 & 0.0 
                                                                            & 0.0 & 0.0 & 0.0 & 0.0 & 0.0
                                                                            & 5.4 & 4.6 & 4.2 & 5.2 & 5.2
                                                                            & 9.8 & 13.9 & 13.2 & 15.0 & 14.2
                                                                            & 5.8 & 11.8 & 13.4 & 15.8 & 14.9\\
                        
    &   \num{0.05}  & 0.0 & 0.0 & 0.0 & 0.0 & 0.0
                        & 0.0 & 0.0 & 0.0 & 0.0 & 0.0
                        & 6.2 & 4.8 & 4.7 & 4.5 & 4.8
                        & 13.9 & 14.1 & 15.7 & 16.2 & 14.2
                        & 13.9 & 14.3 & 13.2 & 14.4 & 15.2\\
        
    &   \num{0.10}  & 0.0 & 0.0 & 0.0 & 0.0 & 0.0
                        & 0.0 & 0.0 & 0.0 & 0.0 & 0.0
                        & 5.8 & 5.6 & 5.5 & 5.9 & 5.7
                        & 15.8 & 14.2 & 14.2 & 15.3 & 12.9
                        & 12.8 & 15.6 & 15.4 & 16.0 & 14.6\\
        
    &    \num{0.15} & 0.0 & 0.0 & 0.0 & 0.0 & 0.0
                        & 0.0 & 0.0 & 0.0 & 0.0 & 0.0
                        & 4.8 & 4.8 & 5.7 & 5.6 & 5.1
                        & 13.9 & 15.6 & 16.7 & 13.3 & 13.5
                        & 16.1 & 16.0 & 14.3 & 15.1 & 17.5\\
        
    &    \num{0.20} & 0.0 & 0.0 & 0.0 & 0.0 & 0.0
                        & 0.0 & 0.0 & 0.0 & 0.0 & 0.0
                        & 5.3 & 5.2 & 5.0 & 4.2 & 4.6
                        & 14.0 & 15.0 & 14.4 & 13.3 & 14.9
                        & 16.1 & 17.1 & 16.9 & 13.8 & 17.2\\
        
        \hline\hline
    \multirow{5}{*}{\rotatebox{90}{$e_1=\num{5e-3}$}}&    \num{0.001}    & 0.0 & 0.0 & 0.0 & 0.0 & 0.0
                                                                            & 0.0 & 0.0 & 0.0 & 0.0 & 0.0
                                                                            & 13.0 & 10.7 & 10.3 & 11.7 & 10.8
                                                                            & 10.3 & 14.3 & 16.8 & 13.5 & 15.4
                                                                            & 4.5 & 12.7 & 13.2 & 13.3 & 12.0\\
        
    &    \num{0.05} & 0.0 & 0.0 & 0.0 & 0.0 & 0.0
                        & 0.0 & 0.0 & 0.0 & 0.0 & 0.0
                        & 11.9 & 10.3 & 11.6 & 11.3 & 8.5
                        & 16.5 & 17.6 & 16.2 & 15.6 & 14.7
                        & 9.8 & 11.5 & 11.9 & 12.3 & 13.0\\
        
    &    \num{0.10} & 0.0 & 0.0 & 0.0 & 0.0 & 0.0
                        & 0.0 & 0.0 & 0.0 & 0.1 & 0.0
                        & 11.0 & 13.2 & 12.4 & 12.5 & 10.9
                        & 14.0 & 15.2 & 16.6 & 16.6 & 15.5
                        & 12.9 & 12.8 & 11.9 & 12.4 & 13.5\\
        
    &    \num{0.15} & 0.0 & 0.0 & 0.0 & 0.0 & 0.0
                        & 0.0 & 0.0 & 0.0 & 0.0 & 0.1
                        & 11.3 & 11.1 & 11.6 & 10.4 & 9.6
                        & 17.6 & 15.4 & 15.1 & 15.5 & 16.7
                        & 11.7 & 14.6 & 14.5 & 12.4 & 14.4\\
        
    &    \num{0.20} & 0.0 & 0.0 & 0.0 & 0.0 & 0.0
                        & 0.0 & 0.0 & 0.0 & 0.2 & 0.0
                        & 10.2 & 11.6 & 11.3 & 11.4 & 10.7
                        & 15.9 & 15.0 & 14.4 & 15.5 & 14.5
                        & 13.0 & 14.0 & 12.8 & 13.5 & 13.8 \\
        
\hline\hline  
    \multirow{5}{*}{\rotatebox{90}{$e_1=\num{1e-2}$}}&  \num{0.001}  & 21.9 & 20.9 & 18.7 & 17.6 & 20.2
                                                                        & 38.5 & 36.5 & 33.3 & 33.6 & 31.9
                                                                        & 23.3 & 32.8 & 34.4 & 36.6 & 34.5
                                                                        & 12.1 & 19.5 & 20.5 & 26.3 & 22.4
                                                                        & 6.2 & 9.6 & 12.4 & 12.9 & 14.1\\
                        
    &    \num{0.05} & 19.0 & 20.8 & 19.7 & 18.3 & 17.5
                        & 31.1 & 32.2 & 32.7 & 32.0 & 29.1
                        & 33.5 & 33.5 & 34.3 & 36.3 & 34.2
                        & 22.8 & 23.4 & 21.4 & 25.8 & 25.7
                        & 11.3 & 12.2 & 12.7 & 10.3 & 14.7\\
                        
    &    \num{0.10} & 18.0 & 22.5 & 19.2 & 20.2 & 15.3
                        & 30.0 & 33.8 & 34.1 & 34.5 & 31.1
                        & 36.3 & 34.7 & 33.9 & 36.3 & 32.7
                        & 23.3 & 22.6 & 23.8 & 25.6 & 25.1
                        & 12.4 & 14.1 & 13.6 & 14.6 & 12.4\\
                        
    &    \num{0.15} & 14.8 & 18.7 & 18.4 & 18.0 & 13.4
                        & 30.4 & 32.0 & 31.9 & 32.1 & 26.6
                        & 33.5 & 35.7 & 35.6 & 33.0 & 32.4
                        & 24.9 & 26.4 & 25.8 & 27.3 & 24.6
                        & 18.6 & 14.8 & 16.0 & 15.2 & 15.2\\
                        
    &    \num{0.20} & 10.6 & 11.3 & 13.3 & 12.4 & 13.6
                        & 26.9 & 26.9 & 25.2 & 26.3 & 24.8
                        & 33.1 & 35.6 & 30.1 & 34.9 & 30.3
                        & 25.6 & 25.5 & 24.3 & 27.3 & 26.6
                        & 15.3 & 17.4 & 17.4 & 17.1 & 14.5\\
       \hline\hline
    \multirow{5}{*}{\rotatebox{90}{$e_1=\num{1.5e-2}$}}&    \num{0.001}  & 48.7 & 54.8 & 61.1 & 63.3 & 58.8
                                                                            & 58.5 & 67.3 & 69.3 & 68.3 & 71.7
                                                                            & 37.2 & 42.3 & 45.0 & 49.7 & 47.7
                                                                            & 24.6 & 25.4 & 25.6 & 26.4 & 28.9
                                                                            & 15.5 & 15.6 & 16.0 & 17.7 & 19.1\\
        
    &    \num{0.05} & 61.7 & 62.8 & 63.1 & 62.4 & 61.6
                        & 68.8 & 69.3 & 68.4 & 70.4 & 69.0
                        & 43.3 & 42.9 & 44.6 & 48.6 & 49.9
                        & 25.6 & 23.7 & 25.3 & 24.1 & 29.1
                        & 15.8 & 16.4 & 16.0 & 17.7 & 17.3\\
        
    &    \num{0.10} & 60.7 & 59.3 & 58.8 & 59.3 & 59.1
                        & 68.1 & 68.2 & 68.8 & 67.7 & 64.9
                        & 49.3 & 49.6 & 51.3 & 46.7 & 51.4
                        & 25.4 & 26.7 & 28.9 & 31.2 & 29.2
                        & 16.8 & 17.1 & 17.9 & 17.4 & 16.2\\
        
    &    \num{0.15} & 58.8 & 57.9 & 55.5 & 55.3 & 55.1
                        & 63.5 & 61.3 & 63.1 & 63.6 & 61.1
                        & 50.5 & 50.8 & 49.6 & 50.6 & 53.5
                        & 29.6 & 32.6 & 37.8 & 29.4 & 29.7
                        & 17.2 & 19.1 & 17.3 & 19.5 & 17.2\\
        
    &    \num{0.20} & 50.9 & 53.8 & 52.7 & 48.6 & 51.4
                        & 60.9 & 58.4 & 59.3 & 55.8 & 57.1
                        & 50.5 & 49.6 & 51.1 & 50.9 & 51.0
                        & 32.5 & 31.3 & 29.2 & 30.4 & 33.1
                        & 19.7 & 20.4 & 21.0 & 18.2 & 19.4\\
       \hline\hline
    \multirow{5}{*}{\rotatebox{90}{$e_1=\num{2e-2}$}}&    \num{0.001}    & 60.4 & 66.1 & 72.0 & 74.4 & 73.9
                                                                            & 74.0 & 74.9 & 77.3 & 77.7 & 79.4
                                                                            & 44.1 & 48.5 & 47.8 & 52.7 & 53.2
                                                                            & 36.0 & 40.0 & 35.9 & 36.7 & 36.4
                                                                            & 24.7 & 24.5 & 24.5 & 24.8 & 24.5\\
        
    &    \num{0.05} & 68.8 & 68.3 & 68.5 & 76.2 & 77.4
                        & 77.7 & 79.0 & 78.7 & 79.1 & 80.2
                        & 50.0 & 48.6 & 58.8 & 49.4 & 51.4
                        & 39.4 & 36.9 & 37.9 & 40.0 & 36.3
                        & 25.5 & 25.0 & 26.6 & 25.8 & 25.9\\
        
    &    \num{0.10} & 76.7 & 76.2 & 75.7 & 78.3 & 76.6
                        & 81.5 & 78.5 & 82.1 & 78.9 & 80.5
                        & 52.2 & 47.8 & 51.0 & 49.7 & 51.9
                        & 39.7 & 39.3 & 40.4 & 40.9 & 37.0
                        & 27.1 & 23.8 & 27.6 & 23.0 & 26.6\\
        
    &    \num{0.15} & 77.8 & 75.3 & 75.3 & 75.0 & 73.1
                        & 78.8 & 80.3 & 80.2 & 77.5 & 81.0
                        & 50.6 & 55.7 & 52.9 & 54.7 & 54.8
                        & 38.0 & 38.6 & 38.5 & 42.7 & 39.0
                        & 24.7 & 25.9 & 23.7 & 27.0 & 24.0\\
        
    &    \num{0.20} & 74.8 & 73.5 & 73.9 & 70.7 & 71.1
                        & 75.4 & 75.0 & 79.1 & 76.9 & 76.4
                        & 57.0 & 58.5 & 55.0 & 53.3 & 52.4
                        & 37.2 & 39.3 & 38.6 & 38.5 & 38.8
                        & 24.0 & 24.9 & 25.9 & 24.5 & 25.9\\
                        \hline  
    \end{tabular}
     \renewcommand{\arraystretch}{1} 
\end{sidewaystable*}

\end{appendix}

\end{document}